\documentclass[prd,aps,twocolumn]{revtex4}

\usepackage{amssymb,graphicx}

\begin{document}

\title{3D simulations of Einstein's equations: symmetric
hyperbolicity, live gauges and dynamic control of the constraints}

\author{Manuel Tiglio$^{1,2}$, Luis Lehner$^1$, and David Neilsen$^1$.}

\affiliation{$1$ Department of Physics and Astronomy, Louisiana State
University, 202 Nicholson Hall, Baton Rouge, LA 70803-4001\\
$2$ Center for Computation and Technology, 302 Johnston Hall, Louisiana State
University, Baton Rouge, LA 70803-4001}

\begin{abstract}
We present three-dimensional simulations of Einstein equations
implementing a symmetric hyperbolic system of equations with dynamical
lapse. The numerical implementation makes use of techniques 
that guarantee 
linear numerical stability for the associated initial-boundary value problem. 
The code is first tested with a gauge wave solution, where rather larger amplitudes and 
for significantly longer times are obtained with respect to other state of
the art implementations. 
Additionally, by minimizing a suitably defined energy for the
constraints in terms of free constraint-functions in the formulation one can
dynamically single out preferred values of these functions for the problem at hand. We apply 
the technique to fully three-dimensional simulations of a stationary black hole spacetime with 
excision of the singularity, 
 considerably extending the lifetime of the simulations. 
\end{abstract}

\maketitle

\section{Introduction}

The construction of an accurate and stable numerical implementation of
Einstein equations in settings such as three dimensional binary black hole collisions
represents a major challenge. Indeed, although significant progress
has been achieved in the last few years (see ~\cite{3dbh} and references therein),  
the goal remains
elusive. In practice, numerical or continuum instabilities often arise,
rendering particular simulations of little use after some time, or at high
enough resolutions. Due to the complicated set of equations one is dealing
with, coupled to the often scarce computational resources, tracking down
the source of problems is usually difficult. Faced with this situation, a
possible strategy is to make use of techniques that systematically control
different aspects of the problem under consideration.  One possibility for
such a strategy can be achieved by proceeding along the following lines:

 First, as has been emphasized in a number of works \cite{ibvp}, by
 choosing a well posed
initial-boundary value problem (IBVP), which is a necessary condition for
a numerically stable implementation (see, for example, \cite{gko,stability}). A symmetric 
hyperbolic system with maximally
dissipative boundary conditions is an example of a system that defines
such a well posed problem. Furthermore, the boundary conditions should
not only define a well posed initial value problem but also must
conform to the physical situation in mind. For instance, they 
should preserve the constraints and have minimal spurious influence on the solution.

Second, by translating the previous analytically-framed considerations
into the numerical arena. That is, by constructing a numerically stable
scheme for the IBVP under consideration. One way of doing so is by
constructing difference operators and imposing the discrete boundary
conditions in a way such that the steps followed at the continuum to
show well posedness can be reproduced at the discrete level \cite{gko,olsson,exc}.

Third, by implementing the equations so that spurious growth in time of the solution
is removed or minimized. A numerically stable implementation
need not free the simulation from errors that at fixed resolution grow 
fast in time (although at fixed time they would go away with resolution).  
There are a number of possible alternatives for this effect to be minimized.

One option is to achieve semi-discrete or discrete {\it strict stability},
i.e. given a sharp energy estimate at the continuum, to discretize in
a way such that the estimate also holds at the semi-discrete or discrete
levels \cite{olsson,sbp2}. In this way, growth in the numerical
solution that is not called for by the continuum system is ruled out. In
this strategy one must count with a sharp energy estimate for the problem
at hand, such as those of \cite{friedichstablemink,chrisklain,rodklain}.

An alternative and/or complementary way is to consider the addition of a
small amount of artificial dissipation (in a way that does not spoil  the
available discrete energy estimates and numerical stability). Sometimes
this is enough to partially or completely rule out errors growing fast
in time, and in addition it helps to control high frequency modes. 

There are other cases, in which a sharp estimate is not available, and
for which the addition of some artificial dissipation does not rule out
 undesired growth in the solution either. This happens quite often in
 evolutions of Einstein's equations in the strong field regime. One
possibility in such a case is  to realize that even though a sharp
energy estimate for the main evolution system might not be available,
 an ideal one for the subsidiary system that describes how the
constraints propagate is trivially obtained \cite{dyn}. Namely,
at the continuum one would like the constraints when perturbed to
(for example) remain constant as a function of time, or decay to zero
\cite{lambda}. Similarly, at the discrete level one may want constraint
violations to remain close to their initial, discretization
value. Once some desired estimate for the constraints is chosen, one
can enforce it in a number of ways.  One of them is to dynamically
redefine the equations during evolution off of the constraints surface \cite{dyn}.

In this paper we present results obtained with a fully non-linear code
that evolves Einstein's equations in a three-dimensional setting, analyzing
implementations of techniques that ensure some of the desired properties
just discussed.  The particular formulation of the equations  that we use is a symmetric
hyperbolic one with a dynamical gauge condition (more precisely, a
slight generalization of the Bona-Masso slicing conditions) presented
in Ref. \cite{st}, and summarized in Section \ref{formulation}.  In Section
\ref{minimization} details on how free constraint-functions in the formulation
can be dynamically adjusted to ensure minimal growth of some energy, or
norm, associated with constraints, is discussed in the context of the
symmetric hyperbolic formulation here used.  Section \ref{preliminar}
briefly summarizes the numerical techniques used in this paper, already presented in Ref. \cite{exc},  
and the details of the test-beds here studied. One of these test-beds is the study of a periodic gauge wave,  
presented in Section \ref{gauge_wave}. There we show that the use of a symmetric hyperbolic formulation 
and a small amount of artificial dissipation suffices to evolve this solution with rather large amplitudes 
and for long times.  
The other test-bed that we study is a non-spinning, stationary black hole with excision 
of the singularity and dynamic minimization of the constraints' growth. A detailed analysis is presented in 
Section \ref{bh}, discussing several issues relevant to the constraint minimization technique and the results of 
fully three-dimensional simulations whose lifetime is considerably extended by making use of this 
technique. Section \ref{final} summarizes and discusses the main 
lessons of this work and points out possible extensions of it.

\section{The symmetric hyperbolic formulation used} \label{formulation}

In this paper we use the symmetric hyperbolic formulation of 
the Einstein equations admitting a dynamical lapse introduced in~\cite{st}.
This system has thirty-four variables, including: the three metric, $g_{ij}$, 
the extrinsic curvature, $K_{ij}$, and the lapse, $N$.  Further,
 variables $d_{kij}$ and $A_i$ are constructed from 
the spatial derivatives of $g_{ij}$ and $N$, respectively,
and introduced as independent variables to make the system first order 
in space.  When all constraints are satisfied these variables satisfy
$d_{kij}=\partial_k g_{ij}$ and $A_i=N^{-1}\partial_iN$.
The evolution equations in this formulation are:
\begin{eqnarray}
\partial_0 g_{ij} &=&  -2K_{ij} \; , \label{gdot}\\
\partial_0 K_{ij} &=& R_{ij} - \frac{1}{N} \nabla_i\nabla_j N - 
  2 K_{ia} K^a_{\; j} + K K_{ij} \nonumber \\
  && + \gamma(x^\mu)\, g_{ij}C + \zeta(x^\mu)\, g^{ab}C_{a(ij)b} \; , 
\label{kdot}\\
\partial_0 d_{kij} &=& -2\partial_k K_{ij} - 2A_k K_{ij}
  + \eta(x^\mu)\, g_{k(i}C_{j)} \nonumber\\
  && + \chi(x^\mu)\, g_{ij}C_k , \label{ddot} \\
\partial_0 N &=& - F(N,K,x^{\mu} ) + S(x^\mu) \label{ndot} \\
\partial_0 A_i &=& -\frac{\partial F(N,K,x^\mu)}{\partial N} A_i 
  - \frac{1}{N}\frac{\partial F(N,K,x^\mu)}{\partial K}\partial_i K\nonumber\\
 && - \frac{1}{N}\frac{\partial F(N,K,x^\mu)}{\partial x^i} 
    + \xi(x^\mu) \,  C_i\, , \label{adot}
\end{eqnarray}
where we define $\partial_0 = N^{-1}\left( \partial_t - \pounds
_{\beta } \right)$.  The Ricci tensor in Eq.~(\ref{kdot}) is written as
\begin{eqnarray*}
R_{ij} &=& \frac{1}{2} g^{ab} \left(
 -\partial_a d_{bij} + \partial_a d_{(ij)b} + \partial_{(i} d_{|ab|j)} 
 - \partial_{(i} d_{j)ab} \right) \\
& & + \frac{1}{2} d_i^{\;\; ab} d_{jab} 
    + \frac{1}{2}(d_k - 2b_k)\Gamma^k_{\; ij}
    - \Gamma^k_{\; lj} \Gamma^l_{\; ik}\; , 
\label{eq:ricci}
\end{eqnarray*}
where $b_j \equiv d_{kij} g^{ki}$, $d_k \equiv d_{kij} g^{ij}$, and
\begin{displaymath}
\Gamma^k_{\; ij} = \frac{1}{2} g^{kl} \left( 2d_{(ij)l} - d_{lij} \right),
\end{displaymath}
Finally, the second order derivatives of $N$ that appear in Eq.~(\ref{kdot})
are calculated as
\begin{displaymath}
\frac{1}{N} \nabla_i\nabla_j N = \partial_{(i} A_{j)} - \Gamma^k_{\;
ij} A_k + A_i A_j\; .
\end{displaymath}

The slicing condition, eq.~(\ref{ndot}), contains two functions,
$F(N,K,x^\mu)$ and $S(x^\mu)$. The function $F$ may be any function of
the lapse, the trace of the extrinsic curvature, $K = g^{ij}K_{ij}$, and
the spacetime coordinates, with the condition that $\partial F/\partial K
> 0$.  The function $S$ is a gauge source function, and is specified {\it a priori}
but in an arbitrary way as a function only of the spacetime coordinates.
This slicing is a generalization of the Bona--Masso slicing conditions,
 obtained by setting $S = 0$ and $F(N,K,x^\mu) = f(N)K$.
Moreover, choosing $S=0$ and $f=N$ gives the harmonic time slicing
condition, or a generalized harmonic condition if $S\ne 0$. The time harmonic slicing condition, or its 
generalized form, is the choice used for all runs in this paper.  Finally, the shift,
$\beta^i(x^\mu)$, must also be specified
{\it a priori} as an arbitrary function of spacetime.

The Einstein equations are a constrained system, and the  evolution 
equations here considered are not only subject to the physical constraints, the Hamiltonian 
and momentum ones, but also non-physical constraints that come from
introducing first-order variables.  The Hamiltonian constraint is
\begin{equation}
C=(R-K_{ab}K^{ab}+K^2)/2,
\end{equation}
where $R = g^{ij}R_{ij}$ and the Ricci tensor given by Eq~(\ref{eq:ricci}).
The momentum constraints, $C_i = \nabla ^aK_{ai} -\nabla_i K$, are
\begin{equation}
C_i = g^{ab}\left( \partial_a K_{bi} - \partial_i K_{ab} \right)
    + \frac{1}{2}(d^k - 2b^k)K_{ki} + \frac{1}{2} d_i^{\;\; ab} K_{ab}\; .
\end{equation}
Finally, the non-physical constraints, $C_{A_i},C_{kij}$ and $C_{lkij}$, are
defined as
\begin{eqnarray*}
C_{A_i} &=& A_i-N^{-1}\partial_iN \; ,\\
C_{kij} &=& d_{kij}-\partial_k g_{ij} \; , \\  
C_{lkij} &=& \partial_{[l}d_{k]ij}.
\end{eqnarray*}

The Einstein equations resulting from the 3+1 ADM decomposition are
only weakly hyperbolic.  However, it is possible to manipulate the
principal part of the equations by adding the constraints in specific
combinations to the evolution equations in order to obtain a
strongly or symmetric hyperbolic system of equations \cite{reula}.  In the system here considered, the 
constraints are added to the RHS of Eqs.~(\ref{gdot}--\ref{adot}), and 
the spacetime constraint-functions $\{\gamma, \zeta, \eta, \chi, \xi \}$ are introduced as
multiplicative factors to the constraints.  Requiring the evolution
system to be symmetric hyperbolic imposes algebraic conditions on 
these factors, as discussed below, and they are not treated as 
completely independent.  Typically these factors
are taken to be constant {\it parameters}, however this restriction is actually not needed 
for strong or symmetric hyperbolicity of the system to hold. Moreover, we wish to exploit some
freedom in choosing these constraint-functions to minimize the effect of constraint
violating modes that may appear in the solution~\cite{dyn}.  Thus, we choose
the factors to be functions of time but constant in space (future work will concentrate in allowing for space dependence):
$\{\gamma(t), \zeta(t), \eta(t), \chi(t), \xi(t) \}$.  Therefore, here we will 
refer to these factors as {\it constraint-functions} rather than {\it parameters}. 

The characteristic speeds of the system are 
$\beta^i n_i$, $\pm N + \beta^i n_i$, $\pm N\sqrt{\lambda_i} + \beta^i n_i$, 
with 
\begin{eqnarray*}
\lambda _1 &=& 2\sigma_{eff}\, ,\\
\lambda _2 &=& 1 + \chi - \frac{1}{2}(1 + \zeta)\eta + \gamma (2-\eta + 2\chi),\\
\lambda _3 &=& -\frac{1}{4}\chi - \frac{1}{8}(3\zeta - 1)\eta - \frac{1}{2}\xi\,.
\end{eqnarray*}
and $\sigma_{eff}  = (\partial F/ \partial K)/(2N)$.

There are two strongly hyperbolic multiple constraint-function 
families with $\lambda_2=1$ and $\lambda_3=1$, which correspond 
to propagation speeds along the light cone and the hypersurface normal.
One such family has three free constraint-functions $\{ \gamma(t) , \zeta(t) , \eta(t) \}$:
\begin{eqnarray*}
\gamma & \neq & -\frac{1}{2}\,, \\
\chi & = & \frac{(1+\zeta )\eta - 2 \gamma (2-\eta)}{2(1+2\gamma )} \,,\\
\xi & = & -\frac{1}{2}\chi - \frac{1}{4}(3\zeta -1)\eta - 2\,,
\end{eqnarray*}
A second family has only two free constraint-functions $\{ \zeta(t) , \chi(t) \}$:
\begin{displaymath}
\gamma = -\frac{1}{2}\, , \quad
\zeta \eta = -2 \, , \quad
\xi= -\frac{1}{2}\chi + \frac{1}{4}\eta - \frac{1}{2} \; .
\end{displaymath}

In this paper we set $\zeta =-1$ to simplify the calculation of the
characteristic variables needed to impose maximally dissipative
boundary conditions.  This gives a symmetric hyperbolic system with one
free constraint-function $\chi(t)$
\begin{equation}
\mbox{Single constraint-function system}
\left\{
\begin{array}{ccc}
\gamma &=& -\frac{1}{2} \\
\zeta &= & -1 \\
\eta &=& 2 \\
\xi &=& -\frac{\chi }{2} \\
\chi &\ne& 0
\end{array}
\right. \label{monoparametric}
\end{equation}
and another symmetric system with two 
varying constraint-functions $\{ \eta(t), \gamma(t) \neq -1/2 \}$:
\begin{equation}
\mbox{Two constraint-function system}
\left\{
\begin{array}{ccc}
\zeta &= & -1 \\
\chi &=& -\frac{\gamma (2-\eta)}{1+2\gamma} \\
\xi &=& -\frac{\chi}{2} + \eta -2 \\
\gamma &\ne & -\frac{1}{2} \\
\eta & & 
\end{array}
\right. \label{biparametric}
\end{equation}
One can show that these families  not only are strongly
hyperbolic as shown in ~\cite{st} but, as mentioned above, 
also symmetric hyperbolic~\cite{sarbach}.

\section{Dynamic control of the constraints' growth.} \label{minimization}

The formulation of the Einstein equations summarized in the previous section is made symmetric hyperbolic 
 by adding constraints to the evolution
equations multiplied by the time-dependent constraint-functions.  Requiring
that the propagation speeds be along the light cones or $t=$constant hypersurfaces normal,
and a further simplification obtained by setting $\zeta=-1$, results in 
two families of equations.   The first has a single free constraint-function,
$\chi$, and the second has two constraint-functions, $\{\gamma, \eta\}$.
Several papers have been presented showing that the long-term stability of 3D numerical 
simulations is 
extremely sensitive to these constraint-functions, even if symmetric hyperbolicity is guaranteed 
independently of the values these  constraint-functions take \cite{sym_stability}. 
 Recently, in~\cite{dyn} a method  to 
dynamically choose these constraint-functions by minimizing the constraint
growth during the evolution has been presented.  We here include a brief summary
of this method, and discuss its particular application to the black hole
runs presented later in this paper.

Consider a system of hyperbolic equations with constraint terms, $C_c$,
written schematically
\begin{equation}
\dot{u}_a = \sum_bA^b(u,t,\vec{x})\partial_b u_a 
    + B_a(u,t,\vec{x}) + \sum_c\mu_{ac} C_c(u,\partial_ju)
\label{linearc}\;, 
\end{equation}
where $u_a$, $B_a$ and $C_c$ are vector valued functions, and $\mu_{ac}$ is
a matrix (generally not square) that is a function of the spacetime.  (Note
 that in this section $C_c$ represents a vector function of general constraint
variables, and not specifically the momentum constraint introduced in the
previous section.)  Here the indices $\{a,b,c\}$ range over the 
size 
of each vector or matrix function, while the indices $\{i,j,k\}$ will label
points on a discrete grid.
We then define an {\it energy} or {\it norm} of the discrete 
constraint variables, e.g.,
\begin{equation}
{\cal N}(t) = \frac{1}{2n_xn_yn_z} \sum_{c}\sum_{ijk} C_c(t) ^2 \label{constraintnorm} \, ; 
\end{equation}
where $n_x,n_y,n_z$ are the number of points in the $x,y,z$ directions, and where 
we have omitted the grid indices $\{i,j,k\}$ to simplify the notation. 
The time derivative of the norm can be calculated using Eq.~(\ref{linearc})
\begin{equation}
\dot{\cal N} = {\cal I}^{hom} 
 + \mbox{Tr} (\mu   {\cal I}^{\mu }) \label{split}
\end{equation}
and therefore can be known in closed form provided the 
matrix valued sums
\begin{eqnarray}
 {\cal I}^{hom} &=& 
\sum_{ijk} \sum_{a,b} \frac{C_a}{n_xn_yn_z}\left[\frac{\partial C_a}{\partial u_b}+
\sum_k\frac{\partial C_a}{\partial D_k u_a}D_k \right] \times
 \nonumber \\
&&  \left[\sum_c(A^cD_cu_b) +B_b\right]  \label{split1}  \\
{\cal I}^{\mu}_{bc} &=&  \sum_{ijk} \sum_{a} 
    \frac{C_a}{n_xn_yn_z} \times  \nonumber \\
&& \left[\frac{\partial C_a} {\partial u_b}+  
 \sum_k\frac{\partial C_a}{\partial D_k u_b}D_k \right]C_c \label{split2}
\end{eqnarray}
are computed during evolution; where $D_i$ is the discrete derivative
approximation to $\partial_i$. We then use 
the dependence of the energy growth on the free constraint-functions 
to achieve some desired behavior for the
constraints, i.e., solving Eq.~(\ref{split}) for $\mu_{ac}$. 
For example, if we choose \footnote{There is a slight abuse of notation here, in the sense that 
$a$ does not denote an index, as before. Similarly, the subscript in $n_a$ indicates that the quantity 
is related to $a$ through Eq.(\ref{a}).}
\begin{equation} 
\dot{\cal N} = -a {\cal N}, \qquad a>0,   \label{edot}
\end{equation}
any violation of the constraints will decay exponentially 
\begin{equation}
{\cal N}(t+\triangle t) = {\cal N}(t)e^{-a\triangle t} \label{decay}\;.
\end{equation}
As discussed in~\cite{dyn}, one good option among many others seems to
be choosing a tolerance $T$ value for the norm of the constraints that
is close to its initial, discretization value, and solving for $\mu_{ac}$
such that the constraints decay to this tolerance value after a given
relaxation time.  This can be done by adopting an  $a$ such that after time
$n_a \triangle t$ the constraints have the value $T$. Replacing ${\cal
N}(t+\triangle t)$ by $T$ in equation (\ref{decay}) and solving for
$a$ gives
\begin{equation}
a(t) = -\frac{1}{n_a\triangle t}\ln{\left(\frac{T}{{\cal N}(t)}\right)} 
\; . \label{a}
\end{equation}
If one then solves 
\begin{equation}
\dot{\cal N} = -a {\cal N} = {\cal I}^{hom} 
   + \mbox{trace}(\mu \times {\cal I}^{\mu }) \label{eq_for_mu}
\end{equation}
for $\mu$, with $a$ given by Eq.(\ref{a}), the value of the norm ${\cal
N}(t+n_a\triangle t)$ should be $T$, independent of its initial value.

In principle, Eq.(\ref{eq_for_mu}) has non-unique solutions, since
the equation is scalar and $\mu_{ac}$ is a matrix. As discussed in
Section \ref{runs}, this non-uniqueness is sometimes crucial for making this a
practical method for controlling growth in the constraints.

We also note that the
technique discussed in this section can be implemented without affecting
symmetric hyperbolicity~\cite{dyn}.

Finally, we now describe how $\dot{\cal N}$ is calculated
for the symmetric hyperbolic families of the Einstein equations used 
in this paper. The time 
derivative of the energy for the constraints is
\begin{equation}
\dot{\cal N} = \tilde{{\cal I}}^{hom} + \chi \tilde{{\cal I}}^{\chi}  
    + \omega {\tilde {\cal I}}^{\omega} + \gamma \tilde{{\cal I}}^{\gamma} 
    +  \xi \tilde{{\cal I}}^{\xi}  + \eta \tilde{{\cal I}}^{\eta} 
\label{energy_dot}
\end{equation}
For the single-function family, Eq.~(\ref{monoparametric}),
$$
\dot{\cal N} = \tilde{{\cal I}}^{hom} + \chi (\tilde{{\cal I}}^{\chi}  
   - \frac{1}{2} {\tilde {\cal I}}^{\omega} ) 
   - \frac{1}{2} \tilde{{\cal I}}^{\gamma} -\tilde{{\cal I}}^{\xi} 
   + 2\tilde{\cal I}^{\eta}  
$$
That is, 
\begin{equation}
\dot{\cal N} = {\cal I}^{hom} + \chi {\cal I}^{\chi}  \label{edot_mono}, 
\end{equation}
with
\begin{eqnarray*}
{\cal I}^{hom} &=& \tilde{\cal I}^{hom} - \frac{1}{2} \tilde{\cal I}^{\gamma} 
    -\tilde{\cal I}^{\xi} + 2\tilde{\cal I}^{\eta} \\
{\cal I}^{\chi } &=& \tilde{\cal I}^{\chi}  
    - \frac{1}{2} {\tilde {\cal I}}^{\omega}
\end{eqnarray*}

The evaluation of $\dot{\cal N}$ as a function of $\chi$
is a two-step process. In order to compute the quantities 
${\cal I}^{hom}$ and ${\cal I}^{\chi}$,
so as to obtain the dependence of the time derivative of the energy in
terms of $\chi$, Eq.(\ref{edot_mono}),
{\it two} evaluations of ${\cal N}$ are required to extract the individual 
contributions. 
For example, the homogeneous term
is obtained performing a set of evaluations with $\chi=0$,
$$
{\cal I}^{hom} = \dot{\cal N}(\chi=0)
$$
Once this term is known, ${\cal I}^{\chi }$ is obtained doing a set of 
evaluations with an arbitrary but  non-vanishing 
$\chi=\chi_0$,
$$
{\cal I}^{\chi } = \frac{\dot{\cal N}(\chi_0) -
  {\cal I}^{hom}}{\chi_0} 
$$
This involves at each step an 
evaluation of the right hand side of the evolution
equations and of the spatial derivatives of such right hand side, and
evaluation of the derivatives of the constraints with respect  to the
main variables and with respect to their spatial derivatives (all of
this at each grid point) The constraints, and therefore their
derivatives,  do not depend on the constraint-functions.  Therefore they need to
be computed only once at any given time.

Similarly for the two-function family (\ref{biparametric}): 
the time derivative of the norm of the constraints is
\begin{equation}
\dot{\cal N} = {\cal I}^{hom} + \chi {\cal I}^{\chi} + \gamma
{\cal I}^{\gamma} +  \eta {\cal I}^{\eta}  \label{edot_bi}
\end{equation}
with [c.f. Eq. (\ref{biparametric})]
$$
\chi = -\frac{\gamma (2- \eta)}{1+2\gamma}
$$
and
\begin{eqnarray*}
{\cal I}^{\hom} &=& -\xi \tilde{{\cal I}}^{\xi} -2 \tilde{{\cal I}}^{\omega} \, , \\
{\cal I}^{\chi} &=& \tilde{{\cal I}}^{\chi} -\frac{1}{2} \tilde{{\cal I}}^{\omega }\, ,\\
{\cal I}^{\gamma} &=& \tilde{{\cal I}}^{\gamma }\, ,\\
{\cal I}^{\eta } &=& \tilde{{\cal I}}^{\eta } + \tilde{{\cal I}}^{\omega }\, .
\end{eqnarray*}

At any given time, four sets of evaluations are needed to numerically
compute the quantities ${\cal I}^{hom}, {\cal I}^{\chi}, {\cal
I}^{\gamma}, {\cal I}^{\eta}$ in Eq. (\ref{edot_bi}). For example, as in 
 the single constraint-function case,  the homogeneous term is obtained through
a set of evaluations with $\eta=0=\gamma$,
$$
{\cal I}^{hom} = \dot{N}_c(\eta=0, \gamma=0)\, . 
$$
Once this is known, ${\cal I}^{\eta }$ is obtained doing a set of
evaluations with an arbitrary but non-vanishing $\eta=\eta_0$, and
$\gamma=0$,
\begin{equation}
{\cal I}^{\eta } = \frac{\dot{N}_c(\eta_0, \gamma=0) -
  {\cal I}^{hom}}{\eta_0} \label{edot_eta}
\end{equation}
Two more sets of evaluations are needed in order to construct ${\cal
I}^{\eta }$ and ${\cal I}^{\gamma }$: given $\gamma_1$ and $\gamma_2$
arbitrary but different, $\gamma_1 \ne \gamma_2 $, it is straightforward
to see that
\begin{eqnarray}
{\cal I}^{\gamma } &=& \frac{ (\gamma _2 +2\gamma_1 \gamma_2 )\dot{N}_c(\gamma_1)
  - (\gamma_2 + 2\gamma_1 \gamma_2 )\dot{N}_c(\gamma_1 ) }{2\gamma_1 \gamma_2 (\gamma_1 -
  \gamma_2)} \label{edot_gamma}\\
{\cal I}^{\chi } &=& \frac{(1+2\gamma_1 )(1+2\gamma_2)}{4\gamma_1 \gamma_2 (\gamma_1 -
  \gamma_2)}\left[ \gamma _2 \dot{N}_c(\gamma_1)
  - \gamma_1 \dot{N}_c(\gamma_2 ) \right] \label{edot_chi}
\end{eqnarray}

The quantities ${\cal I}^{\eta}$, ${\cal I}^{\chi}$, ${\cal I}^{\gamma}$
thus obtained are independent of what values $\eta_0,\gamma_1,\gamma_2$
are used in Eqs.~(\ref{edot_eta},\ref{edot_gamma}, \ref{edot_chi}). We
make use of this fact to perform a non-trivial test of
self-consistency in our simulations. Namely, during evolution we
construct these quantities  ${\cal I}^{\eta}$, ${\cal I}^{\chi}$,
${\cal I}^{\gamma}$  using, at each time step, several different values
of $\eta_0,\gamma_1,\gamma_2$, and we check that the result is, indeed,
independent of that choice. We proceed in a similar way with the single-constraint-function family.

\section{Numerical methods and test problems} 
\label{preliminar}

In this section we introduce the numerical methods that we use,
and then outline two physical problems, the gauge wave and a
Schwarzschild black hole,  that we will analyze in
this paper. These spacetimes will be used in our numerical implementation of
Einstein equations written in the first order form detailed
in section II --equations (\ref{gdot}) through (\ref{adot})--.

\subsection{Numerical method}

We use numerical techniques based on the energy method for hyperbolic
equations~\cite{gko}.  
This method allows one to identify numerically  
stable discretizations by construction for initial-boundary value problems for linear, symmetric
hyperbolic systems.  While we focus here on nonlinear Einstein equations,
we note that some numerical instabilities in the Einstein system are also
observed in the linear regime. Methods that are known to 
be numerically stable for the linearized Einstein equations thus function both
as a foundation and guide for moving to the nonlinear problem.
Our numerical scheme uses second order spatial
difference operators that satisfy summation by parts, as well as an extension
of the standard Kreiss--Oliger dissipation operator that takes into account the presence of (inner 
and outer) boundaries. This operator, which we call $Q_d$, 
is added to the right-hand side of the evolution equations, 
$
\dot{u} = (\ldots) + Q_d$, 
with a free (non negative) multiplicative parameter
$\sigma$. Paper~\cite{exc} describes in detail this operator; here to fix ideas we
include it for  the non-excision case. It is, on each direction, 
\begin{eqnarray} 
Q_d u_0 &=& -2\sigma \Delta x D_+^2 u_0,\nonumber\\ 
Q_d u_1 &=& -\sigma \Delta x (D_+^2 -2D_+D_-)u_1,\nonumber\\ 
Q_d u_i &=& -\sigma (\Delta x)^3(D_+D_-)^2 u_i,
           \; \mbox{ for } i=2,\ldots,N-2 \nonumber\\ 
Q_d u_{N-1} &=& -\sigma \Delta x (D_-^2 -2D_+D_-)u_{N-1}, \nonumber\\ 
Q_d u_N &=& -2\sigma \Delta x D_-^2 u_N. \label{eq:KOdiss21} 
\end{eqnarray} 

Maximally dissipative 
boundary conditions are imposed numerically
through projections that are orthogonal in the linear case. 
We use third order Runge-Kutta to integrate
the equations in time.
The computational domain consists of an 
uniform Cartesian grid ($\Delta x= \Delta y = \Delta z = \Delta$).  The
black hole simulations employ a cubical inner boundary to excise the 
singularity from the computational domain.
Unless otherwise stated, the simulations presented throughout the
paper use a dissipative parameter $\sigma =0.03$ and Courant factor
$\lambda =0.5$. This choice for $\sigma $ is motivated by the fact that, at least for the second order wave equation written in 
first order form, it gives the maximum Courant factor allowed by von-Neumann stability (see \cite{exc}).  
For additional information on our numerical
scheme see~\cite{exc}.

Boundary conditions are specified via maximally dissipative boundary
conditions where needed. These are introduced by finding all
the incoming modes and setting the time derivative of these to
zero. Maximally dissipative boundary conditions can be written in
the form 
\begin{equation} 
u^{+} = L u^{-} + g(t,x^A)\, ,
\label{dis}
\end{equation} 
where $L$ is ``sufficiently small,'' such that an energy
estimate for the IBVP can be derived for symmetric hyperbolic systems
in more than one dimension (see, e.g., \cite{kreiss_lorenz}.). The function  $g$
is an {\it a priori} but  arbitrary function of the spacetime coordinates in the boundary and time.
The combination of maximally dissipative boundary conditions and a
symmetric hyperbolic evolution system define a well-posed initial-boundary
value problem.  However, these boundary conditions in general are {\em
not} constraint-preserving.  For the evolutions presented here, setting
the time derivative of {\em all} the incoming modes to zero is consistent
with preserving the constraints at the boundary because the exact solution
is known.  However, this will not hold in general and future work will
concentrate on evolutions with constraint-preserving boundary conditions.

\subsection{Gauge waves}

We first test our numerical method by studying a gauge wave defined by
\begin{equation}
ds^2 = e^{A \sin(\pi (x-t))} (-dt^2 + dx^2) + dy^2 + dz^2 \, ,
\end{equation}
which corresponds to a coordinate transformation in the $(x,t)$ plane,
of flat spacetime.
The analytic solution for the gauge wave is obtained by
setting the gauge source function to zero, $S(x^\mu)=0$, in Eq.~(\ref{adot}):
$-N^2K = \partial_t N$ and the shift $\beta^i=0$. We
adopt periodic boundary conditions to simplify the analysis by
eliminating possible boundary effects.

\subsection{Black Holes}

We then examine in some detail tests with  Schwarzschild
spacetime, which describes a static non-spinning black hole.  The
singularity inside the black hole is excised, which restricts
the possible slicings (surfaces of constant time) that we consider
to those that smoothly penetrate the black hole horizon, such that
we can place an inner boundary inside of the horizon. In the present 
work we consider a Schwarzschild black hole in Kerr-Schild (KS) ---or ingoing
Eddington-Finkelstein--- coordinates.
In these coordinates, the metric of the spacetime is given by the line
element:
\begin{equation}
ds^2=-N^2  dt^2 + g_{ij}(dx^i + \beta^i dt)(dx^j+
\beta^j dt) \, ,
\end{equation}
where,
\begin{eqnarray}
N &=& \left( \frac{r}{r+2m} \right)^{1/2} \, , \\
\beta^i&=& \frac{2m}{r+2m} x^i \, , \\
g_{ij} &=&\delta_{ij} + \frac{2m}{r} \frac{x^i x^j}{r^2} \, .
\end{eqnarray}
The gauge source function $S$ is read off from the exact solution.

\section{Gauge wave simulations} \label{gauge_wave}

The gauge wave is a simple, non-trivial numerical 
test problem, as it is free of boundaries, the amplitude of the fields
can be controlled by  a single parameter, and it does not lead to any
singularity. This solution is used to
compare the performance of different implementations of Einstein's
equations~\cite{appleswithapples}.  Despite its simplicity at the
analytical level, this test illustrates the challenges
associated with the numerical implementation of Einstein equations~
\cite{appleswithapples,bruegmanlast,pavlinkaprivate,kidderprivate}.
In particular, it is often observed that for amplitudes $A \geq 0.01$, the
numerical solutions display exponential growth and loss of convergence.

Since in the cases discussed in this paper 
the analytical solution is known, the convergence factor can be defined as
\begin{equation}
C(F) = ||F_{\triangle}-F_{analytical}||_2/||F_{\triangle/2}-F_{analytical}||_2 \, ;
\end{equation}
with $F$ the variable under consideration,  $F_{\triangle}$ and $F_{analytical}$
its numerical (at resolution $\Delta$) and analytical solutions,  respectively.

In the present context, the gauge wave is used for two purposes.
First, it allows us to probe the stability of our numerical method,
showing its advantages and present limitations.  Second, it
sheds light on possible sources of instabilities or spurious
growth often encountered in these tests. 
In particular, we are able to establish that {\it
constraint violations}, if any, grow quite slowly, allowing one to
accurately follow the system for thousands of crossing times, even for
large values of the amplitude parameter $A$.
We have ran our simulations for a wide range of values for $A$, observing
qualitatively the same results. 

Note that one can readily show  that the constraints, $C$, $C_i$, and $C_{ijkl}$,
for the gauge wave in this formulation are satisfied to round-off
level if the variables depend solely on $(t,x)$. That $C$, $C_i$ are satisfied
follows directly from the fact that  $g_{ij} = (f(x)- 1) \delta_x^i \delta_x^j 
+ \eta_{ij}$,
$K_{ij} = g(x) \delta^i_x \delta^j_x$, $d_{ijk} = h(x) \delta_x^i \delta_x^j 
\delta_x^k$;
with $\{g(x),f(x),h(x)\}$ arbitrary functions. Note that: 
(i) at any $t=const$ hypersurface , $g_{ij}$ describes a flat metric 
and so $R_{ij}=0$ analytically. Furthermore, at the numerical level all terms
in $R_{ij}$ cancel, including those describing the truncation errors in the 
derivatives. 
(ii) $K^2-K_{ij}K^{ij}$ and $K_{ij}-g_{ij}K$ are zero algebraically and so,  
coupled to (i), $C$, $C_i$ are satisfied numerically to round-off level. 
(iii) $C_{ijkl}$ 
is defined by the conmutator of two derivatives; since the only non-trivial 
components
are in the $x$ direction, the only surviving one is $C_{xxxx}=\partial_{[x} 
d_{x]xx}$ which is trivially satisfied.
Consequently, the value of the constraint-function multiplying them in the
right hand side of the evolution equations do not play a significant role. 
This observation holds for the variables depending solely on $(t,x)$, as
mentioned, this is indeed the case throughout our runs.
Therefore, for these tests we adopt the two-constraint-function family 
formulation with fixed values for the constraint-functions: $\eta = \gamma = 
0$.

We concentrate on two non-linear cases with relatively large amplitudes:
$A=0.5$ and $A=1$.  For example, when  $A=0.5$, $g_{xx}$ ranges over the
interval $[0.6,1.7]$, and over $[0.37,2.72]$ when $A=1$.  In \cite{sbp2},
an independent analysis and code for the linear equations around the
gauge wave was presented, illustrating the tendency of the numerical
solution to grow exponentially unless some small amount of dissipation is
added to the RHS. 

The computational domain is chosen to be $ [-1,1]$, and is represented
on a uniform grid with spacing $\triangle = 2/(N-1)$. The gauge wave 
is really a one dimensional problem, with a non-trivial dependence only on 
 $(x,t)$. The solution is constant in the $(y,z)$ coordinates, thus  only 
a few points are need to represent the field in these directions. 
We therefore use $n_x= (80 p + 1)$ points with $p=1,2,4$ in the $x$ direction,
 and five points in each of the $(y,z)$ directions (though tests were performed
 with uniform grids, equally spaced in all directions obtaining exactly the
 same results as expected). In the gauge wave tests, we chose a Courant
 factor of $1/4$ so as to make a more direct comparison with similar tests
 presented in the literature (see for instance \cite{appleswithapples,bruegmanlast,pavlinkaprivate,kidderprivate}).

\subsubsection{Amplitude $A=0.5$}

\begin{figure}[ht]
\begin{center}
\includegraphics*[height=6cm]{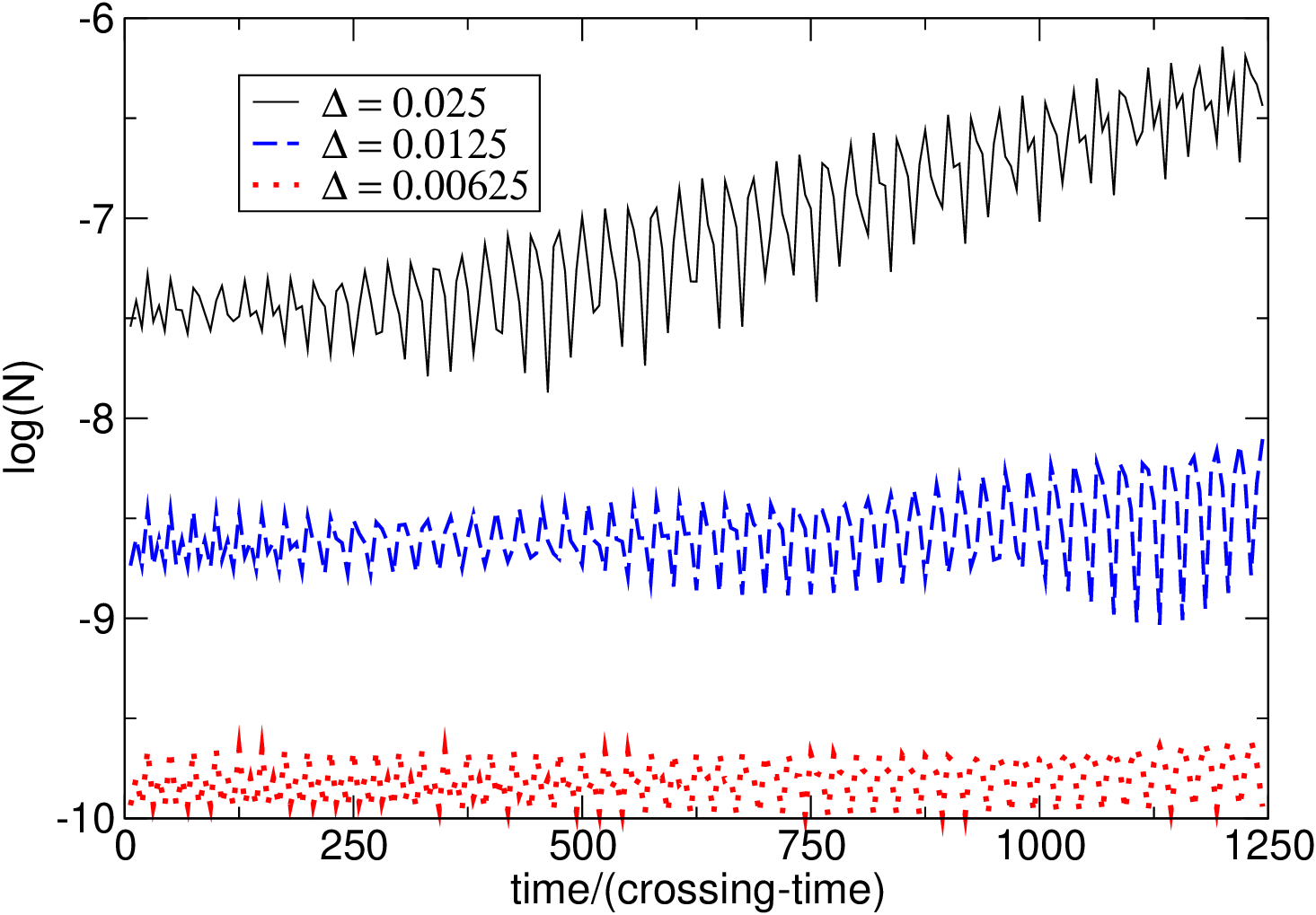}
\caption{This figure shows the logarithm of the energy for the 
constraints, in simulations of the gauge wave with amplitude $A=0.5$.  
Three different resolutions are  used, the coarsest 
run exhibits a slow growth which is negligible at higher resolutions.}
\label{logenergy05}
\end{center}
\end{figure}

\begin{figure}[ht]
\begin{center}
\includegraphics*[height=6cm]{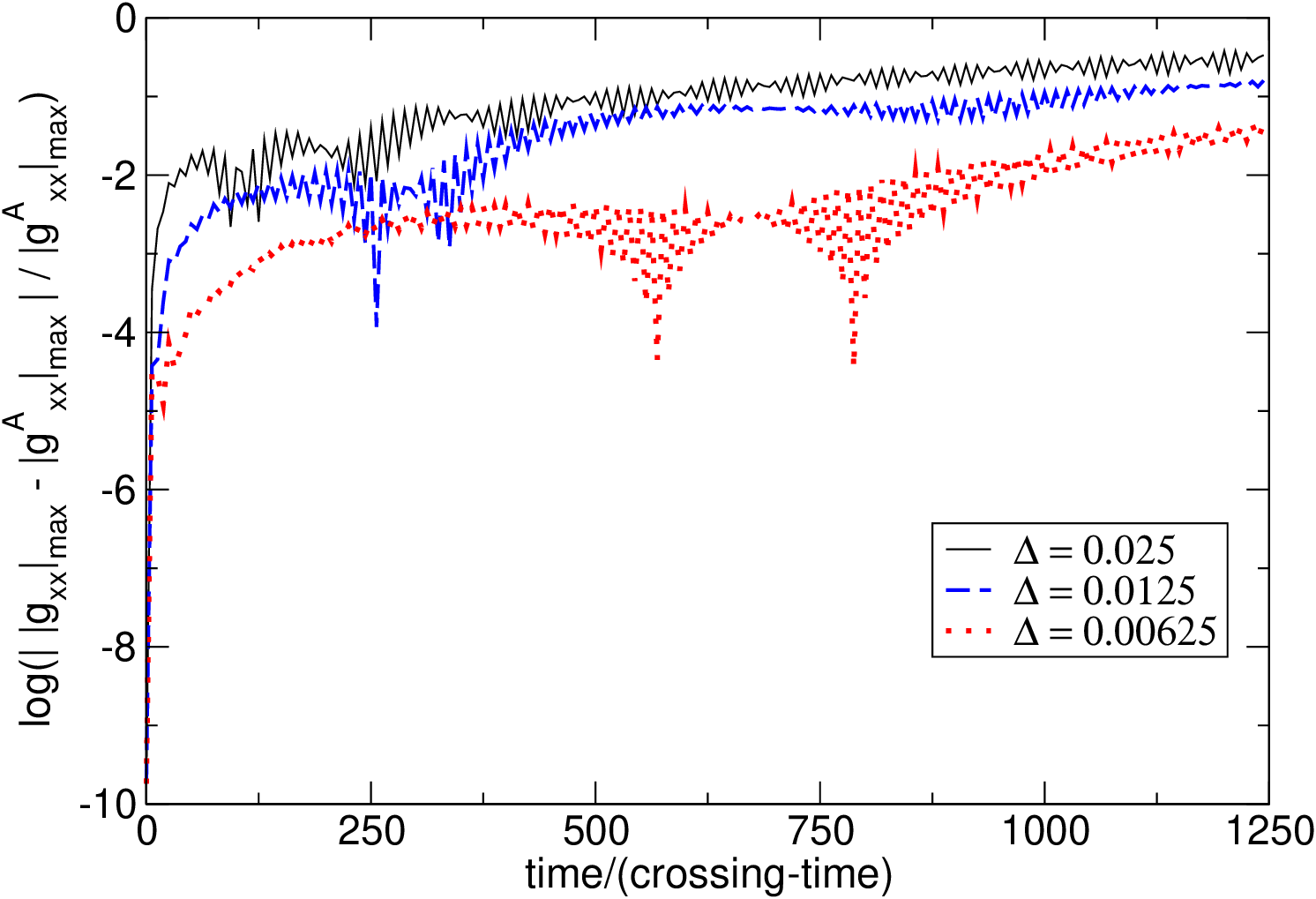}
\caption{The logarithm of the relative error in the maximum value attained 
by $g_{xx}$ (compared to 
its exact value) versus number of crossing times 
for the simulations of Figure~\ref{logenergy05}.}.
\label{log_g05}
\end{center}
\end{figure}

\begin{figure}[ht]
\begin{center}
\includegraphics*[height=6cm]{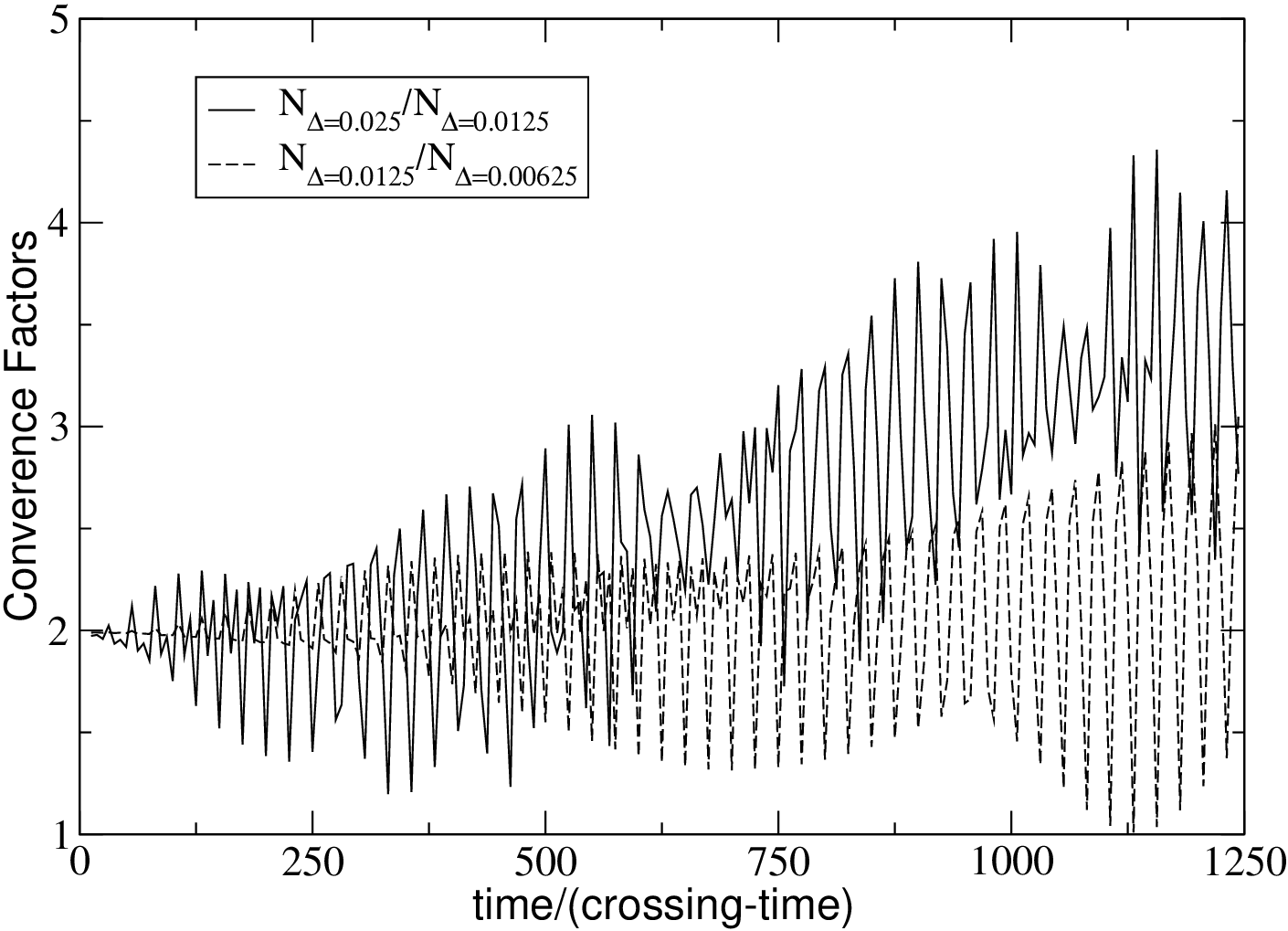}
\caption{Convergence of the constraints energy to zero as a function of the 
number of crossing times, for 
the simulations of Fig.~\ref{log_g05} and Fig.~\ref{logenergy05}. This
figure shows the convergence factors obtained 
by dividing the energy found at different resolutions in consecutive pairs.
The oscillations observed are a product of phase velocities differing
at different resolutions.}
\label{con_A05}
\end{center}
\end{figure}

Figures \ref{logenergy05}, \ref{log_g05} and  \ref{con_A05}
illustrate our results for the gauge wave with amplitude $A=0.5$.
Figure~\ref{logenergy05} shows for three different resolutions the
logarithm of the energy for the constraints as a function of time, for
$1250$ crossing times. The lowest resolution shows a slowly growing
behavior, but this is considerably diminished as better refined ones
are considered. Note that even for the coarsest resolution, the energy
remains smaller than  $10^{-6}$.

In this type of simulation, phase differences between the numerical
and exact solutions typically cause the error to vary in an oscillatory
way: the error going back to a small value after some time, when the
numerical solution achieves a phase difference of $2\pi$ relative
to the exact one~\cite{stability}.  To give a phase-independent
indication of the errors with respect to the exact solution, in Figure
\ref{log_g05} we show the relative error in the maximum attained by $g_{xx}$,
compared to the exact value.  A slow growth is observed.  In particular, the error in the maximum of
the function is $\simeq 0.07$ even after $1250$ crossing times for the
finest discretization. Fig.~\ref{log_g05} also explicitly shows that the
amount of dissipation is very small, as the amplitude is not damped
even in very long runs. The crossing of lines in Fig.~\ref{log_g05}
{\em does not} imply lack of convergence, since one is not necessarily
comparing the field at the same points. (For an explicit convergence illustration
of the solution itself, we present some results for the more stringent $A=1$ case).
Figure \ref{logenergy05} explicitly shows convergence of the constraints to zero, 
while Figure~\ref{con_A05}
 displays the associated convergence factors for those simulations. The latter 
are obtained by dividing the energy for different resolutions in pairs (i.e., ${\cal N}(\Delta_1)/{\cal
N}(\Delta_2)$ and ${\cal N}(\Delta_2)/{\cal N}(\Delta_3)$). For a
second order accurate code this convergence factor should be two in
the convergent regime. Figure~\ref{con_A05} shows that second order
convergence is lost after some time, but this is expected for such long
simulations, owing to accumulation of truncation error.  However, the
convergence factor gets closer to two when computed with the two highest
resolutions.

\subsubsection{Amplitude $A=1$}
%

\begin{figure}[ht]
\begin{center}
\includegraphics*[height=6cm]{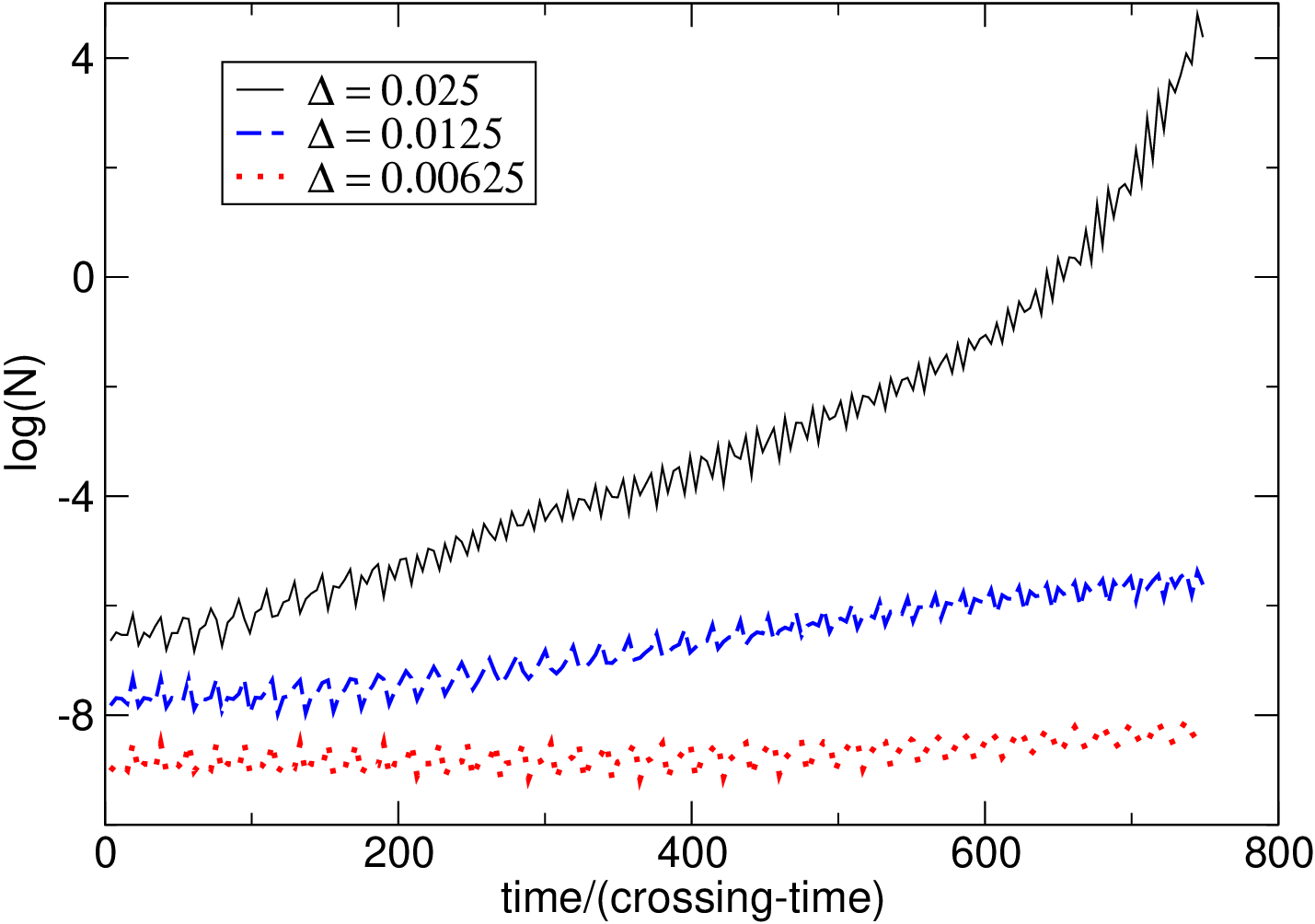}
\caption{Logarithm of the energy for the constraints for the $A=1$ gauge wave 
simulations. The coarsest resolution exhibits a marked growth after $600$ 
crossing times, though with higher resolutions this effect
 becomes negligible.}
\label{logenergy1}
\end{center}
\end{figure}

\begin{figure}[ht]
\begin{center}
\includegraphics*[height=6cm]{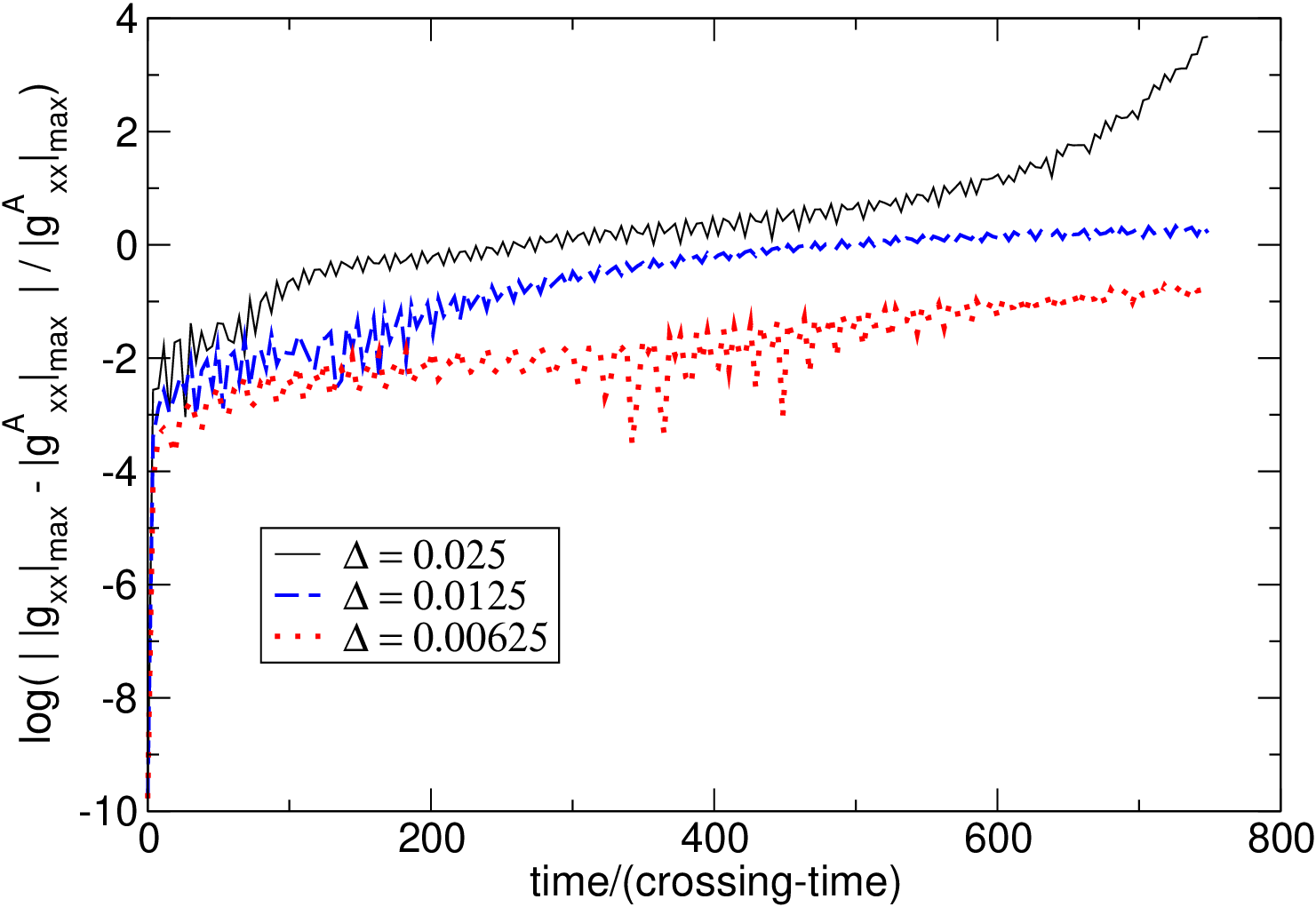}
\caption{The logarithm of the relative error in the maximum attained by 
$g_{xx}$ for the simulations of Fig.~\ref{logenergy1}. The
 errors exhibit a slow growth which diminishes with resolution.}
\label{log_g1}
\end{center}
\end{figure}

\begin{figure}[ht]
\begin{center}
\includegraphics*[height=6cm]{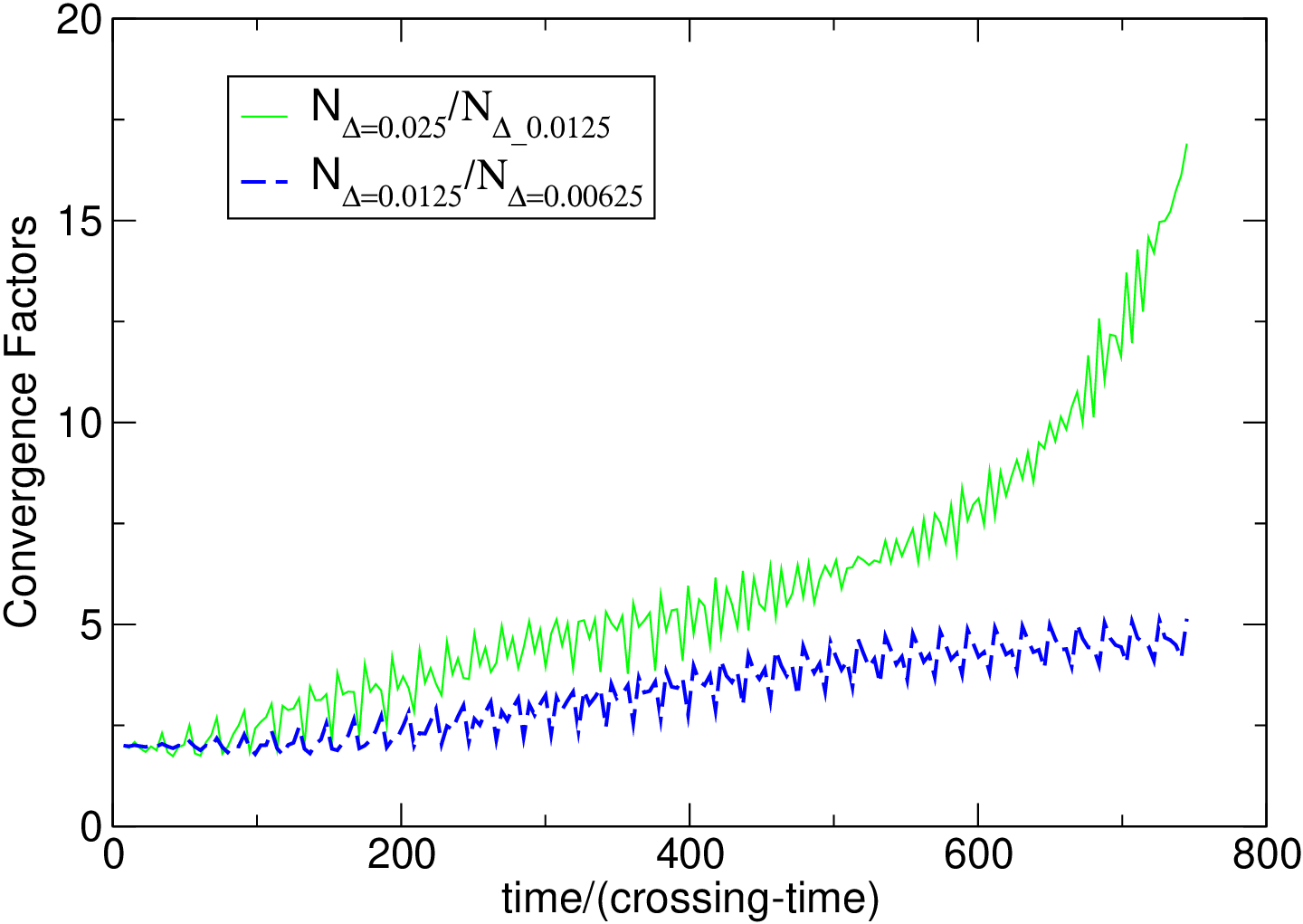}
\caption{Convergence of the constraint energy to zero, for the 
simulations of Figures~\ref{logenergy1} and 
\ref{log_g1},  obtained by dividing the energy for each
resolution in consecutive pairs. As in Figure \ref{con_A05}, after some time the 
convergence factors deviate from the expected value of two, but this difference diminishes when 
 the two highest resolutions are used to 
compute the convergence factor.}
\label{con_A1}
\end{center}
\end{figure}

Increasing the amplitude of the gauge wave introduces some complications
when comparing to lower amplitude runs {\em at the same resolutions.} For
example, at the coarsest resolution ($\Delta = 0.025$) the fields and
the energy grow considerably: after $750$ crossing times the energy is
of order $10^4$. Clearly, even in the convergent regime, errors of
this magnitude mean the numerical solution is of little use.  Simulations  with errors below
ten percent last until about $600$ crossing times for this particular grid
resolution.  To demonstrate the effect of resolution on the quality
of the solution, we take $750$ crossing times as the end-point of our
simulations.  Figures~\ref{logenergy1}, \ref{log_g1} and  \ref{con_A1}
illustrate the results.  The energy of the constraints, as shown
in Figure~\ref{logenergy1}, shows marked exponential growth, with a
large growth rate for the coarsest resolution.  However, increasing the
resolution diminishes the growth rate considerably, and simulations can
be extended for at least $1500$ crossing times (at which point we simply
stopped the simulations, with small errors).

\begin{figure}[ht]
\begin{center}
\includegraphics*[height=6cm]{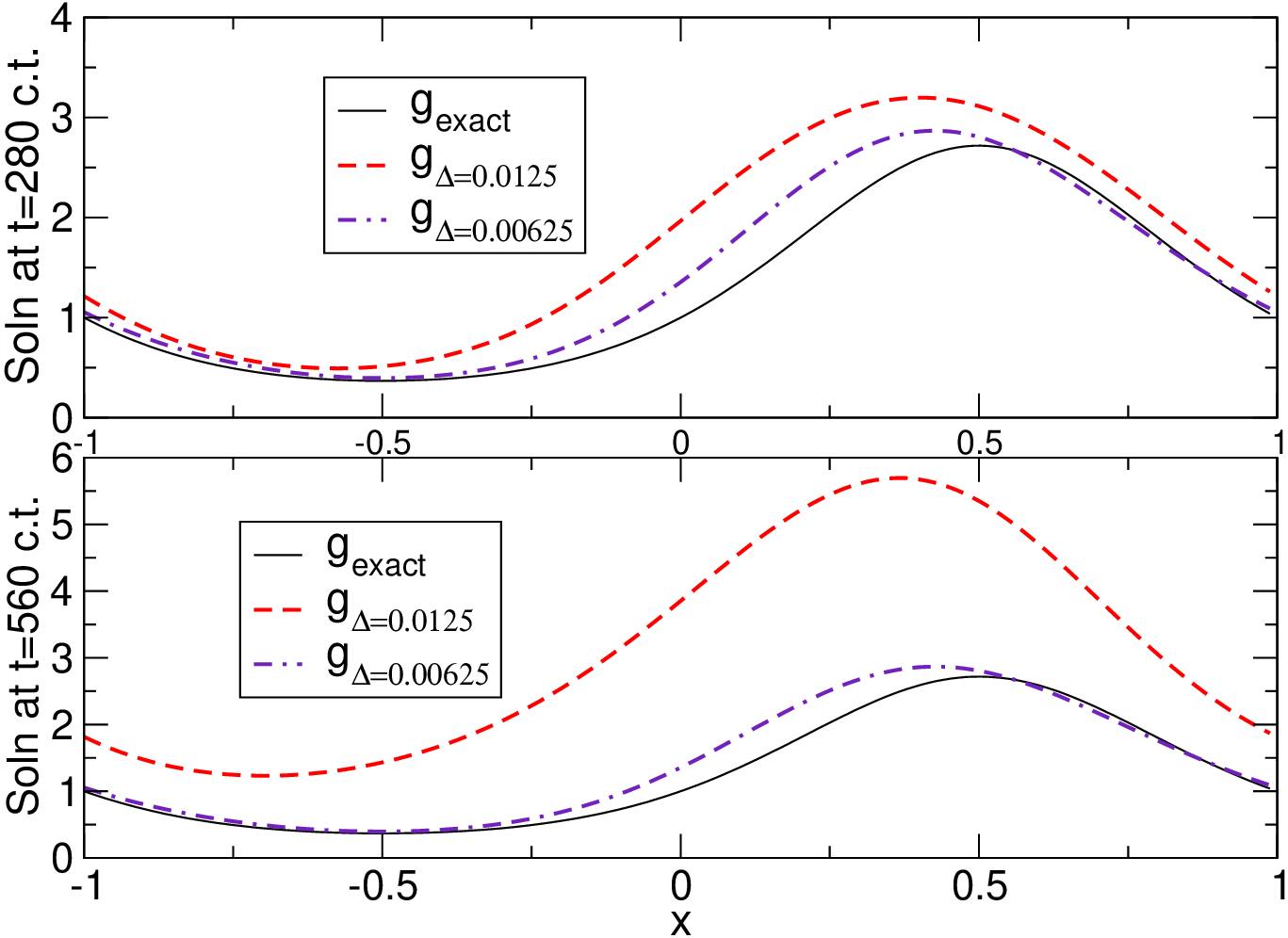}
\caption{Snapshots at $280$ and $560$ crossing times displaying the
exact solution and the numerically calculated ones for the middle ($\triangle=0.0125$)
and fine ($\triangle=0.00625$) resolutions. Convergence to the exact solution is evident
as resolution is increased.}
\label{snap}
\end{center}
\end{figure}

\begin{figure}[ht]
\begin{center}
\includegraphics*[height=6cm]{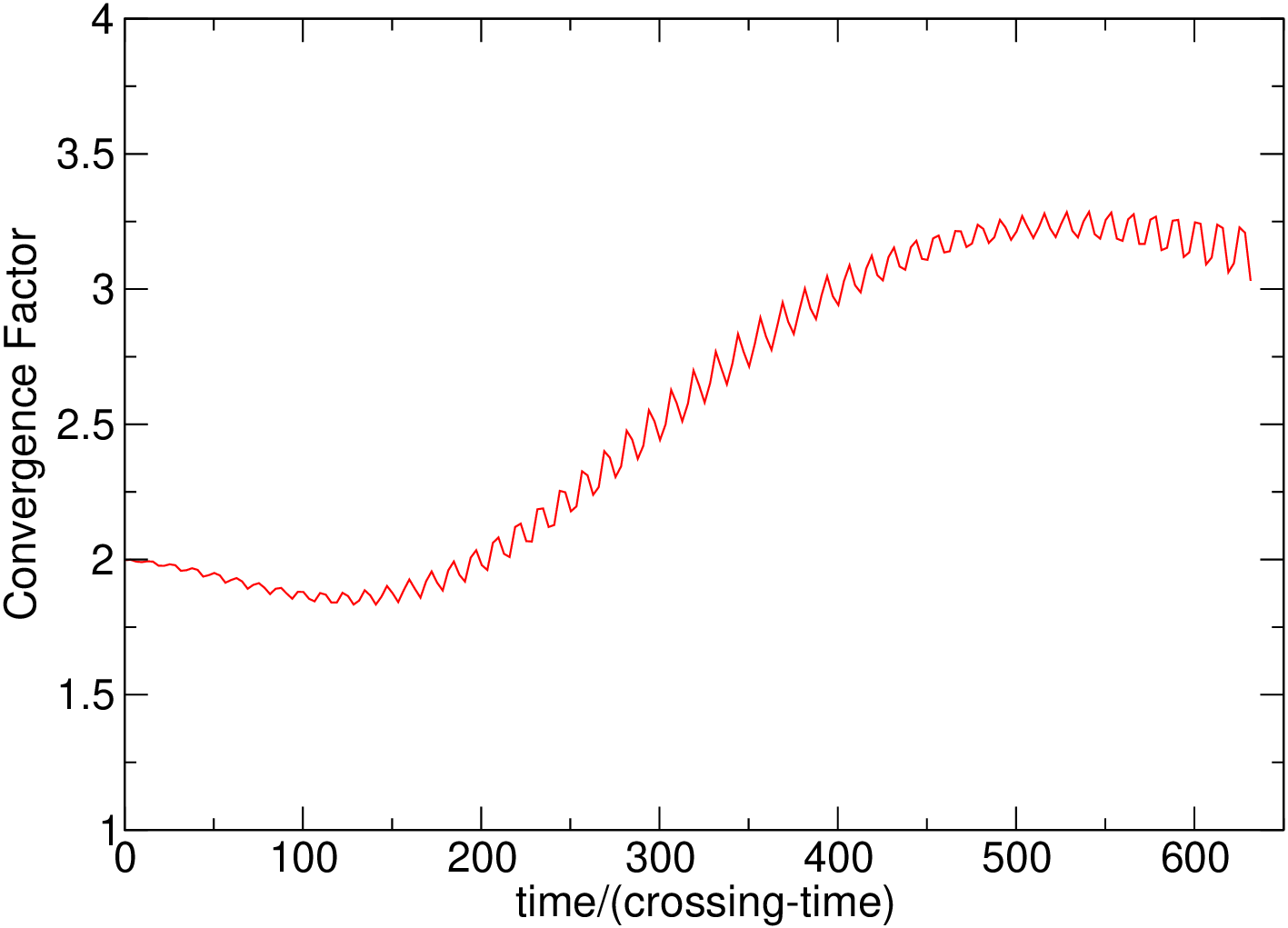}
\caption{Convergent factor calculated with the numerical solutions for the middle and fine
resolutions. At late times acumulation of errors makes the obtained value deviate from its 
expected one.}
\label{snapconvergefactor}
\end{center}
\end{figure}

\begin{figure}[ht]
\begin{center}
\includegraphics*[height=6cm]{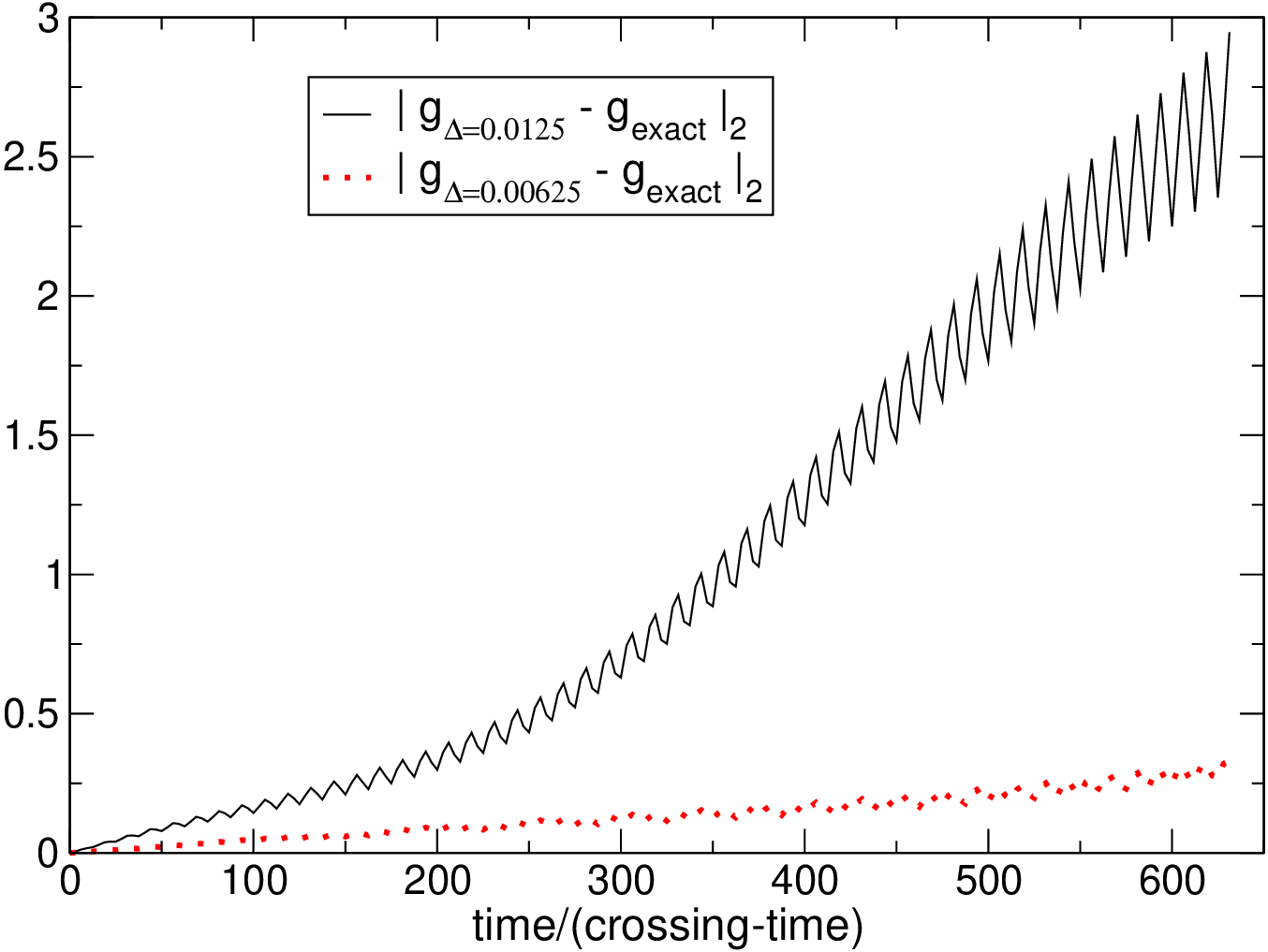}
\caption{$L_2$ norm of error in the numerical solution. Although a faster than linear
growth is observed for the middle resolution, this effect converges away as a finer resolution
is considered.}
\label{snapconverge}
\end{center}
\end{figure}

Figure~\ref{log_g1} shows the relative error in the maximum attained
by  $g_{xx}$ for the high amplitude gauge wave $A=1$.  Again, a rapid raise in
the error is clearly seen for the coarsest resolution, but this effect
is less noticeable at higher resolutions. At $750$ crossing times, the
error in the maximum of the function is less than ten percent.  Finally,
the convergence of the code is explicitly illustrated in Fig.~\ref{con_A1},
which shows the convergence factors obtained by taking the energy for the constraints 
at different resolutions and dividing them in pairs. The order of convergence of the constraints to zero is close to two for some time, and the lenght of this time increases with resolution.
Finally, to illustrate the overall quality of the obtained numerical
solution for different resolutions, figures \ref{snap}, \ref{snapconvergefactor} and
\ref{snapconverge} display explicit comparisons between the exact
solution and the middle and fine resolutions ($p=2,4$). Figure \ref{snap} presents
snapshots of the solution at $280$ and $560$ crossing times, clearly both drifts in phase
and amplitudes are observed for the middle resolution while for the finer one
the difference is mainly observed in the
 phase. Despite these differences, the solution obtained is converging
to the exact one. Figure \ref{snapconvergefactor} provides the convergence
factor calculated with the two resolutions. As before, the factor deviates from its expected 
value after a while, but this improves with resolution. Finally
figure \ref{snapconverge} shows the $L_2$ norm of the difference between the exact and
numerical solutions. Clearly even after $600$ crossing times, the fine resolution stays
close to the exact solution and does not exhibit growth faster than the expected
linear one.

\subsection{Observations}

As mentioned above, the constraints
$C$, $C_i$, and $C_{ijkl}$ are initially satisfied to the level of round-off
in the gauge wave tests. For grids with at least $n_x=161$ points,
these constraints remain quite small throughout the runs, even for the 
high amplitude case, $A=1$.  This has several consequences: 
\begin{itemize}

\item First, most hyperbolic formulations differ, among other things,
in how the constraints are added to the right hand side of the
equations~\cite{reula}. Since the constraints themselves stay
negligible small, we expect the conclusions found with this test should
be applicable to most hyperbolic formulations with the same choice of lapse and shift. Namely, that the use
of symmetric hyperbolic systems and numerical techniques guaranteeing
stability at the linear level, plus the addition of a small amount of
dissipation stabilizes the problem. For instance, without the use of
dissipation, for the case $A=1$ and $n_x=161$ points we can follow the
system for about ten crossing times before the errors in the numerical
solution become of order one. With a small dissipative term, on the
other hand, the system after $1200$ crossing times does not yet exhibit
errors of order one.

\item Second, the minimization technique presented in 
Section~\ref{minimization} explicitly makes use of  
non-homogeneous terms (in
the free constraint-functions) in the expression for the energy growth in order to
minimize it.  However, in the current case these non-homogeneous terms are
considerably smaller than the homogeneous contribution. 
Therefore, one would need to use huge constraint-function values
for them to play a role in minimizing the growth, which would require
the use of an extremely small Courant factor. Note that the formulation used employs 
the addition of the constraints $C$, $C_i$ and $C_{ijkl}$ but not
$C_{ijk}$ and $C_{A_i} = \partial_i N/N - A_i$. The latter ones are {\it not} satisfied
to round-off level but just to truncation level initially. As a result one could
have considered a constraint energy which includes them or a modification
of the formulation employed by the addition of these constraints in a suitable manner.
We have not explored these options in the present work as in the black hole case the
non-homogeneous terms in the constraint energy 
are not negligible as in the gauge wave case. Indeed, this is what would be expected
 for generic scenarios.

\end{itemize}

\section{Black hole Simulations} \label{bh}

We now turn our attention to the Schwarzschild black hole spacetime. 

In this case discretization errors (that is, due to finite grid spacing)
make the constraints start-off at non-negligible values and so the
minimization technique can be used to pick up preferred formulations for
this problem. Before applying the minimization procedure we examine the
system with fixed constraint-functions with the goal of understanding some specific
issues.  As in the rest of the paper, the simulations of this section are done
with the two-constraint-function  family of formulations discussed in Section II
(the reason for not using the single-constraint-function family is explained in
Section \ref{why_not}); and the  fixed values for $\gamma$ and $\eta$
here used are, in the absence of any other obvious choices, $\gamma = 0 = \eta$.
Recall that this two-constraint-function  family is symmetric hyperbolic for {\em
any} values of $\eta$ and $\gamma$, in particular for  $\gamma = 0 = \eta$

Next we study the influence of the position of the inner boundary;
the outer boundaries (their influences being discussed later in the
paper) are kept at the same position,  $\pm 5M$ in all the runs of this
section. The outer boundaries are chosen quite close on purpose, to make
the turn-around time for the runs shorter for the rather detailed 
analysis of this section. However,
it should be clear that the points here made
 are quite independent on the particular position for the
outer boundaries chosen. Later, when applying the minimization of the
constraints, we will choose several different values for the position
of the outer boundary.

\subsubsection{Inner boundary and the outflow condition}

Black hole excision is usually based on the assumption that an inner
boundary (IB) can be placed on the domain such that information from
this boundary does not enter the computational domain.  The boundary
is supposed to be contained inside the black hole and {\em also} to be 
purely outflow, i.e., all modes
propagate off of the grid at the boundary.  This requirement places
strenuous demands on cubical excision for a Schwarzschild black hole
in Kerr-Schild, Painlevee-Gullstrand or the Martel-Poisson \cite{martelpoisson} 
coordinates:  the cube must be
inside $0.37M$ in each direction.  This forces one to excise very close to
the singularity, where gradients in the solution can become very large,
requiring very high resolution near the excision boundary to adequately
resolve the solution.  Finally, we note that this requirement follows
directly from the physical properties of the Schwarzschild solution in
these coordinates, and is independent of the particular formulation of
the Einstein equations~\cite{sbp2}.

With our current uniform Cartesian code, we are not able to provide the
resolution required to adequately represent the Schwarzschild solution
close to the singularity.  While we are actively working on solutions
to this problem, currently our only practical alternative is to place the inner
boundary inside the event horizon, but outside the region specified
by the outflow condition, i.e., the solution has incoming modes on the
inner boundary.  One could attempt to provide data for the incoming modes
on the excision boundary.  However, there is no general theory well-posed
problems when the principal part rank, i.e., number, of zero speed modes on the boundary,
is not constant.  Moreover, ill-posed problems for such
configurations are known.  Thus we simply do not apply boundary conditions
to the incoming modes on the inner boundary, resulting formally in an
ill-posed problem~\cite{tadmor}.  In this section, however, we argue that
errors from this inconsistency do not prevent us from learning much about
the numerical properties of our formulation for black hole spacetimes.

\begin{figure}[ht]
\begin{center}
\includegraphics*[height=6cm]{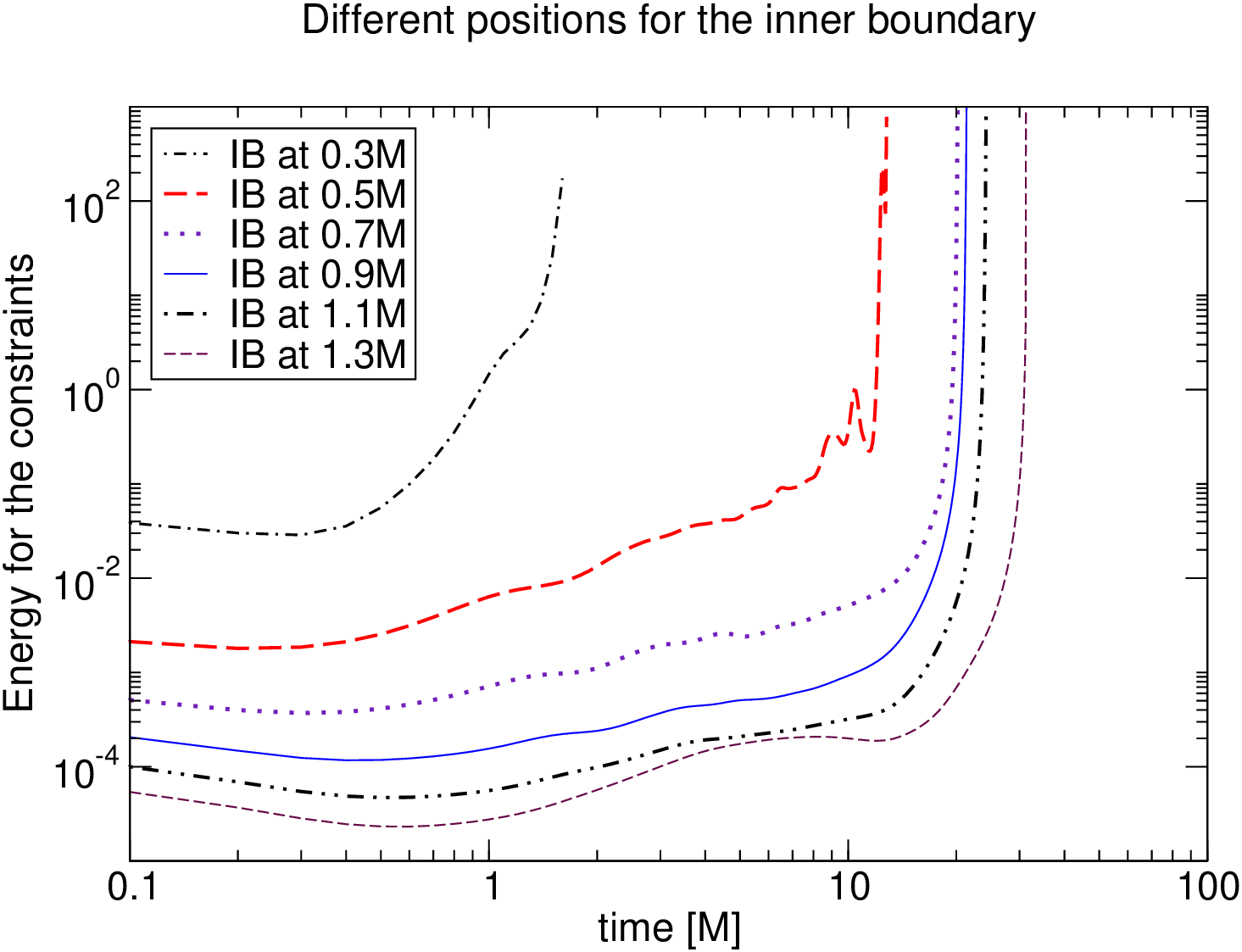}
\caption{Black hole simulations, with fixed outer boundaries (at $5M$), and different position 
for the inner boundary (IB). The resolution is $\Delta = M/5$. }.
\label{inner} 
\end{center}
\end{figure}

Figure (\ref{inner}) shows the results of simulations with different
positions for the inner boundary (IB), with $\triangle = M/5$, obtained without imposing
boundary conditions at the IB. The IB at $0.3M$ (i.e. half the length of a cube
centered at the origin) corresponds to a purely
outflow boundary.  At the other extreme, the inner boundary at $1.3M$
gives an inner boundary that penetrates outside the event horizon.
The cases between $0.3M$ and $1.3M$ correspond to IB inside the black
hole, with inflow portions.  All runs shown in Fig.~\ref{inner}, except
the first with IB at $0.3M$, should have convergence problems, since not
giving boundary conditions is inconsistent with the structure of the
characteristic modes. However, one also sees that at this resolution,
placing the inner boundary so close to the singularity causes the code
to run a factor of ten less, presumably because of lack of resolution.

\begin{figure}[ht]
\begin{center}
\includegraphics*[height=6cm]{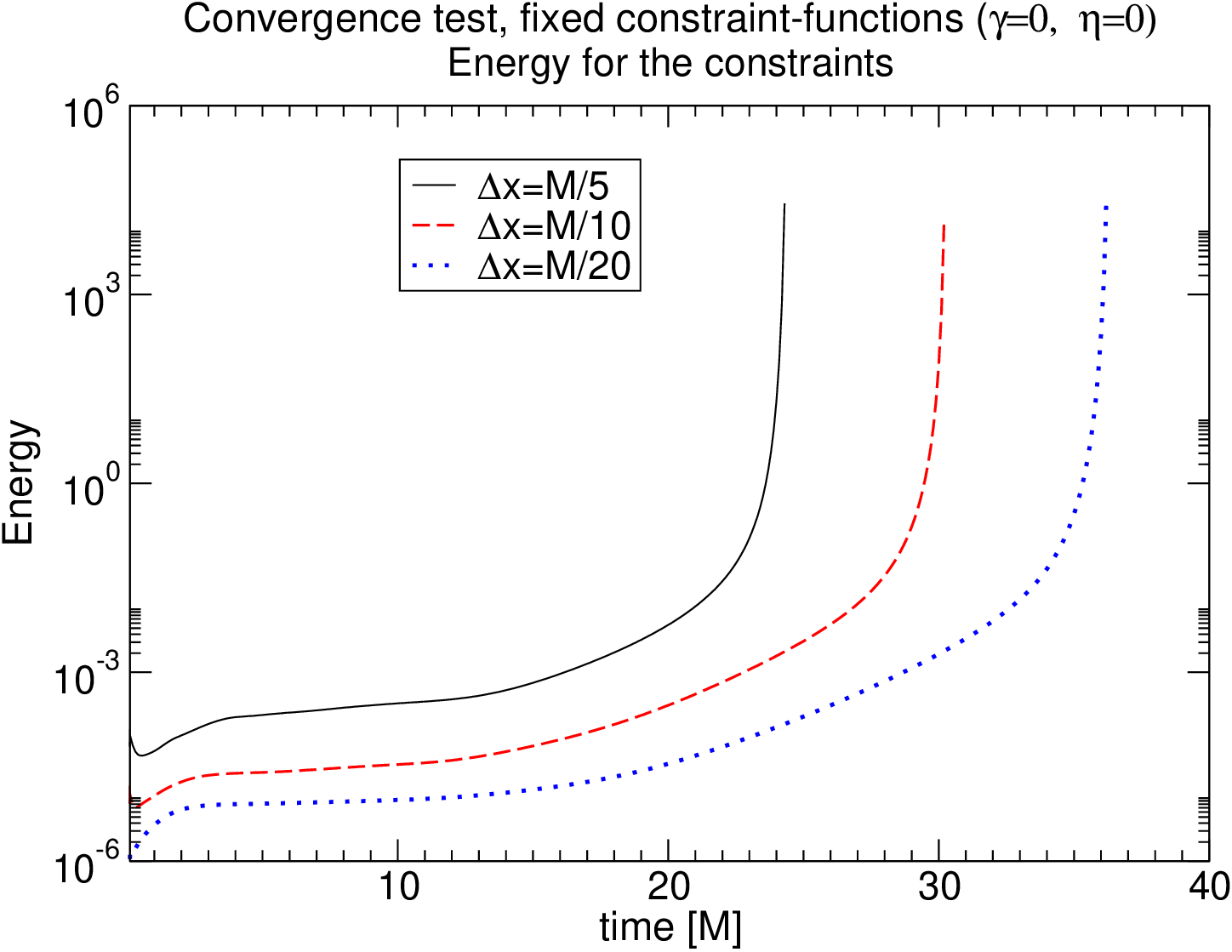}
\includegraphics*[height=6cm]{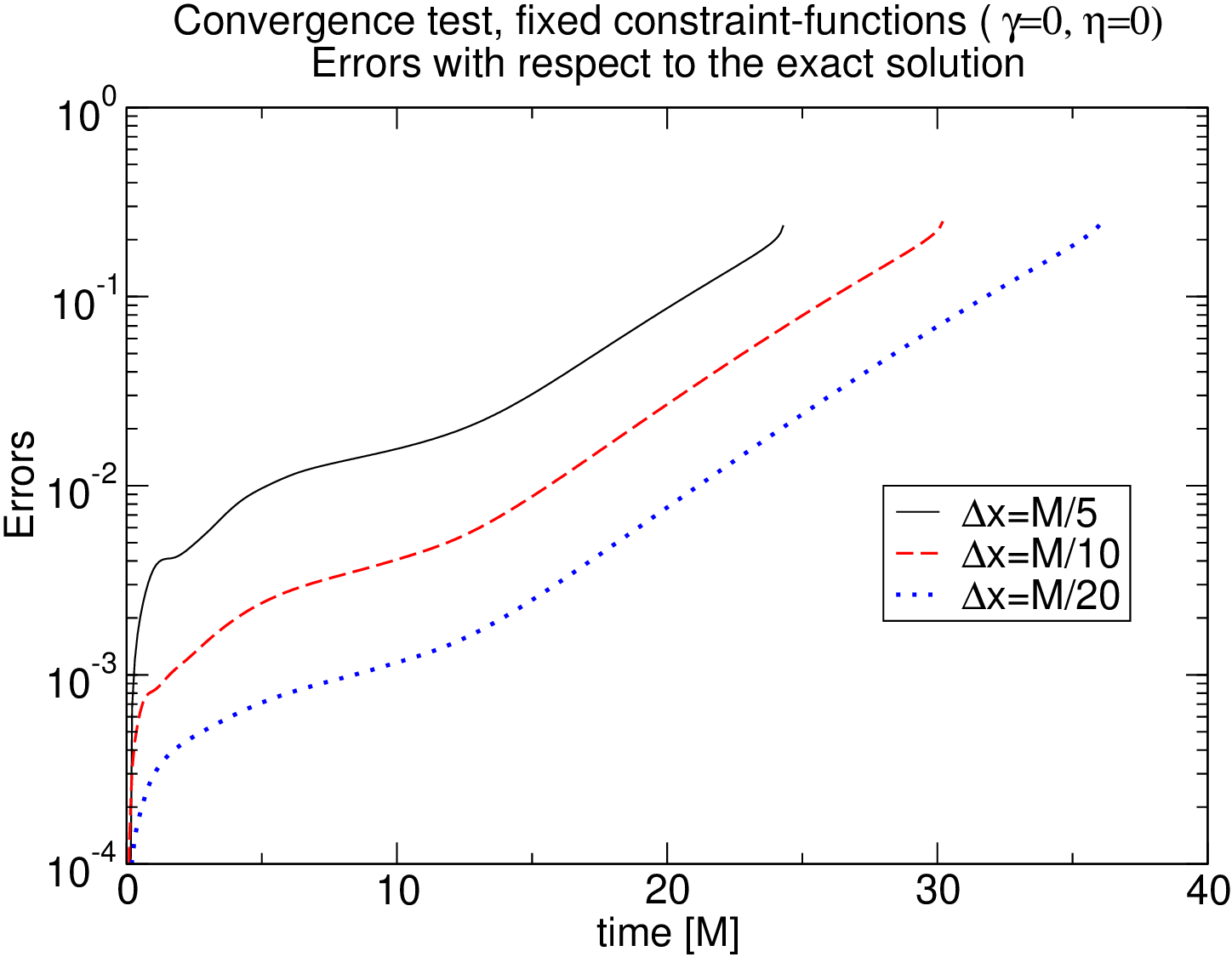}
\caption{Two-constraint-function  family, with fixed values $\gamma=0=
\eta$, inner and outer boundaries at $1.1M$ and $5M$, respectively.}
\label{convergence1} 
\end{center}
\end{figure}

Figure (\ref{convergence1}) shows a convergence test for a configuration
test with a the excision boundary at $1.1M$ with resolutions 
$\triangle = M/5$, $M/10$, $M/20$.  From here on the errors shown in the plots are those of the numerical 
solution $u_n$ relative and with respect to 
the exact one $u_e$; more precisely, the $L_2$ norm of  
$$
\left( \frac{\sum (u_{n}-u_e)^2}{\sum u_e^2} \right)^{-1/2} \;,
$$
where the sum is over the components of the vector valued functions $u$.

This test can give some indication of 
possible numerical stability problems.   While the solution diverges  in all
cases, the fact that the code appears to converge in the short term
indicates that at these resolutions, and for these run times,
the expected instability owing to improper boundary placement appears
to grow slower than other unstable modes in the solution.  Thus we
can still obtain valuable information about the solution and its
numerical properties.
Finally, we emphasize that the inconsistent
inner boundary probably leads to convergence difficulties that could be
detected with more extensive tests, such as
performing a Fourier decomposition of the numerical
solution, and a convergence test frequency by frequency likely would make
the numerical instability manifest, see for instance \cite{stability}.

\subsection{Dynamic minimization: preliminary discussion} \label{preliminar2}

In this section we examine several issues that arise in the
constraint energy minimization procedure.

\subsubsection{Why one should use the two-constraint-function family} \label{why_not}

\begin{figure}[ht]
\begin{center}
\includegraphics*[height=6cm]{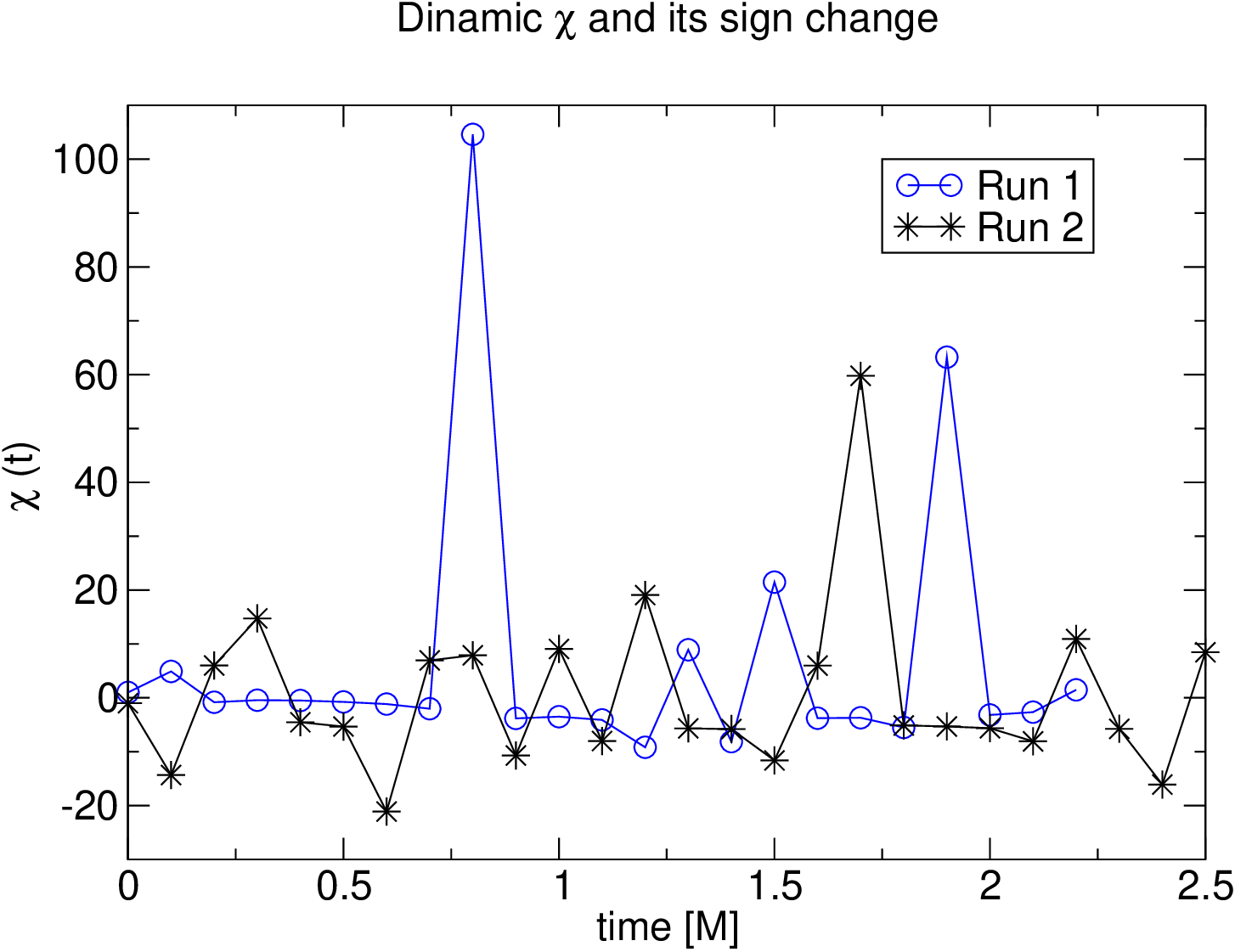}
\caption{Generally $\chi$ would need to change sign in
  order to achieve certain control on the constraints. Therefore, it
  would not have a continuous limit when 
$\triangle t \to 0$. 
For this reason, the two-constraint-function formulation is used throughout 
 this paper.}
\label{chi} 
\end{center}
\end{figure}

We consider only the two-constraint-function family of formulations of the equations
for all constraint minimization runs.  The single-constraint-function formulation is
inadequate because, as mentioned above, the free constraint-function $\chi$ may not
equal zero.  Thus $\chi$ may only be negative or only positive during
an entire run to be a continuous function of time.  This considerably
limits the power of the dynamic minimization technique, since in order
to control the constraints one might need at some time a positive value
of $\chi$ and at some other time a negative one. Indeed, this occurs as
Figure~(\ref{chi}) shows. For this resolution and
location of inner and outer boundaries 
 the initial discretization error for the constraints leads to a value
of the energy of
\begin{equation}
{\cal N}(0)=0.99925 \times 10^{-4} \label{tol5m}
\end{equation}
Figure \ref{chi} thus shows two evolutions, one with a
negative seed value, $\chi =-1.0 $, and a tolerance value $10^{-5}$.
The second run has a tolerance value of $10^{-4}$, and as seed value
$\chi=1.0$.  In both cases  $n_a=1$.  In both cases $\chi$ changes sign,
indicating that this can be expected in general, and that  there is
no continuous interpolation for $\chi(t)$ in the limit $\triangle =0$
such that $\chi\neq 0$.

\subsubsection{The accuracy of a semi-discrete picture.}

\begin{figure}[ht]
\begin{center}
\includegraphics*[height=6cm]{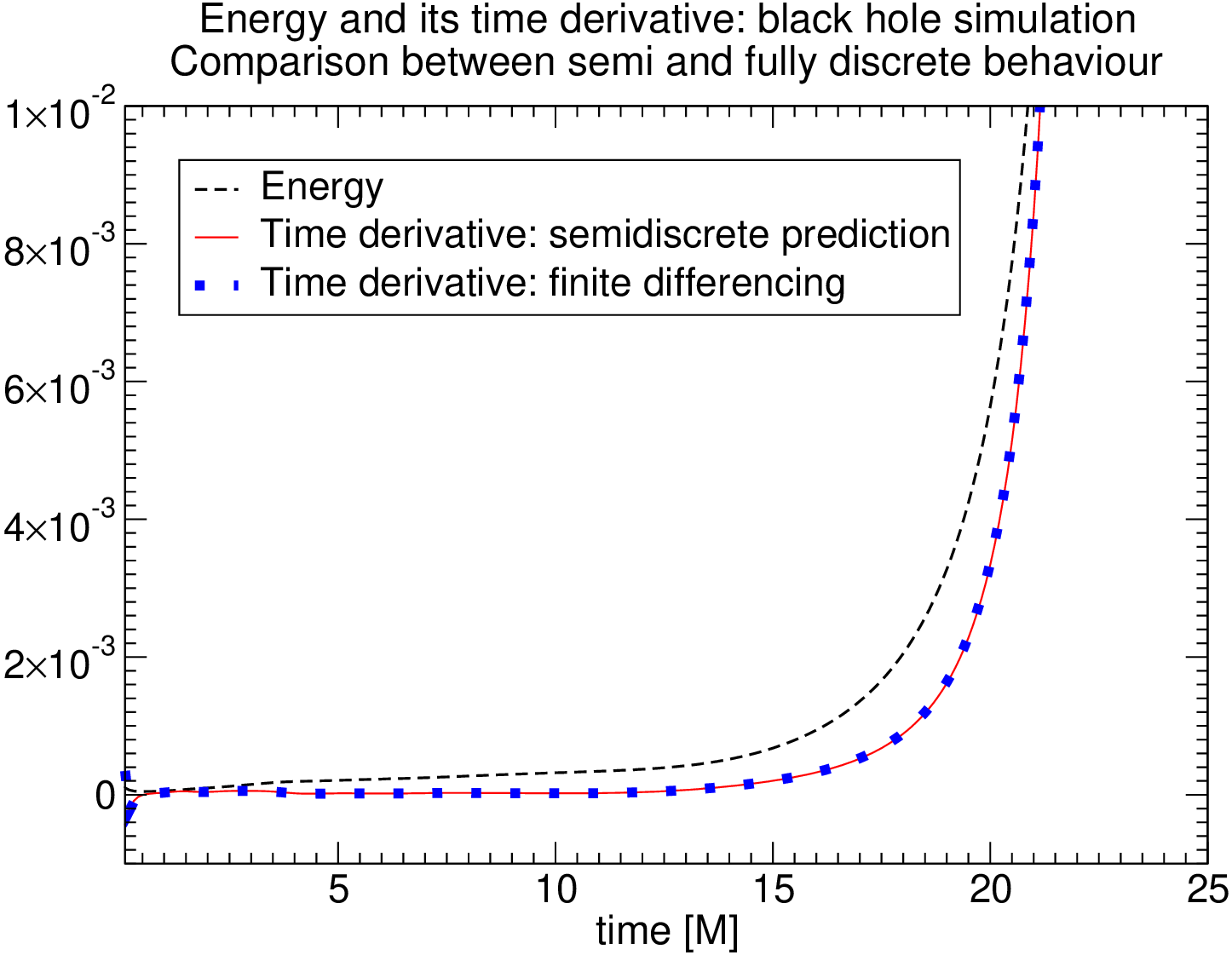}
\caption{Energy for the constraints and its time derivative, $\dot{\cal N}$,
computed through the semi-discrete prediction and through numerical 
differentiation. The remarkable agreement between the two indicates that the semi-discrete analysis used to
calculate the constraint minimization is faithful representation of  the fully
discrete evolution.}
\label{energy} 
\end{center}
\end{figure}

The constraint minimization method, as described in Section \ref{minimization},
is based on {\em semi-discrete} equations, where the spatial derivatives
are discrete, but time continuous.  While a fully discretized method could
be developed, we simply use the semi-discrete analysis here.  In the limit
$\Delta t \rightarrow 0$, one naturally expects this semi-discrete analysis to be a perfect 
description of the fully discrete scheme.  
Here we verify that the fully discrete evolutions are, indeed, very well  
approximated by the semi-discrete analysis, cf.  Eq.~(\ref{split}), even for
rather large Courant factors, such as $\lambda=0.5$ that we use for the black hole runs of
 this paper.

Figure~\ref{energy} shows results of an evolution with inner boundary
at $1.1M, \lambda=0.5, \sigma=0.03$, outer boundaries at $\pm5M$, and
$\triangle = M/5$.  There are two curves for the time derivative of the
energy, $\dot{\cal N}$.  The first curve is obtained via the semi-discrete
prediction, Eq.~(\ref{split}), with the matrix valued integrals $({\cal
I}^{hom},{\cal I}^{\mu})$ computed during evolution.  The second curve is
obtained from second order, centered differences of the energy, $\cal N$.
Both curves agree remarkably well, which indicates that the semi-discrete
expression for $\dot{\cal N}$ captures the dynamic constraint behavior
in the fully-discrete simulation with a Courant factor of $\lambda =0.5$.

\begin{figure}[ht]
\begin{center}
\includegraphics*[height=6cm]{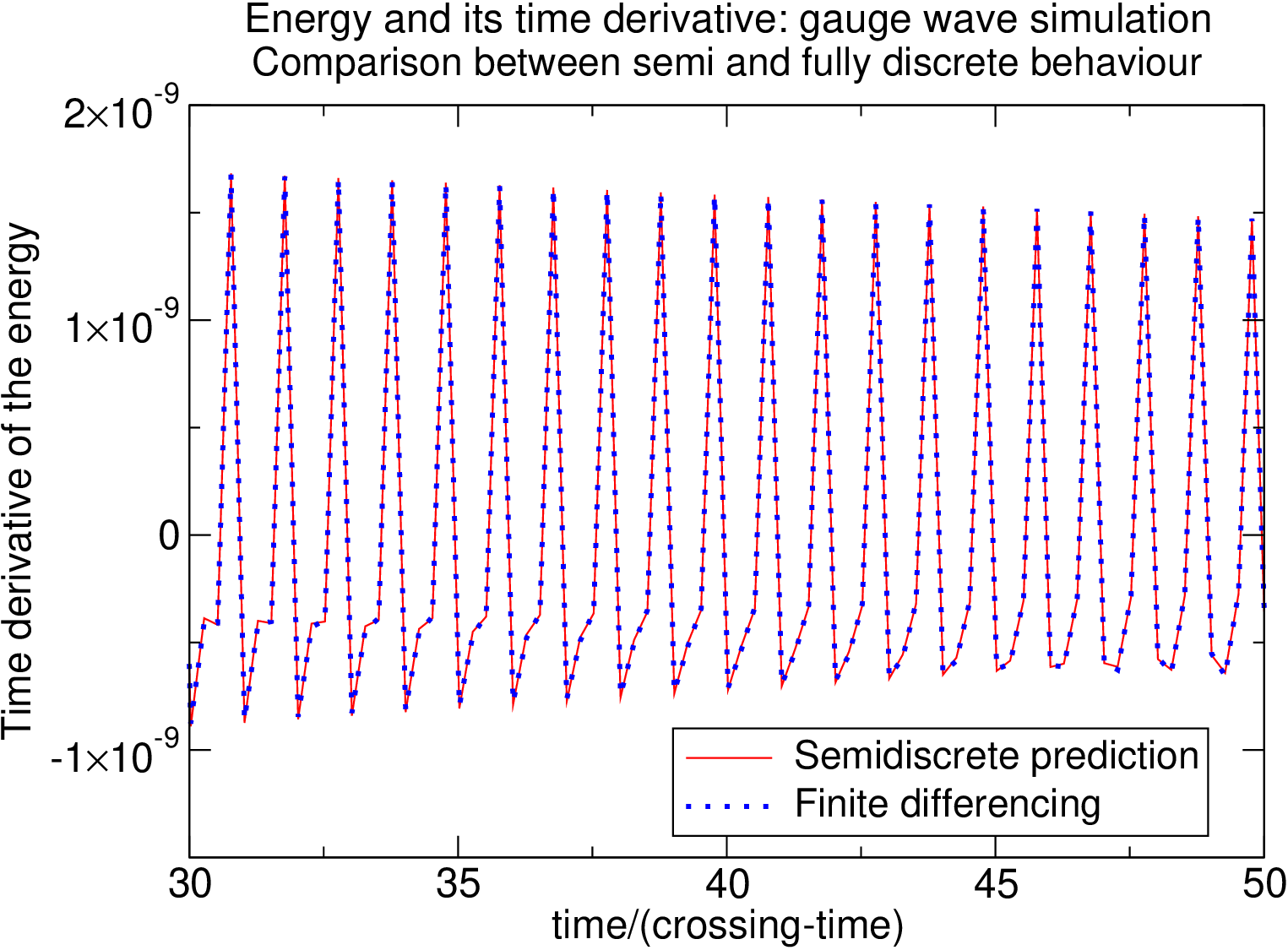}
\caption{Time derivative of the energy of the constraints
computed through the semi-discrete prediction and through numerical 
differentiation for the gauge wave case. The agreement of the two curves
is evident.}
\label{energywave} 
\end{center}
\end{figure}

As another illustration of this, figure ~\ref{energywave} shows the
results obtained for one of the gauge wave tests presented above:  
$A=0.5$, $161$ points, and $\lambda =0.25$.
For a clean visualization of the agreement we show here a period of time
between $30$ and $50$ crossing times. The agreement of the curves is evident.

\subsubsection{Practical questions regarding constraint minimization} 
\label{every}

\begin{figure}[ht]
\begin{center}
\includegraphics*[height=6cm]{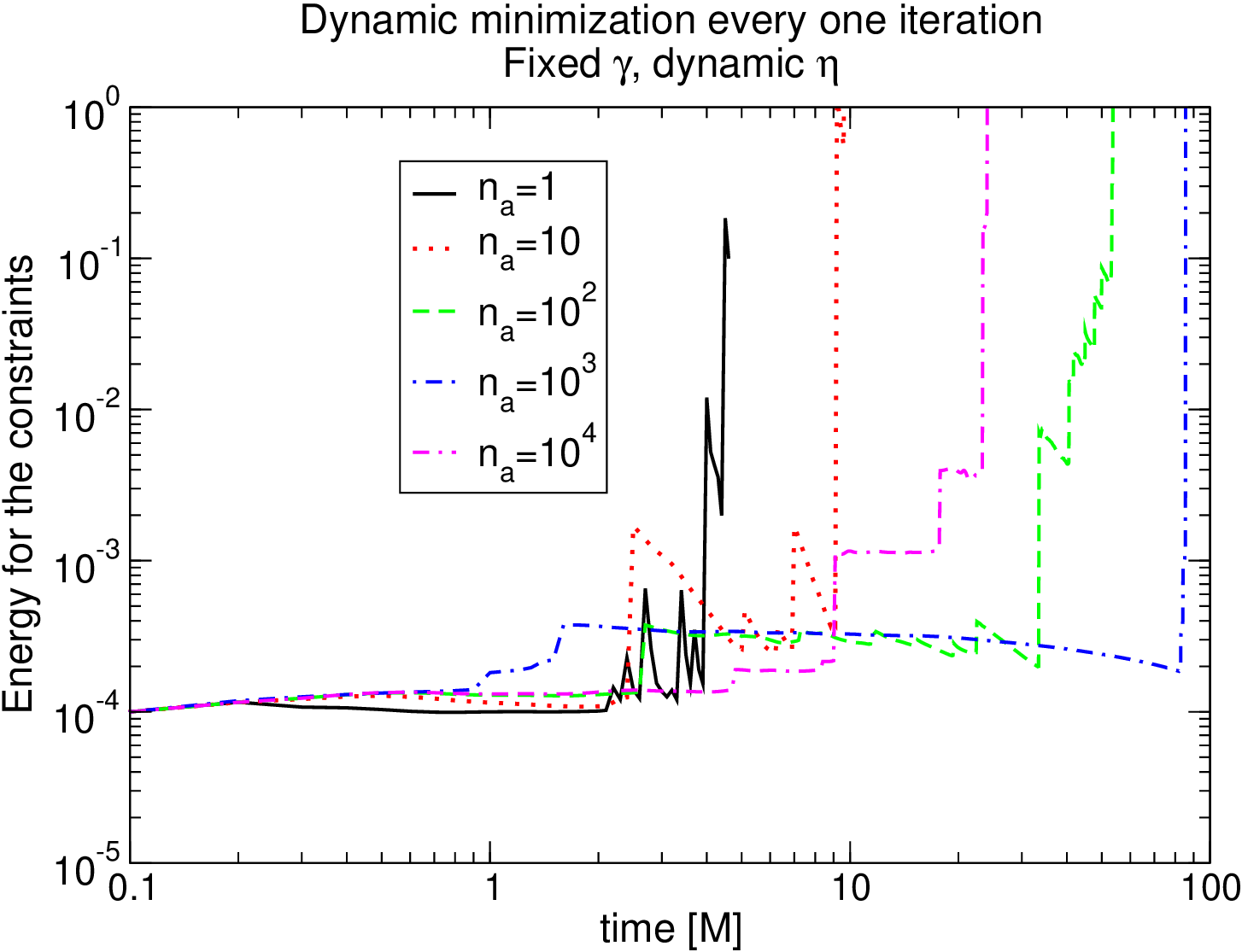}
\includegraphics*[height=6cm]{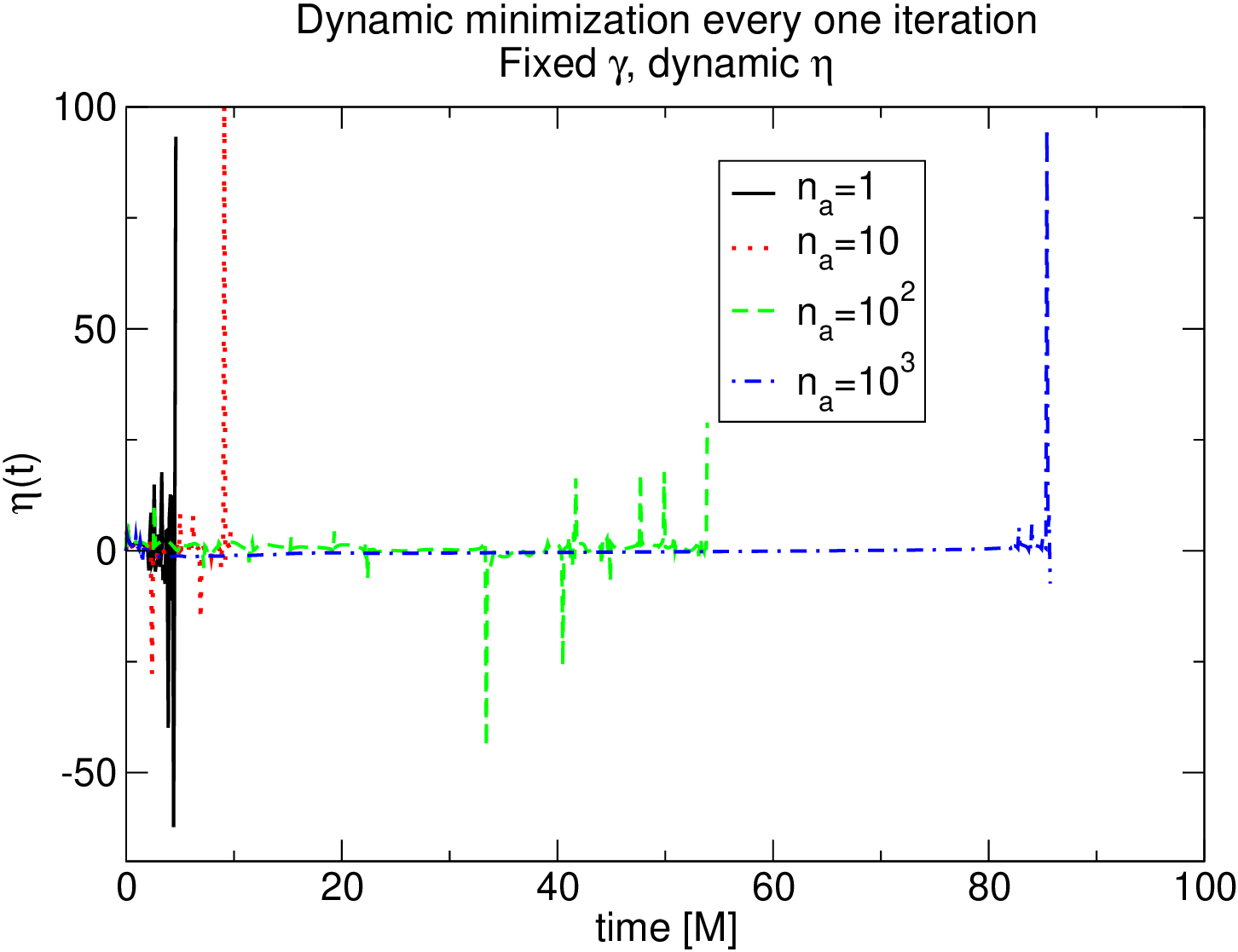}
\caption{The plots in this figure and those in Fig. \ref{energy1_every_it_10}  show the effects of varying the frequency
of performing the constraint minimization, as well as the dependence on $n_a$. The constraint-function $\eta$ for $na=10^4$ 
is not shown as the run crashes very early and the scale in the time axis for the associated plot is not logarithmic.}
\label{energy1_every_it_1} 
\end{center}
\end{figure}

We now consider some practical questions that arise when performing the
constraint minimization method, namely: how often should the minimization
be performed?, and how fast should the constraint-functions multiplying the
constraints in the RHS be allowed to change?  The minimization procedure
can be computationally expensive, making it advantageous to perform the
analysis infrequently if possible. (Note that the evaluation of the constraint
energy require knowledge of the right hand side of the equations, so {\it each} evaluation
takes about as much computer time as is required in a time-step).

 The second question relates to how
fast the constraint-functions are allowed to vary by setting $n_a$,  which is used
to estimate how many steps are required for $\cal N$ to relax to the
tolerance value, $T$.  Furthermore, can one trade the frequency of the
constraint minimization with different values of $n_a$?  For example,
is performing the minimization at every iteration with a large value of
$n_a$ equivalent to performing the minimization every certain number of
steps with a smaller value of $n_a$?. We will see that it is preferable
to perform the minimization at every time step.

\begin{figure}[ht]
\begin{center}
\includegraphics*[height=6cm]{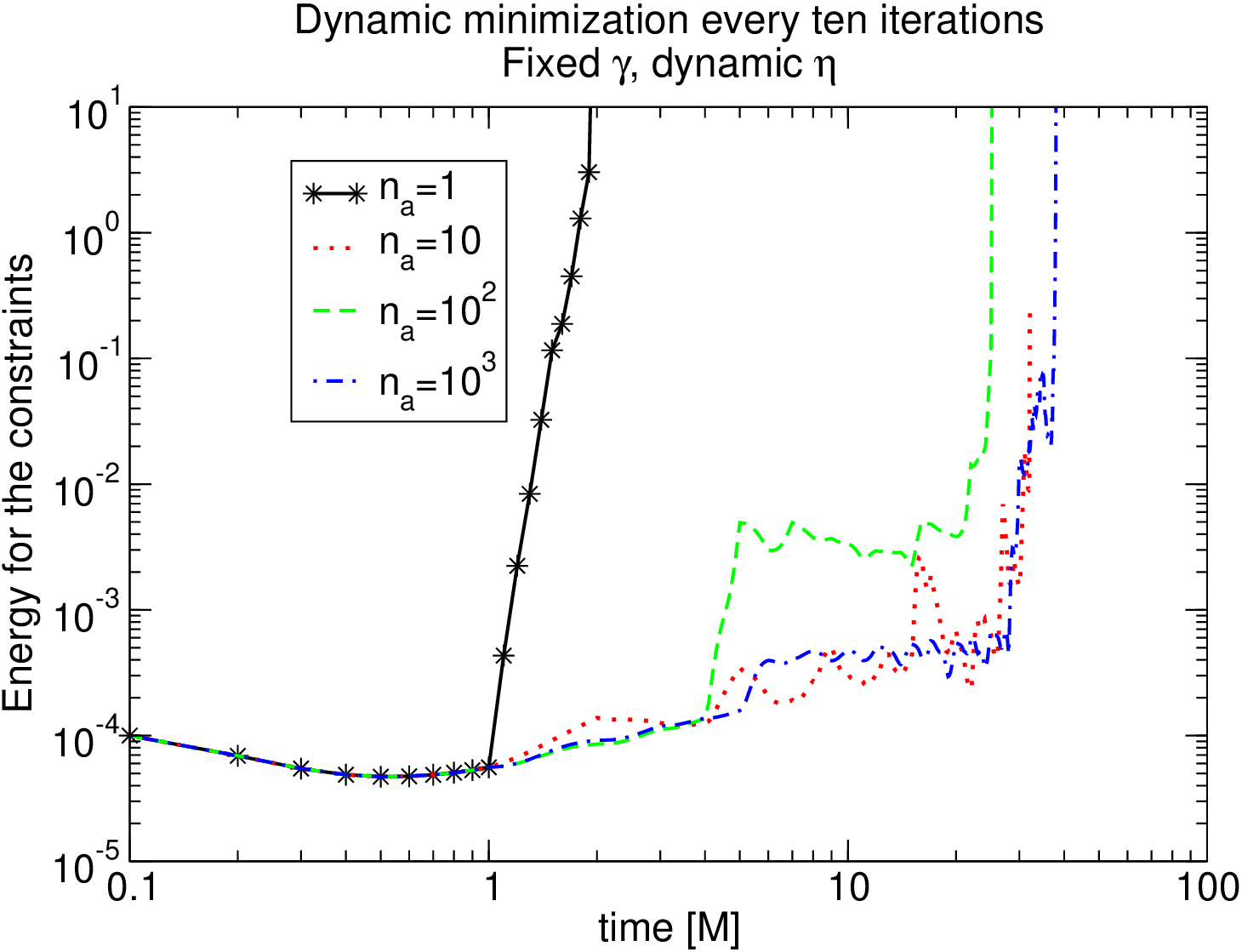}
\includegraphics*[height=6cm]{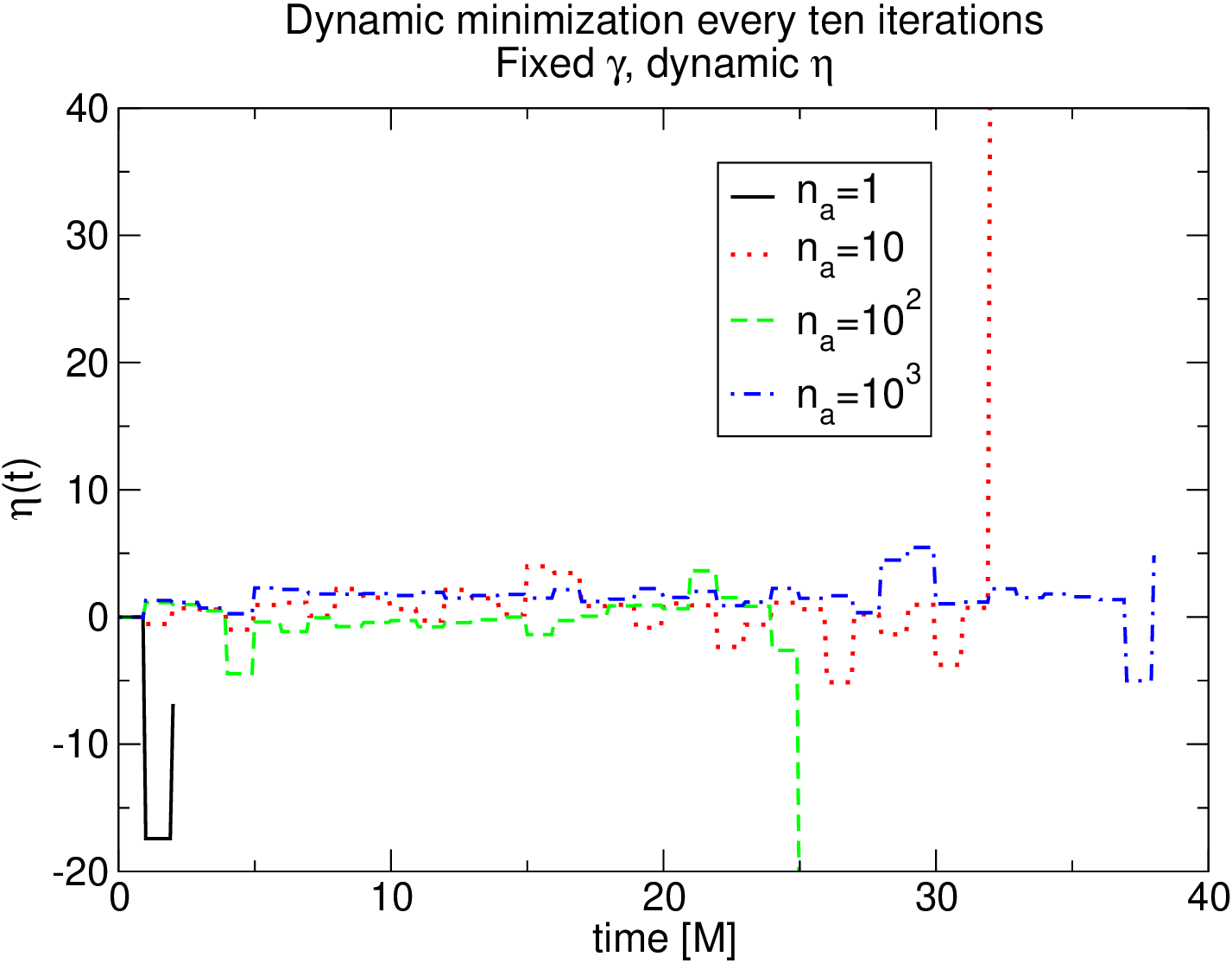}
\caption{This plot shows simulations as those of 
Fig.~(\ref{energy1_every_it_1}), except that here the minimization is done 
every $10$ iterations. Comparing it with the previous plot, it seems
clear that performing the minimization at every iteration seems a better
option}.
\label{energy1_every_it_10} 
\end{center}
\end{figure}

Figure~(\ref{energy1_every_it_1}) shows the results of runs with the same
numerical constraint-functions as in the previous subsection: outer boundaries at
$\pm 5M$, inner boundary at $1.1M$, $\sigma =0.03$, $\lambda =0.5$, $51^3$
points, and a tolerance value $T=10^{-4}$, performing the minimization
at every time step, but now with different values of $n_a$. The constraint-function
$\gamma$ is fixed to $\gamma =0$, and the minimization of the constraints
is applied using $\eta$, as described in Section~\ref{minimization}.
The constraint energy, $\cal N$ is shown in the upper panel, and
$\eta(t)$ in the lower panel.  With a fixed Courant factor, small values
of $n_a$ are problematic because large and fast variations in $\eta(t)$
are allowed and, as shown in the figure, do occur.  On the other hand,
large values of $n_a$ can let the energy grow too much.  Compare now
Figure~(\ref{energy1_every_it_10}), where the minimization is now
applied every ten time steps. As can be seen, it is better to apply the
minimization at every time step.  Otherwise the energy appears to grow
too fast between recalculations of the constraint-functions.  Finally,
we note that this behavior may be model dependent, and in other scenarios
it may be possible to use the minimization less frequently.  In this paper, however, 
we perform the minimization after every time iteration \footnote{At every full time iteration. That is, 
we do not perform the minimization at the intermediate timesteps of the Runge-Kutta integration, we  
have not explored this possibility.}.

\subsubsection{Sensitivity to the tolerance value}

\begin{figure}[ht]
\begin{center}
\includegraphics*[height=6cm]{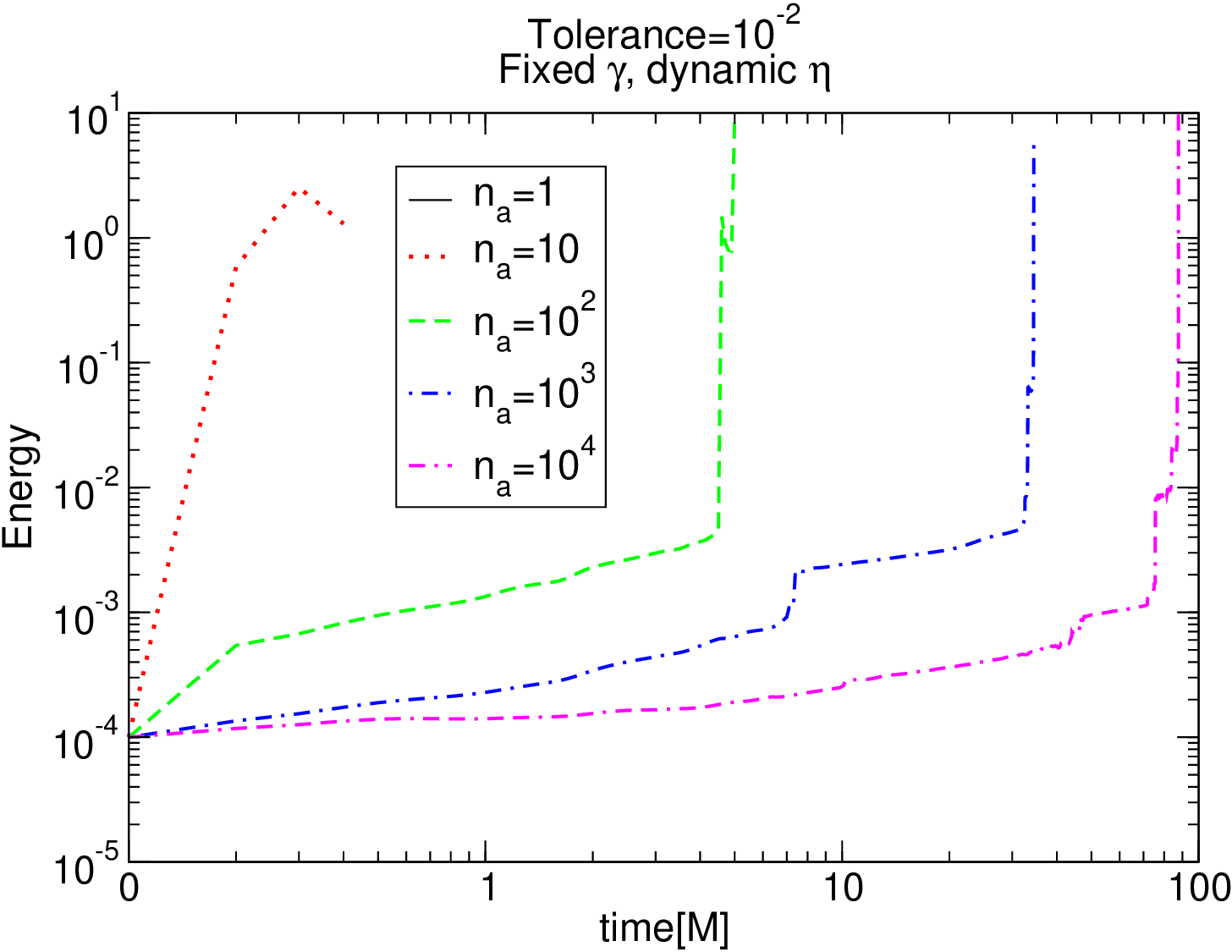}
\includegraphics*[height=6cm]{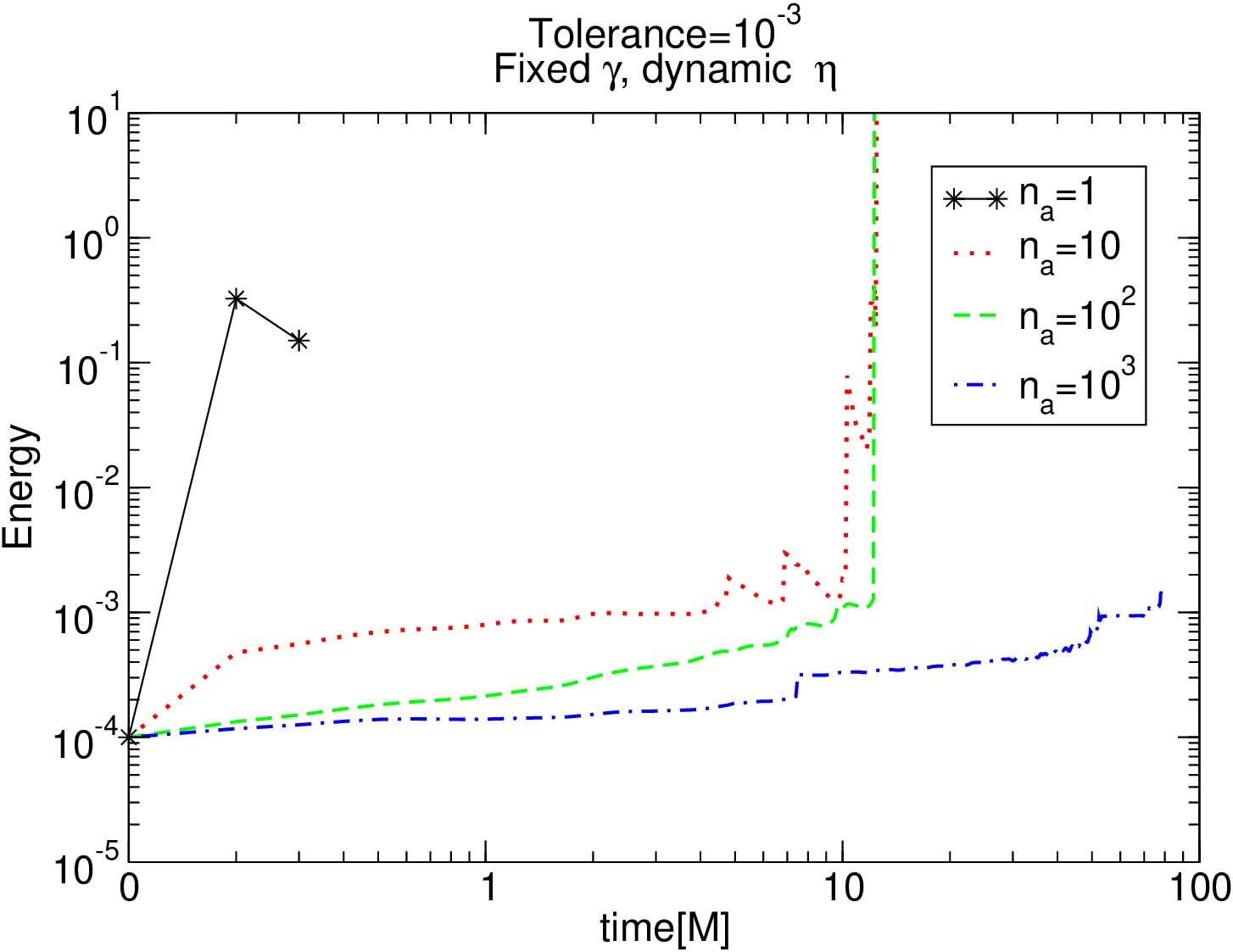}
\caption{This Figure and Figure~\ref{tolerance2} demonstrate the
influence of the tolerance value, $T$, on the results of the constraint
minimization.  The two-constraint-function formulation is used, with $\gamma =0$
and $\eta(t)$ chosen dynamically.  The results are not sensitive to the
chosen tolerance value, provided that one avoids large variations in
$\eta(t)$ by using appropriate values of $n_a$. However, notice that
``appropriate'' values of $n_a$ naturally keep
the value of the energy close to its initial, discretization value (here
given by $0.99925 \times 10^{-4}$. Therefore, keeping the constraint energy near
its initial discretization value appears to give the best results. 
This figure shows runs with $T=10^{-2}$ and $T=10^{-3}$.
Fig.~\ref{tolerance2} shows runs with $T=10^{-4}$ and $T=10^{-5}$.}
\label{tolerance} 
\end{center}
\end{figure}

\begin{figure}[ht]
\begin{center}
\includegraphics*[height=6cm]{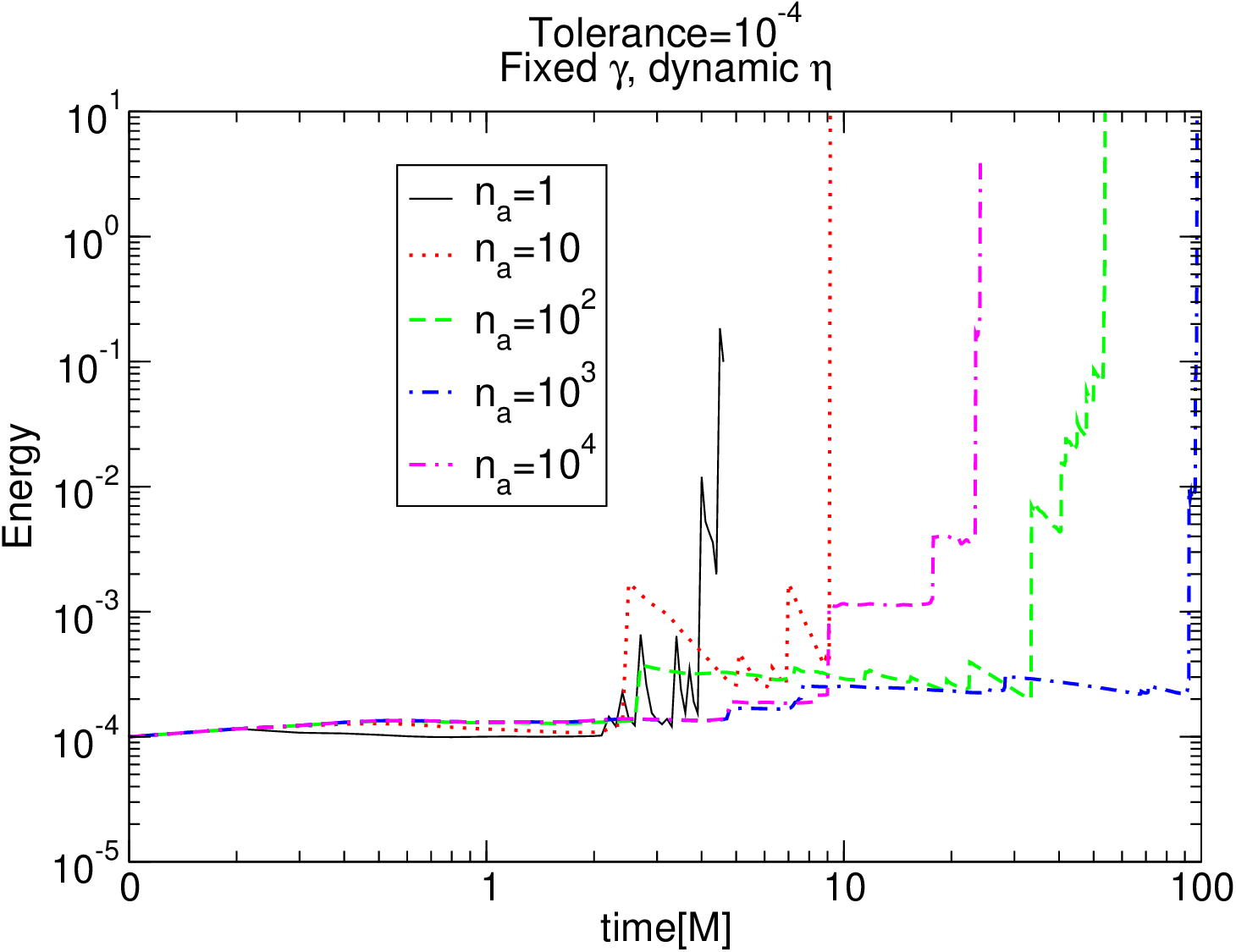}
\includegraphics*[height=6cm]{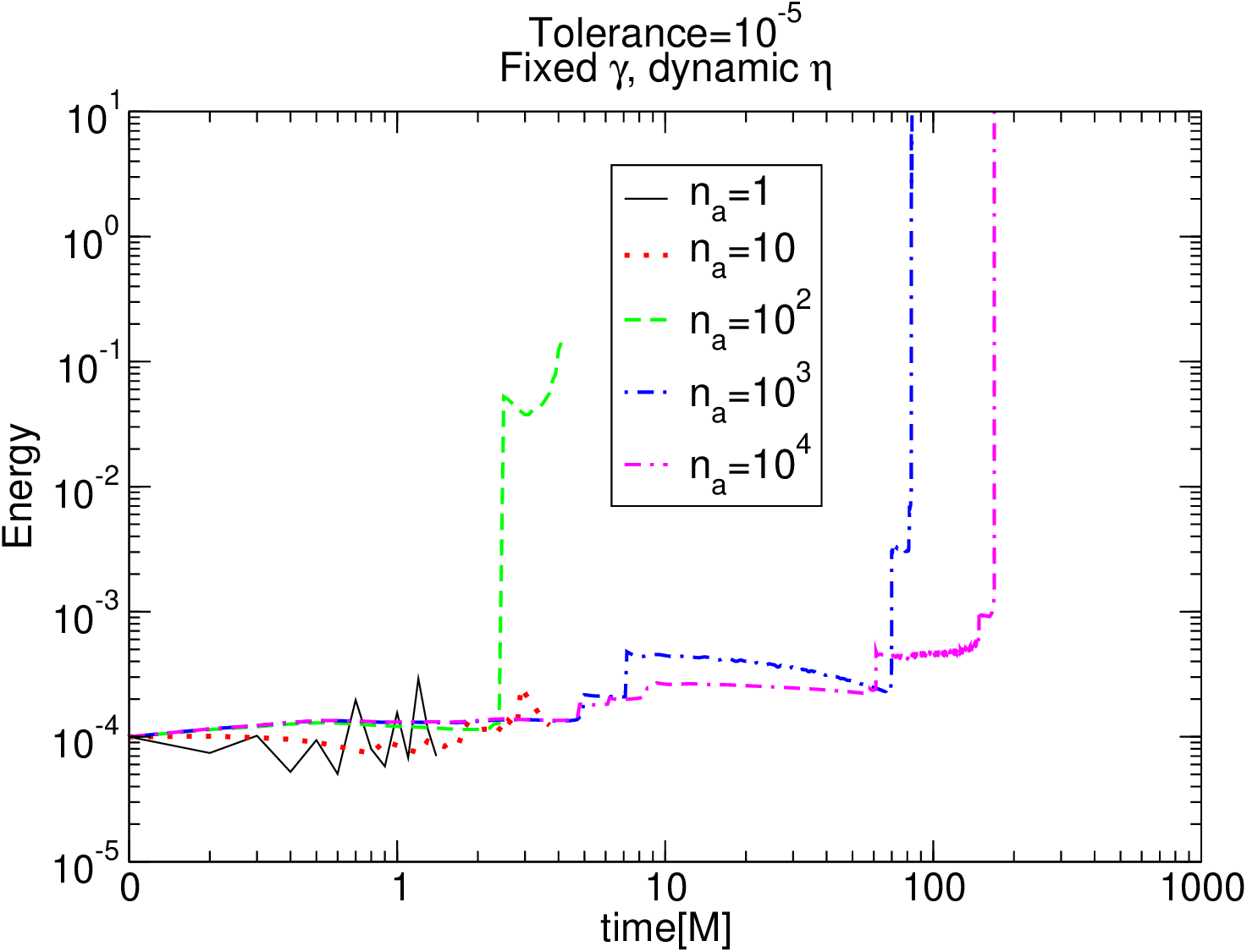}
\caption{This figure compares results for $T=10^{-4}$ and $T=10^{-5}$.
See Fig.~\ref{tolerance} for additional information.}
\label{tolerance2} 
\end{center}
\end{figure}

We choose constraint-functions in the constraint minimization procedure such that the
energy, $\cal N$, decays to a tolerance value, $T$, after certain number
of time steps, $n_a$.  In this section we discuss reasonable choices for
$T$, and discuss its the influence on the final solution.  There are
some reasons to believe that a value close to the initial discretization
error is a good choice~\cite{dyn}.  Here we present additional evidence
for this by comparing simulations with different values of $T$ and $n_a$.

Figures~\ref{tolerance} and \ref{tolerance2} show $\cal N$ as a
function of time, for $T=10^{-2},10^{-3},10^{-4},10^{-5}$ respectively,
each one with $n_a=1,10,10^2,10^3,10^4$.  It can be seen that one can
indeed perform similarly choosing a tolerance value that is below the
discretization error by using an appropriate value for $n_a$.  But notice that
the longest runs are obtained when $\cal N$ naturally has a value near
the initial discretization error.   Even if $T$ is very small, a large value
for $n_a$ allows the constraints-functions to change only very slowly, resulting in
slow changes of $\cal N$ towards $T$.  In summary, a combination of $T$
and $n_a$ that keeps $\cal N$ near the initial discretization error appears
to give the best results.

\subsubsection{Limitations of the constraint-minimization procedure} 
\label{why}

\begin{figure}[ht]
\begin{center}
\includegraphics*[height=6cm]{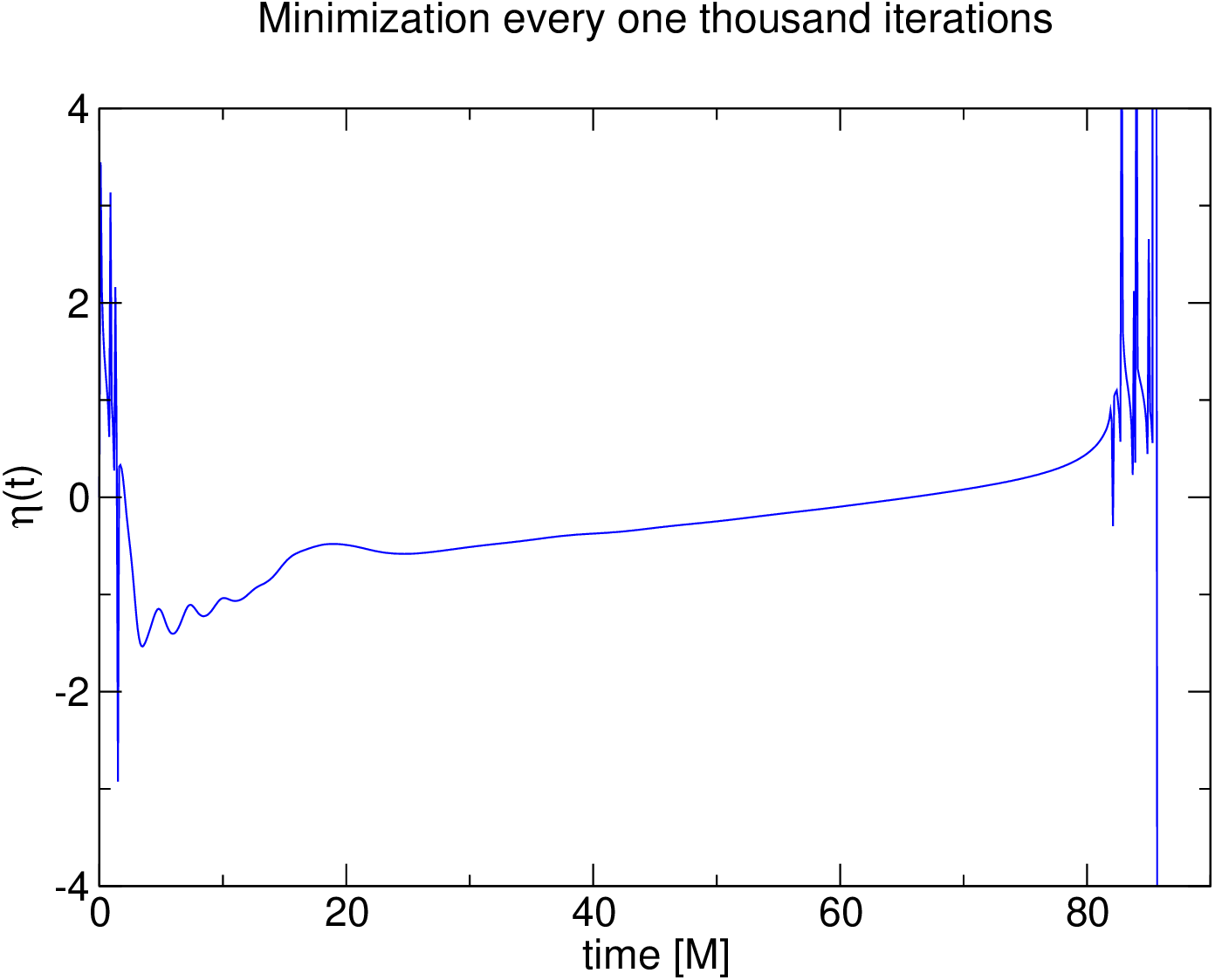}
\includegraphics*[height=6cm]{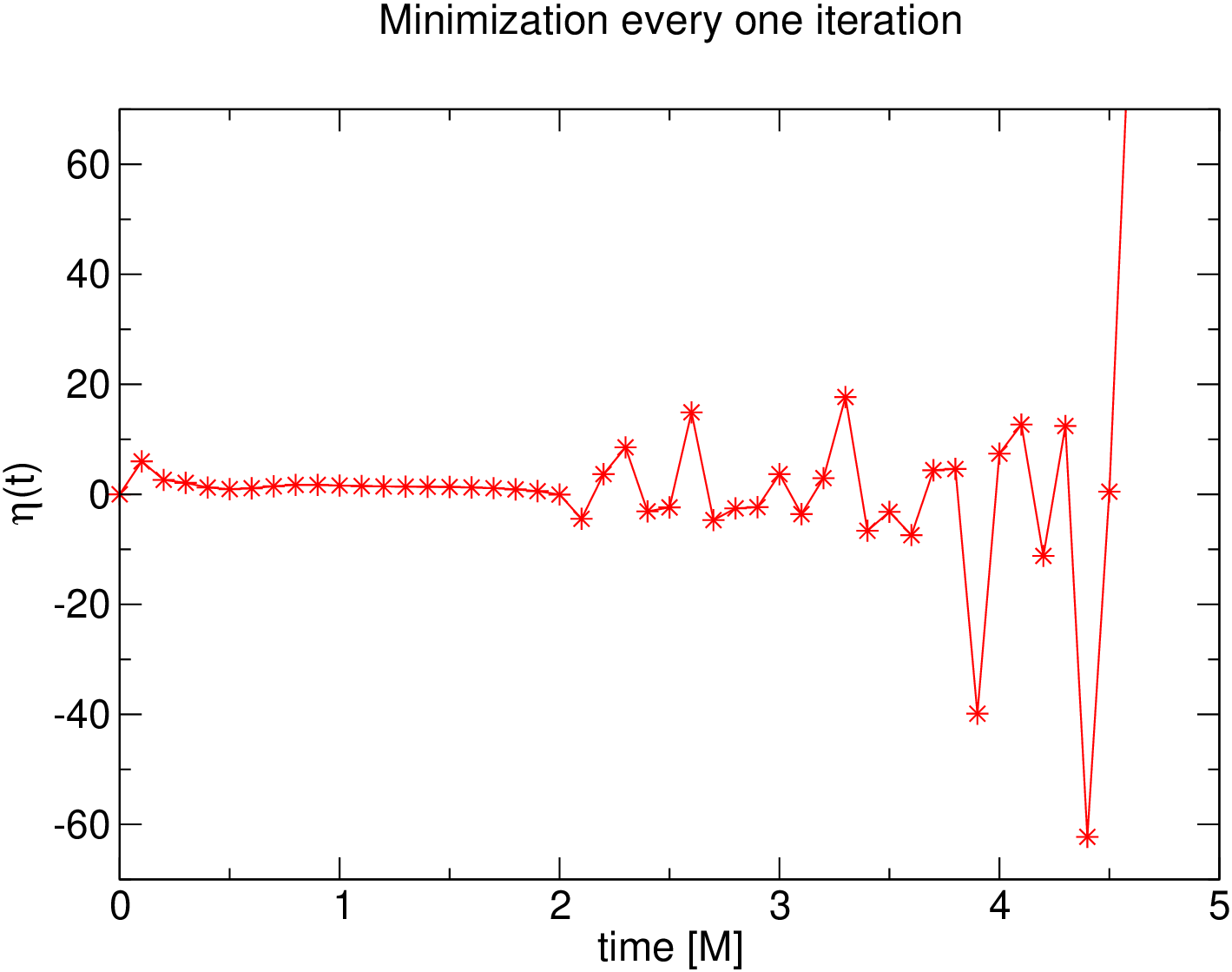}
\caption{These plots show the associated $\eta(t)$ for the $n_a=1,1000$
  cases of the previous plot, in more detail. Notice
the scale in the $n_a=1$ case, $\eta $ changes a lot per time step. On
  the other hand, notice how in the $n_a=10^4$ case $\eta(t)$ changes
  slower but still changes, and eventually it needs to take very large
  values in order to control the constraints.}
\label{etas}
\end{center}
\end{figure}

Fig.~(\ref{energy1_every_it_1}) shows that the constraint minimization
gives an improvement of five to ten times in the lifetime of the
simulations, compared to the results shown in Fig.~(\ref{inner}), the
method does not prevent the eventual code crash when  $\cal N$ is very large.
Given that the constraint minimization method is designed to prevent
this, we consider possible reasons why the method eventually fails.
Operationally, the failure seems to occur because large scale variations
of $\eta$ on ever decreasing time scales are required at late times.
It is not possible for the code to resolve such variations in the
fields with a fixed Courant factor.  This was partly analyzed in
Section~\ref{every}, where we discussed the dependence of $\eta(t)$
on $n_a$ (see Figure~\ref{etas}).  Furthermore, when $n_a$ is large,
$\cal N$ does not return immediately to a value near $T$, but changes
are affected slowly.  Thus, if $\cal N$ grows on time scales of a large number of steps, 
the minimization method is unable to halt the growth.
Having  evidence of why the code crashes, we now can attempt improvements
o the method, as discussed in the next section.

\subsection{Two dimensional minimization and numerical results} 
\label{runs}

In this section we exploit the freedom in the two-constraint-function family to
extend the lifetime of black hole simulations.
The tolerance value for the energy is chosen to be a value roughly
one order of magnitude larger than the initial discretization error, and
$n_a$ is set to either $10^2$ or $10^3$. 
The boundaries are placed at $5M$, as in the runs discussed previously, 
and also at $10M$ and $15M$.

We exploit the fact that there are two free constraint-functions to achieve not
only a given tolerance value, but also to minimize the change in the
constraint-functions $\eta(t), \gamma(t)$, to prevent fast variations in these
constraint-functions, as explained below. Thus, with two constraint-functions to achieve
the desired tolerance value, we can impose an additional condition
to minimize the variation in the constraint-functions from one time step to the
other one.  The motivation for this condition comes from the discussion
in the previous section on the limitations of the constraint minimization
technique, where large oscillations in the constraint-functions were needed in order
to keep the constraints under control. Therefore, it seems reasonable
at this stage to conjecture that the lifetime would be extended even
more if one was able to apply the constraint minimization in a way such
that fast variations in the constraint-functions are not needed. As shown below,
this conjecture seems to be correct.
Therefore,  within all the constraint-functions that achieve the desired 
energy growth for the constraints, we choose at time step $n+1$ 
the pair that minimizes the quantity
\begin{equation}
\triangle := \left[ \eta(n+1) - \eta(n) \right ]^2 
     + \left[ \gamma (n+1) - \gamma(n) \right]^2 \label{triangle}
\end{equation}

To apply this condition, consider that Eq. (\ref{edot_bi}) indicates that
  $\dot{\cal N}$ is non-linear
in $\gamma $ but linear in $\eta$, allowing one to solve for $\eta$
such that $\dot{\cal N} = -a{\cal N}$,
\begin{equation}
\eta  = \frac{ - ( a{\cal N} + {\cal I}^{hom} +
  {\cal I}^{\gamma}\gamma)(1+2\gamma )  + 2\gamma
{\cal I}^{\chi}}{{\cal I}^{\eta} (1+2\gamma ) +
 \gamma {\cal I}^{\chi}}  \label{eta_tent}
\end{equation}
where, as in Section III, $a$ is given by Eq.(\ref{a}).

A set of values for $\gamma$ is chosen within some arbitrary, large interval. For each
$\gamma$ the corresponding $\eta$ given by Eq.~(\ref{eta_tent}) is
computed, and the pair $(\eta, \gamma )$ that minimizes $\triangle $
defined in Eq. (\ref{triangle}) is chosen. As explained in Section, the
constraint-functions $(\gamma, \eta )$ can take any value, except for $\gamma =
-1/2$, value for which the equations are singular. Therefore, there are
two ``branches'', we have only explored the one associated with $\gamma <
-1/2$, by using as seed values $\eta =0, \gamma=-1$ and restricting the
minimization procedure to  values $\gamma < -1/2$.
As discussed next,  almost an extra order of magnitude in the
lifetime of the simulations can be obtained, and the run that
initially lasted for ~$10M$ without the minimization of the constraints,
now runs for around $700M-1000M$ (without any symmetry -- bitant, octant, or of any other type-- 
imposed).

\subsubsection{Boundaries at $5M$.}

\begin{figure}[ht]
\begin{center}
\includegraphics*[height=6cm]{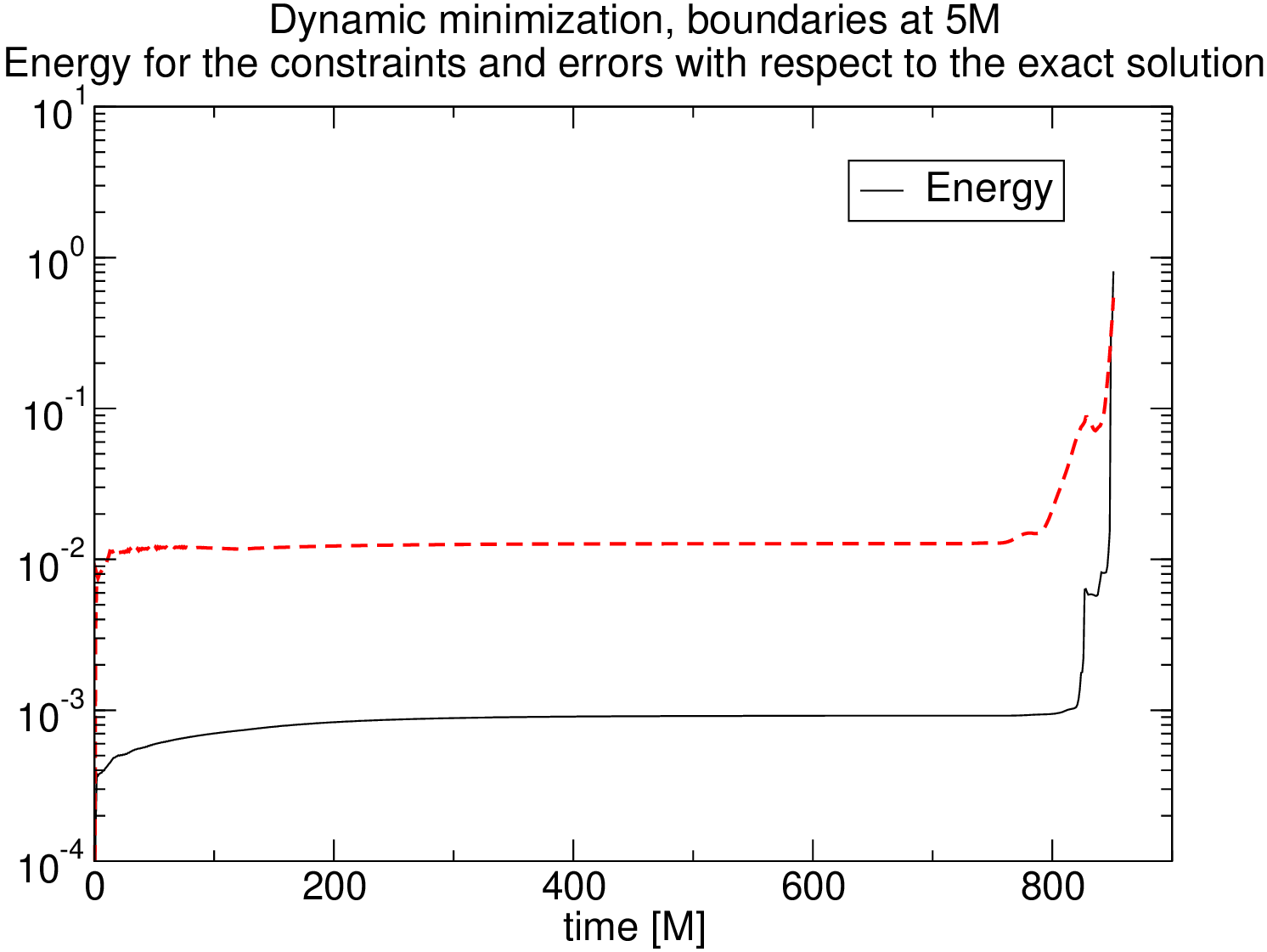}
\includegraphics*[height=6cm]{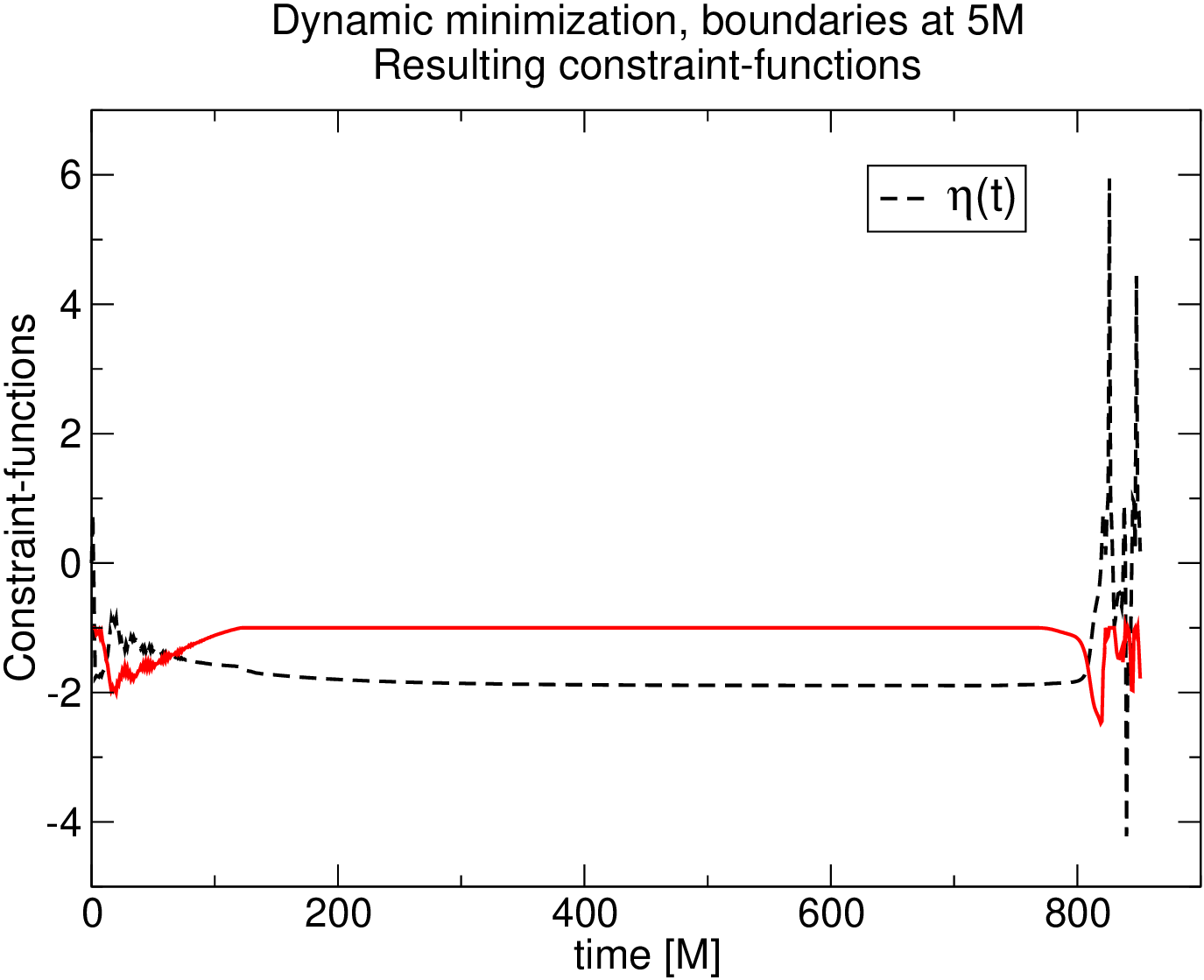}
\caption{Dynamic minimization done with boundaries at $5M$, $\triangle =M/5$, 
$T=10^{-3}$, and $n_a=10^3$.  }
\label{bi_dyn_bound5m} 
\end{center}
\end{figure}

\begin{figure}[ht]
\begin{center}
\includegraphics*[height=6cm]{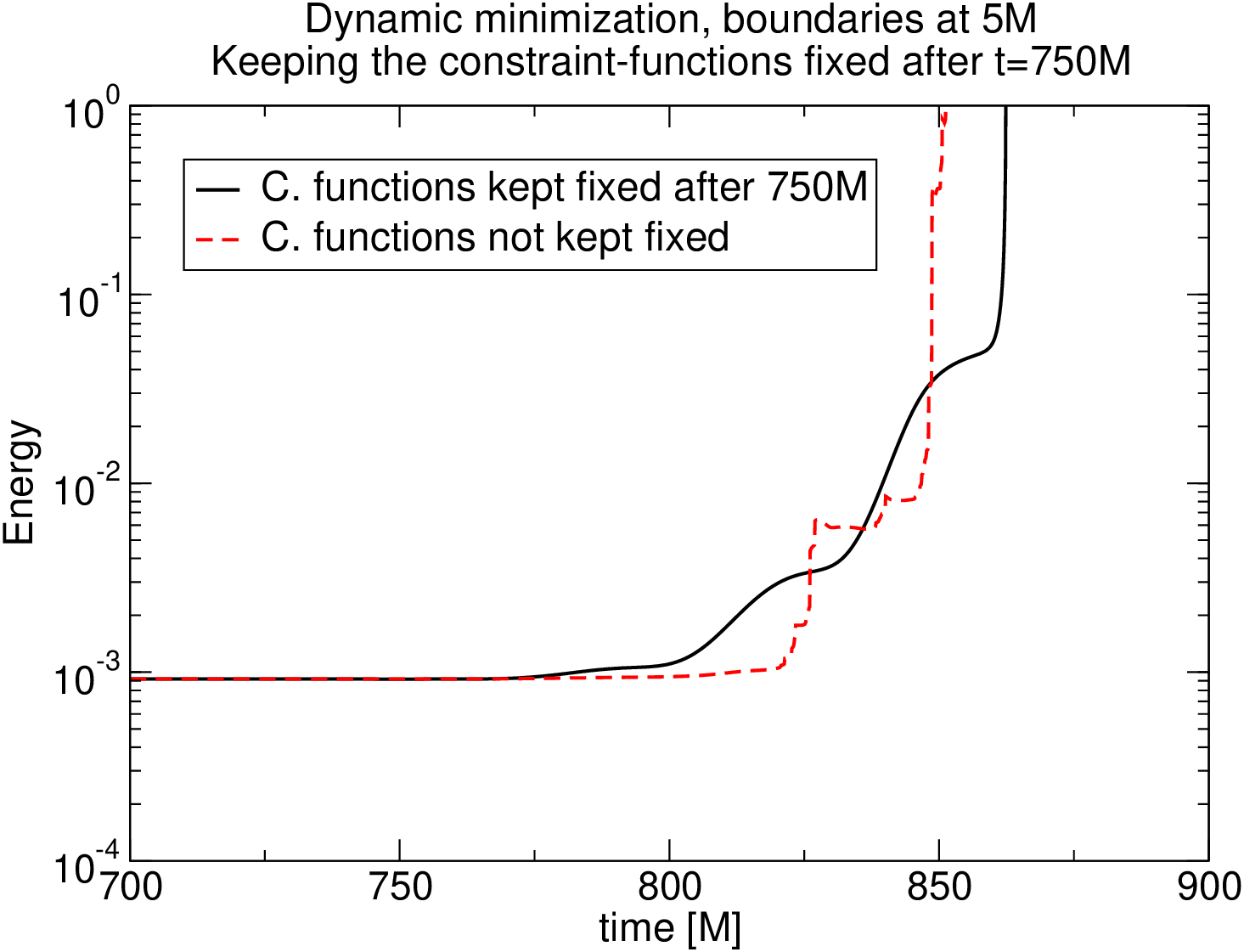}
\caption{Same as previous Figure, but keeping the constraint-functions constant 
  after $750M$. The figure compares the resulting energy
  for the constraints with that of the previous figure (shown at late
  times only,  since because of the setup the runs are identical up to $t=750M$).}
\label{bi_dyn_bound5ma} 
\end{center}
\end{figure}

\begin{figure}[ht]
\begin{center}
\includegraphics*[height=6cm]{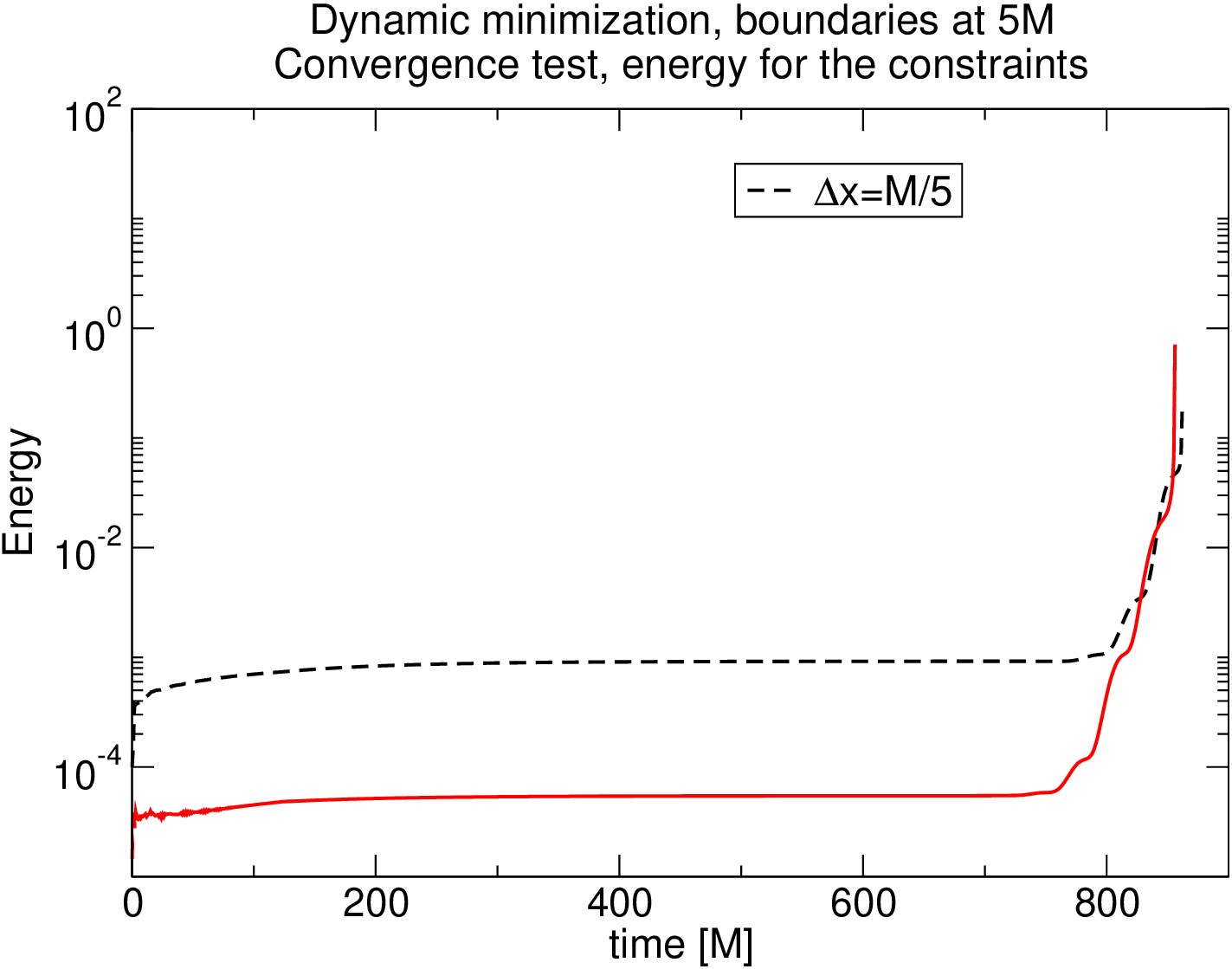}
\includegraphics*[height=6cm]{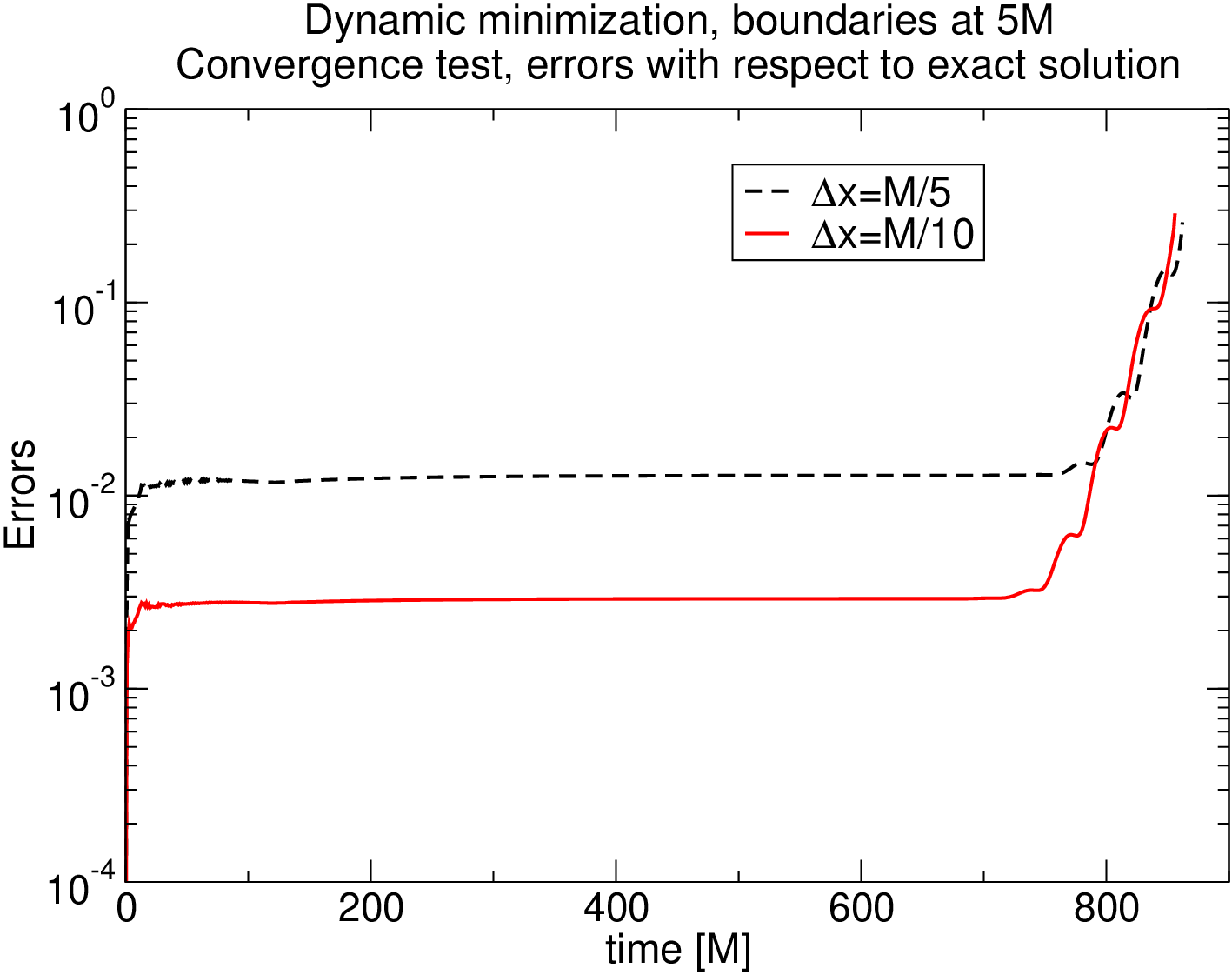}
\caption{Convergence test, with dynamic constraint-functions kept constant after $750M$.}.
\label{bi_dyn_conv_bound5m} 
\end{center}
\end{figure}

Figure~\ref{bi_dyn_bound5m} shows the results of a simulation with
resolution $\triangle =M/5$, $T=10^{-3}$ (close to the initial discretization
error value, given by Eq. \ref{tol5m} ) and $n_a=10^3$.  The figure shows that as the code crashes
the constraint-functions start having large and fast variations. A natural question
that this raises is whether these variations are a cause or consequence
of the code crashing. For reasons discussed below, they appear to be
a consequence.  Figure~\ref{bi_dyn_bound5ma} shows the same run as
that shown in Fig.~\ref{bi_dyn_bound5m}, except that the minimization
is stopped at $750M$ (at which point the constraint-functions are,
 nearly constant), and from there on the last value of the constraint-functions
is used; namely,
\begin{equation}
\eta = -1.88, \gamma = -1.00 \label{pars5m}
\end{equation}
The code still crashes at roughly the same time. Therefore, the 
variations in the constraint-functions observed in
Fig. \ref{bi_dyn_bound5m} do not cause the code to crash, but appear
to be a consequence of other instabilities.
Figure~\ref{bi_dyn_conv_bound5m} shows a convergence test with two
resolutions ($\triangle =M/5$, $\triangle =M/10$), and keeping
the constraint-functions (obtained from the $\triangle =M/5$ resolution run) fixed 
after $750M$.

\begin{figure}[ht]
\begin{center}
\includegraphics*[height=6cm]{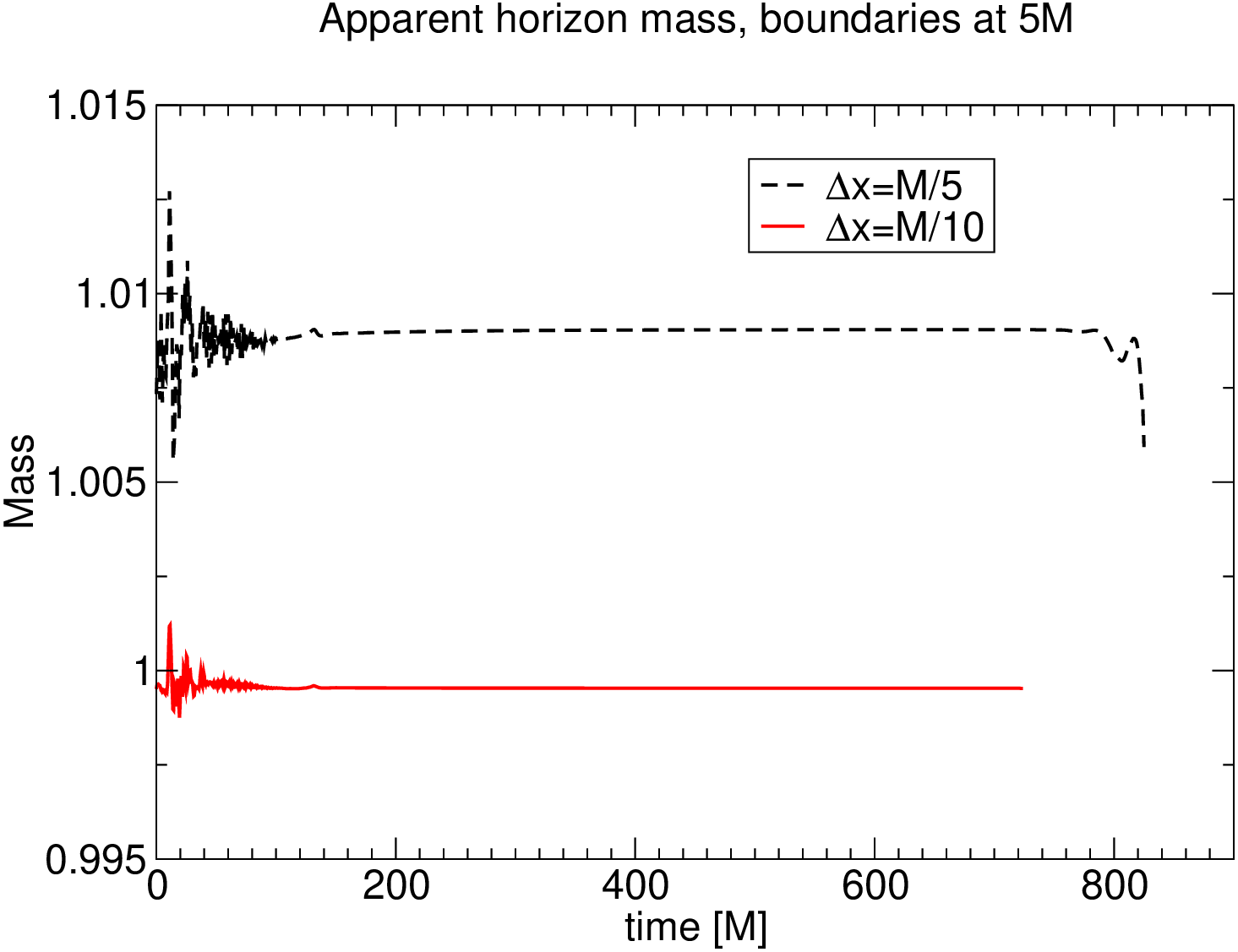}
\caption{Apparent horizon mass, with dynamic constraint-function values. 
The run with higher resolution ran out of 
computing time (in the equivalent plots of the previous Figure an apparent horizon 
was not searched for).}
\label{ah} 
\end{center}
\end{figure}

Another measure of the error in the solution is the mass of the apparent
horizon. Figure~\ref{ah} shows this mass using the dynamic constraint-functions
obtained from the simulation of Figure~\ref{bi_dyn_bound5m} and
by running at each iteration  Thornburg's apparent horizon 
finder~\cite{thornburg}. For the coarsest resolution, the initial value of
the mass, as given by the horizon finder, is $1.007M$. 
Compared to this
value, the initial oscillations have a relative error of less than one
percent. After some time, the mass approximately settles down to a value
that is around $1.009$, which corresponds to an error of the order of
one part in one thousand.  For the higher resolution, the initial value
of the mass as given by the horizon finder is $0.99951$. With respect
to this value the initial oscillations are at most of the order of one
part in one thousand, and at late times the apparent horizon mass settles down to $0.99953$,
which corresponds to a relative error of one part in $10^5$.

\begin{figure}[ht]
\begin{center}
\includegraphics*[height=6cm]{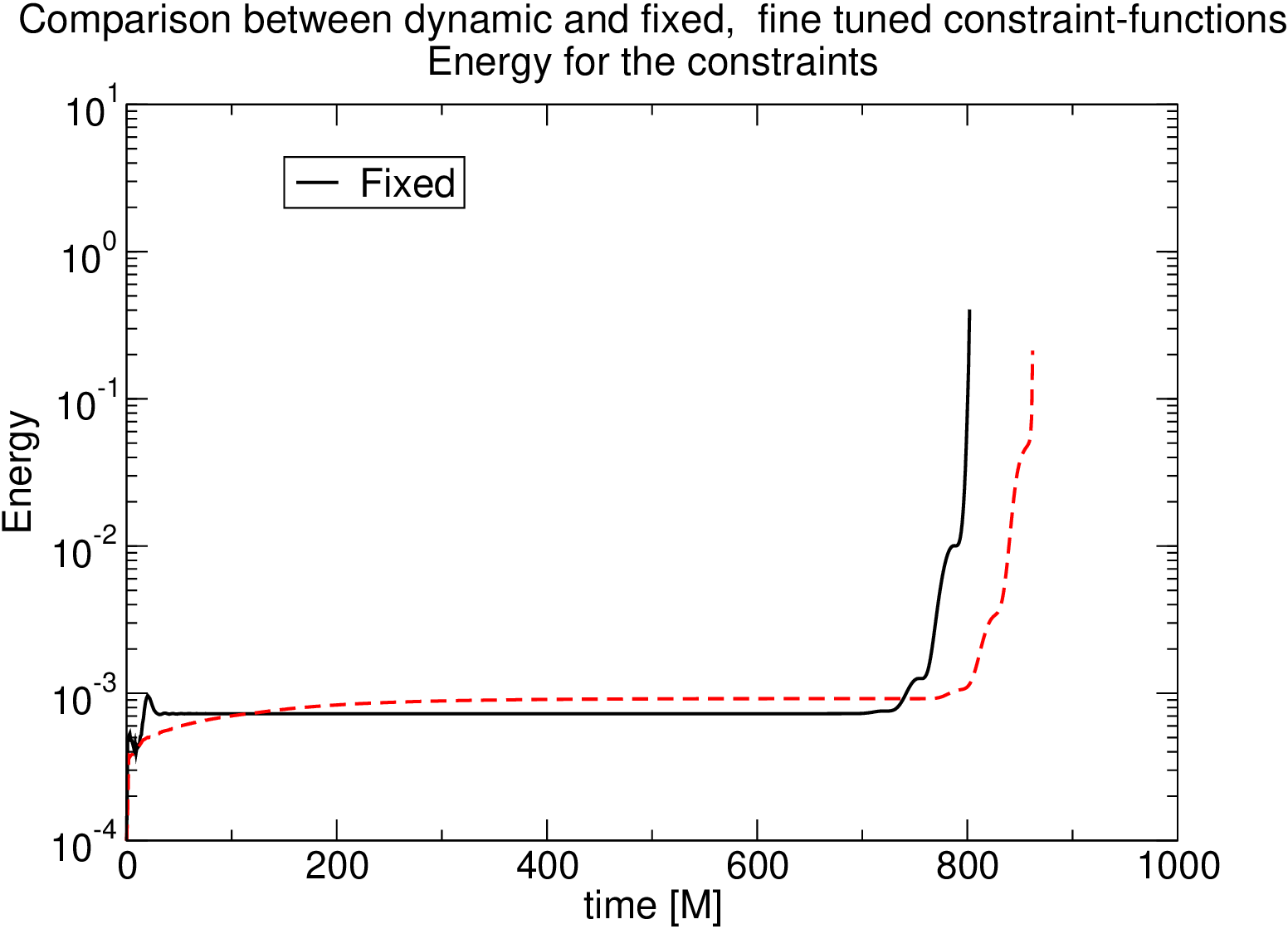}
\includegraphics*[height=6cm]{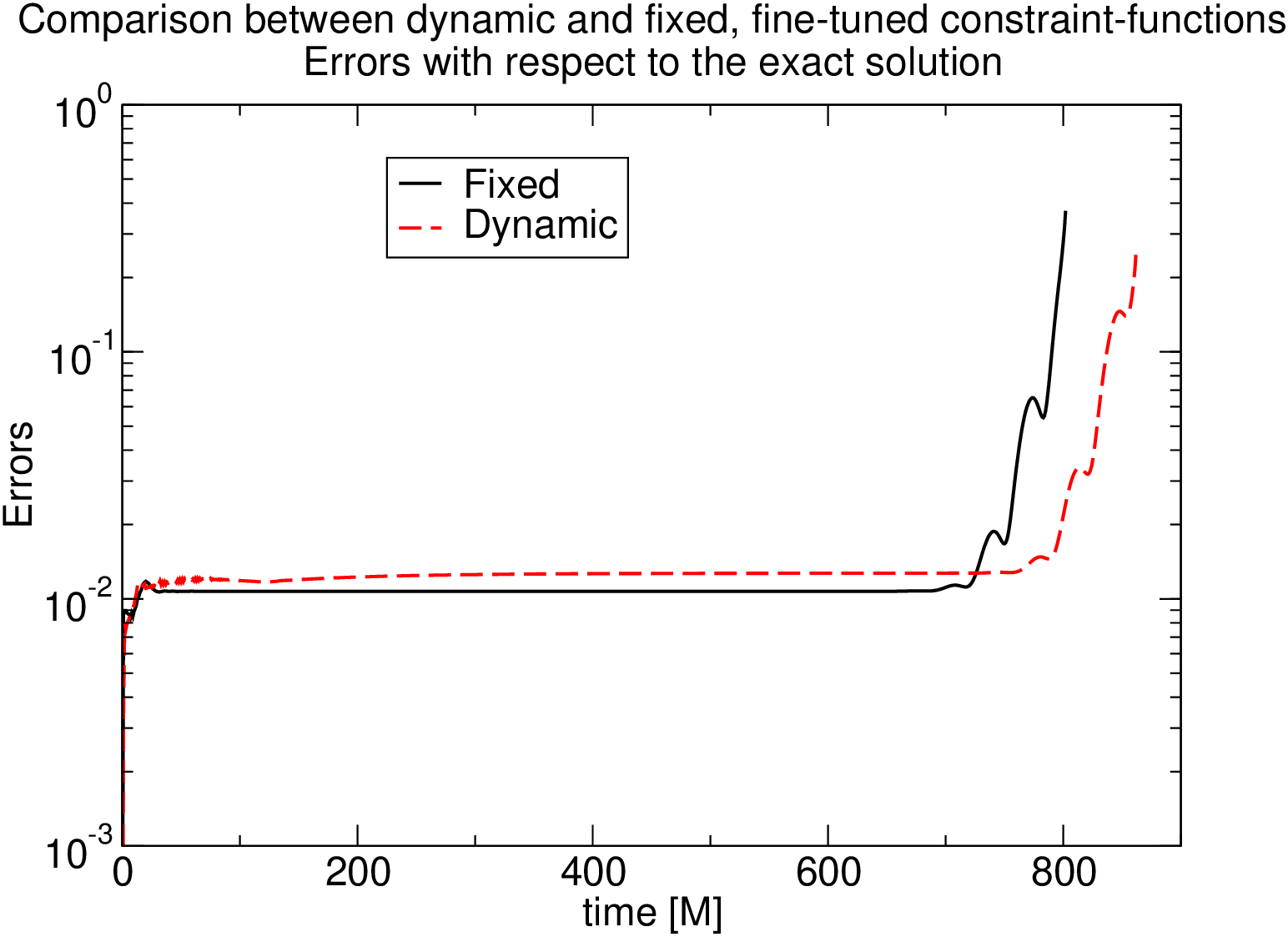}
\caption{Comparison between dynamic and fixed, fine-tuned
constraint-function values given by the asymptotic values at which they 
approach [cf. Eq. (\ref{pars5m})].}
\label{comparison} 
\end{center}
\end{figure}

Even though the constraint-functions do not settle down completely to
a stationary value, they oscillate very little. One question that this
raises is how does the code perform if one fixes these values given 
by Eq.~(\ref{pars5m}) from the very beginning.
The plots in figure~\ref{comparison}  make this comparison.
Interestingly, the run with dynamSic minimization
runs slightly longer, even when the constraint-functions after some time are
essentially constant, and even though the solution being evolved
is stationary at the analytical level.  This shows that the dynamic
minimization not only requires little experimentation  but also seems
to be effective in that it naturally allows for variations in time that
accommodate to the variations of the numerical solution.

\subsubsection{Boundaries at $10M$.}

\begin{figure}[ht]
\begin{center}
\includegraphics*[height=6cm]{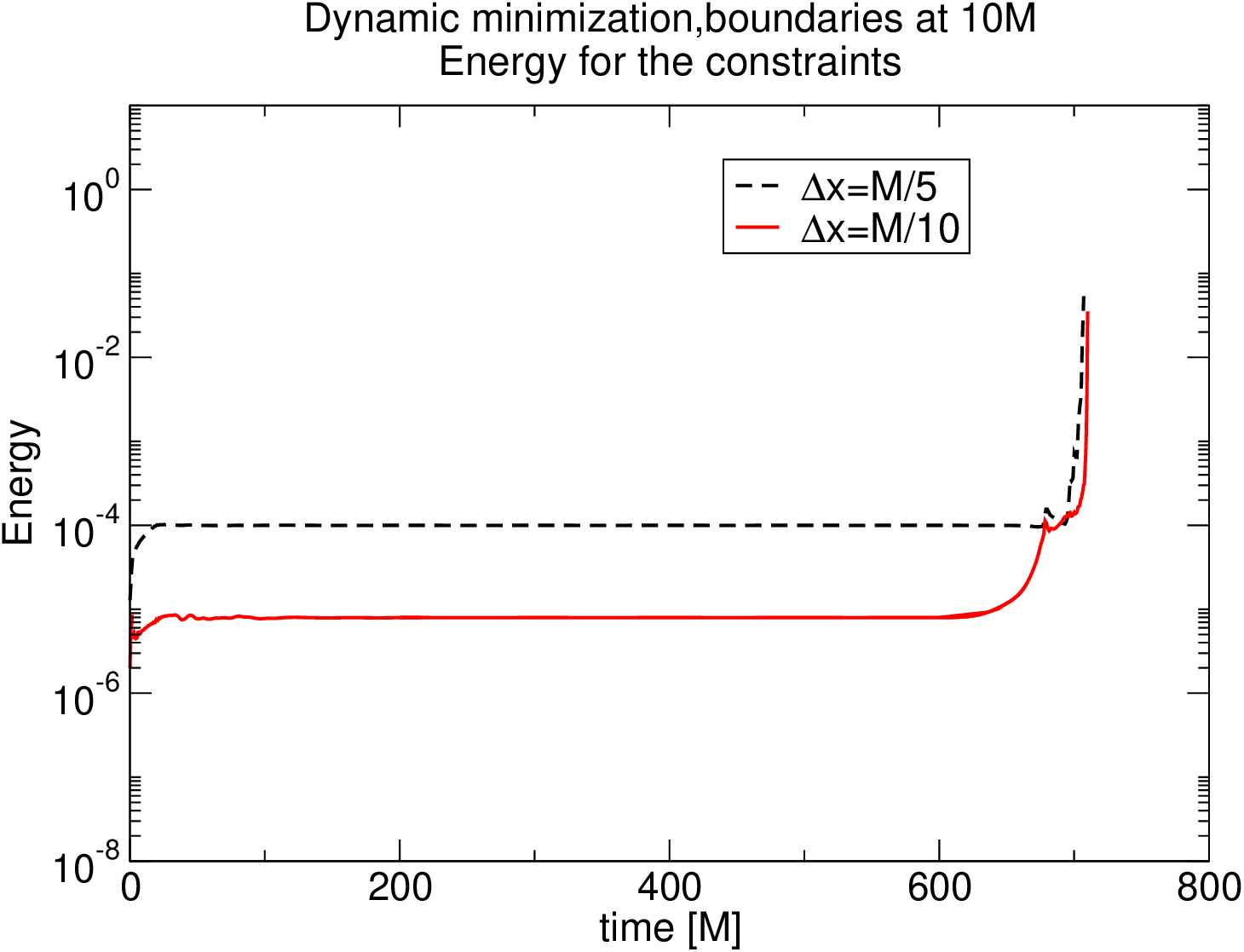}
\includegraphics*[height=6cm]{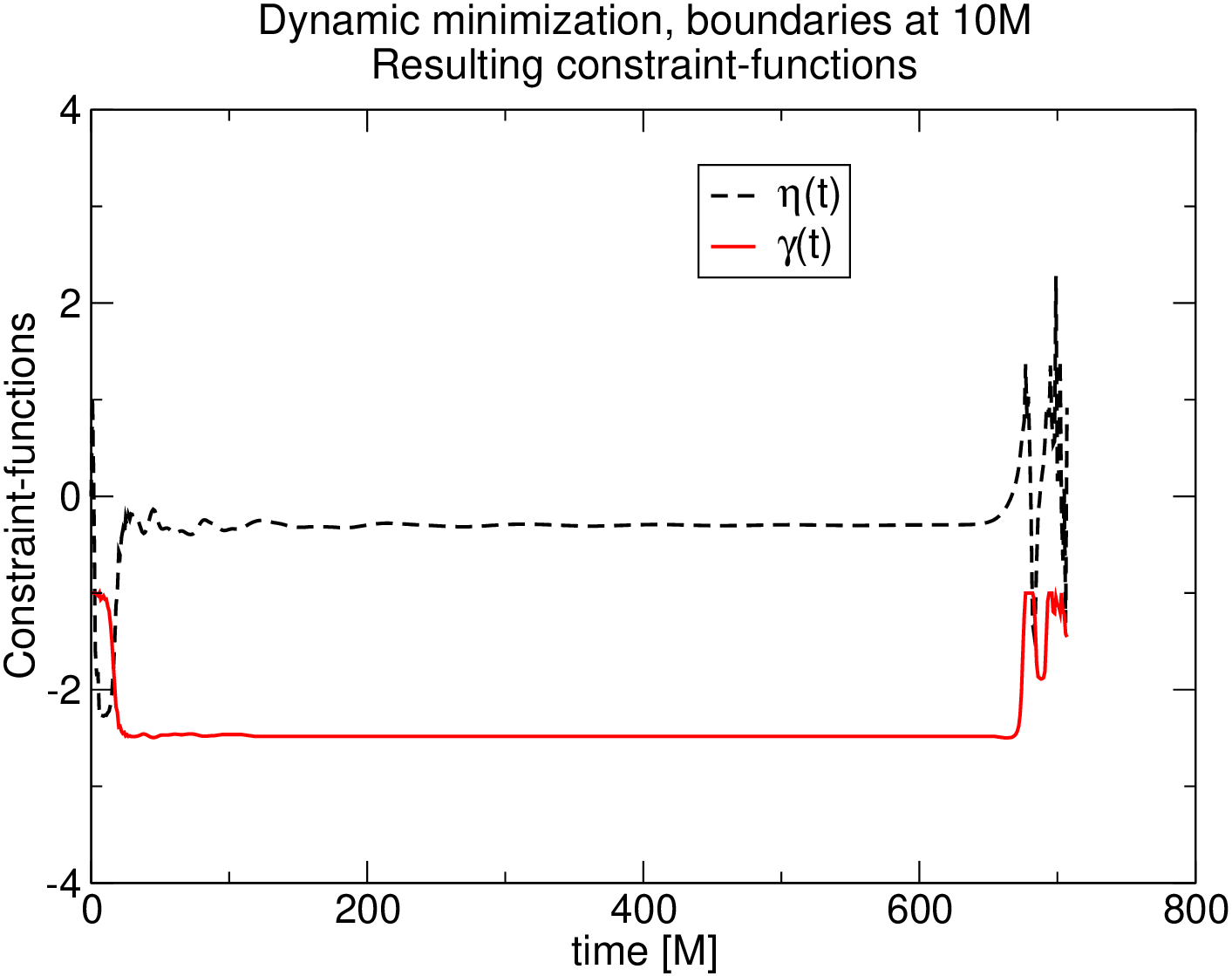}
\caption{Dynamic minimization done with boundaries at $10M$, $\triangle =M/5$, $T=10^{-4}$, 
and $n_a=10^2$. }
\label{bi_dyn_bound10m} 
\end{center}
\end{figure}

\begin{figure}[ht]
\begin{center}
\includegraphics*[height=6cm]{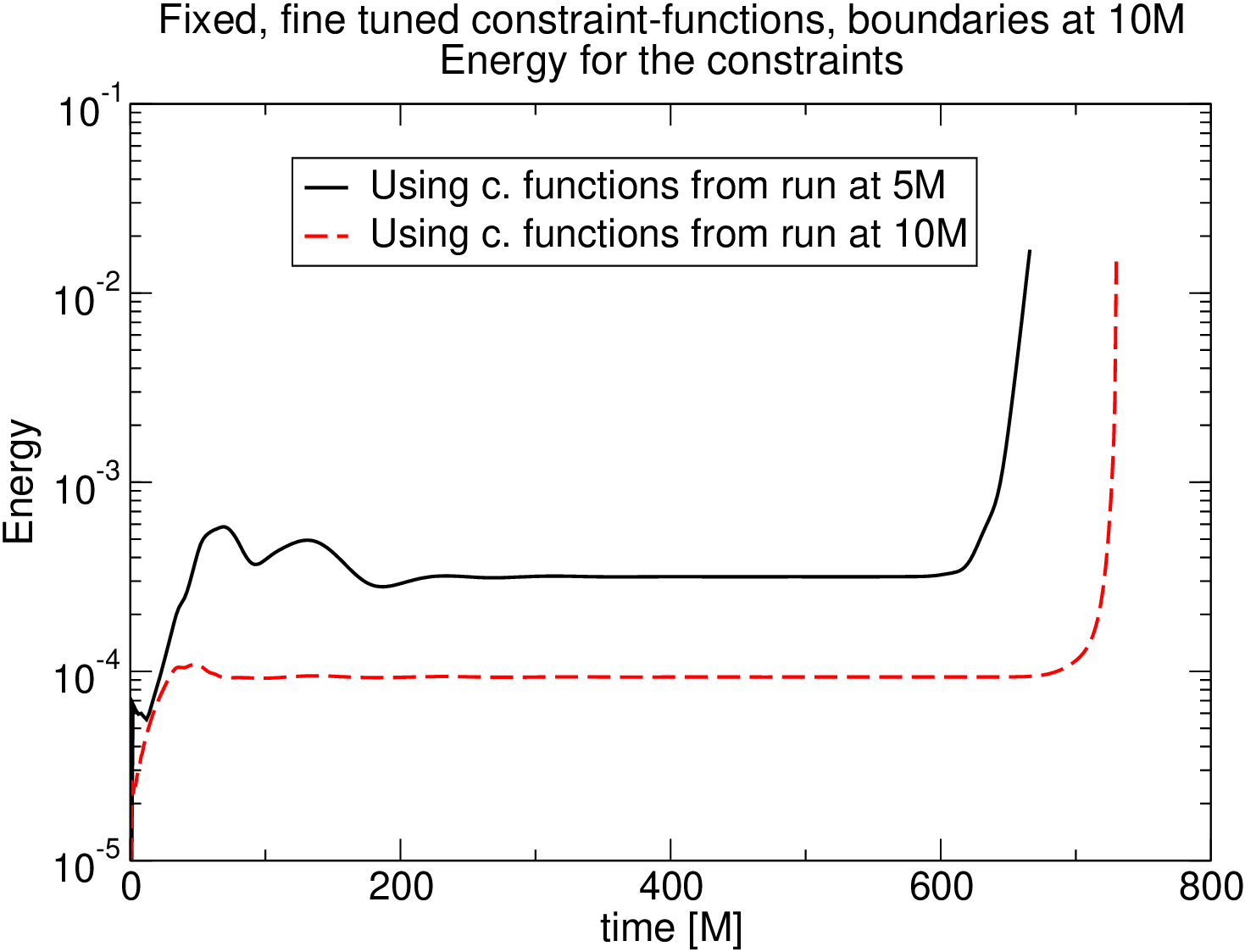}
\caption{Running with boundaries at $10M$, using fixed, but fine tuned 
constraint-functions, obtained from runs with boundaries at $5M$ and $10M$.}
\label{comparisonb10m} 
\end{center}
\end{figure}

We now examine data from a configuration equivalent to that of the
previous section, except that now that the boundaries are at $10 M$.
The initial discretization value for the energy is 
${\cal N}(0) = 1.2845\times 10^{-5}$, and $T=10^{-4}$, and $n_a=10^2$ are chosen.
As seen in the previous case, the constraint-functions eventually settle into
oscillations about fixed values,
\begin{equation}
\eta= -2.96 \times 10^{-1}, \gamma = -2.48, \label{pars10m}
\end{equation}
as shown in Figure~\ref{bi_dyn_bound10m}.  These steady-state
values are quite different from the previous configuration with
boundaries at  $5M$, see Eq.(\ref{pars5m}). 
This raises the question of what
would happen if one ran with boundaries at $10M$, and fixed constraint-functions
given by Eq.(\ref{pars10m}) in one case and Eq.(\ref{pars5m}) in the
other. Figure~\ref{comparisonb10m} makes this comparison. As
expected, the constraint-functions obtained from the run with dynamic minimization
and boundaries at $5M$ do not perform as well as those obtained with
boundaries at $10M$. However, even using the constraint-functions obtained from
the simulation with boundaries at $5M$ is much better than using
a naive choice (say, $\gamma=\eta=0$, which for the resolution
of Fig.~\ref{comparisonb10m} runs for less than $30M$, as shown in
Fig.~\ref{convergence1} ).

\begin{figure}[ht]
\begin{center}
\includegraphics*[height=6cm]{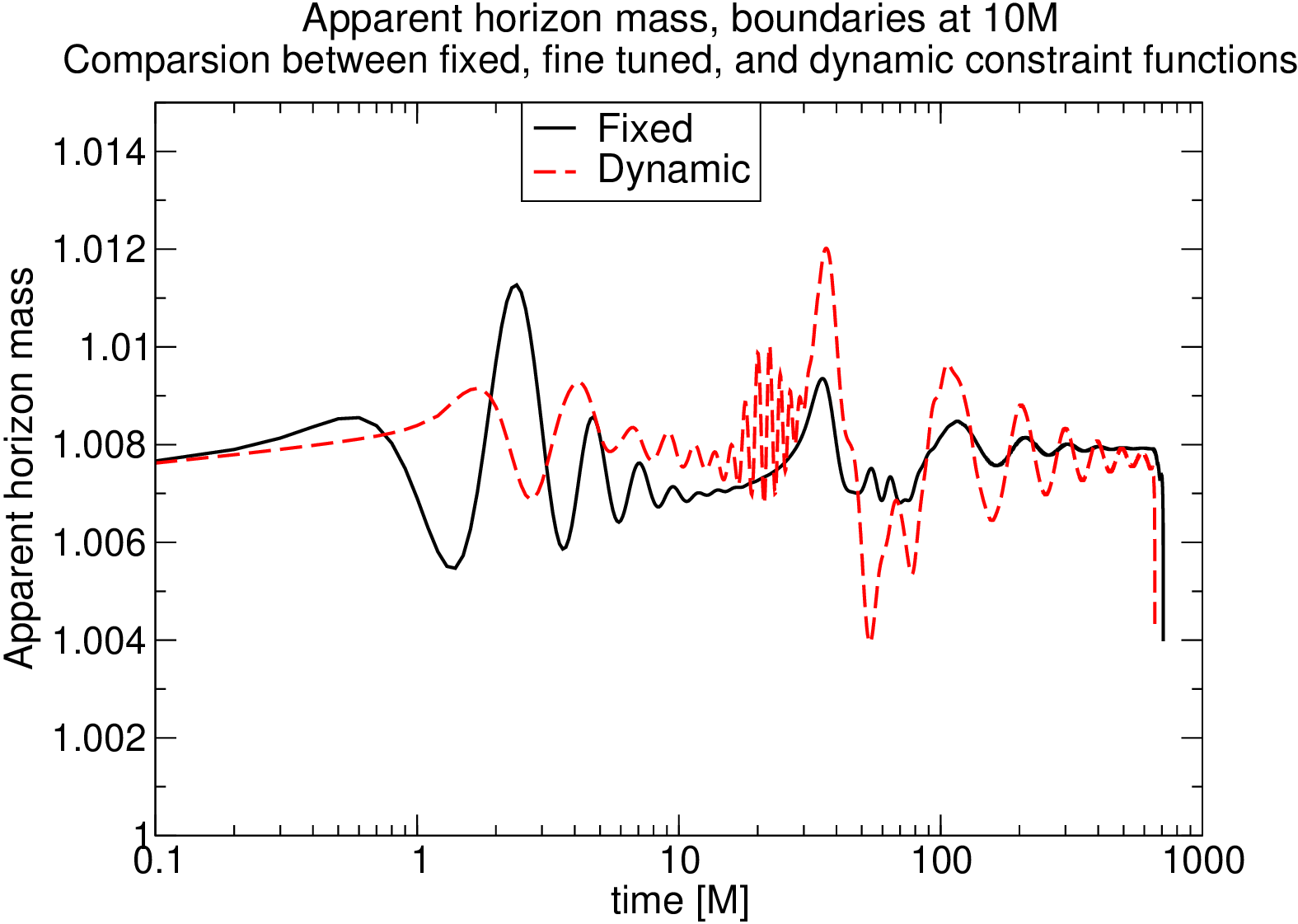}
\caption{Apparent horizon mass for the simulation of Figure 
\ref{bi_dyn_bound10m} and the simulation of 
Fig. \ref{comparisonb10m} with constraint-functions given by Eq.(\ref{pars10m}). 
A logarithmic scale in time is used in order 
to show the oscillations in more detail. The oscillations are not caused 
by time variations in the constraint-functions, since 
they are also present in the fixed-constraint-functions case.}
\label{ah_b10m} 
\end{center}
\end{figure}

Figure~\ref{ah_b10m} shows the apparent horizon mass, for the
simulations of Figures~\ref{bi_dyn_bound10m}, and one simulation
from Fig.~\ref{comparisonb10m} with constraint-function values given by
Eq.(\ref{pars10m}). In both cases the resolution is coarse,
$\triangle =M/5$. As for the case with boundaries at $5M$ and with the same
resolution, the errors are less than one percent when compared to the
mass given  by the initial data.  From Fig.~\ref{ah_b10m} one can also
see that the oscillations in the mass do not seem to be caused by the
time variation of the constraint-functions, as they are still present in the case
in which fixed  constraint-functions are used.

\subsubsection{Boundaries at $15M$.}

\begin{figure}[ht]
\begin{center}
\includegraphics*[height=6cm]{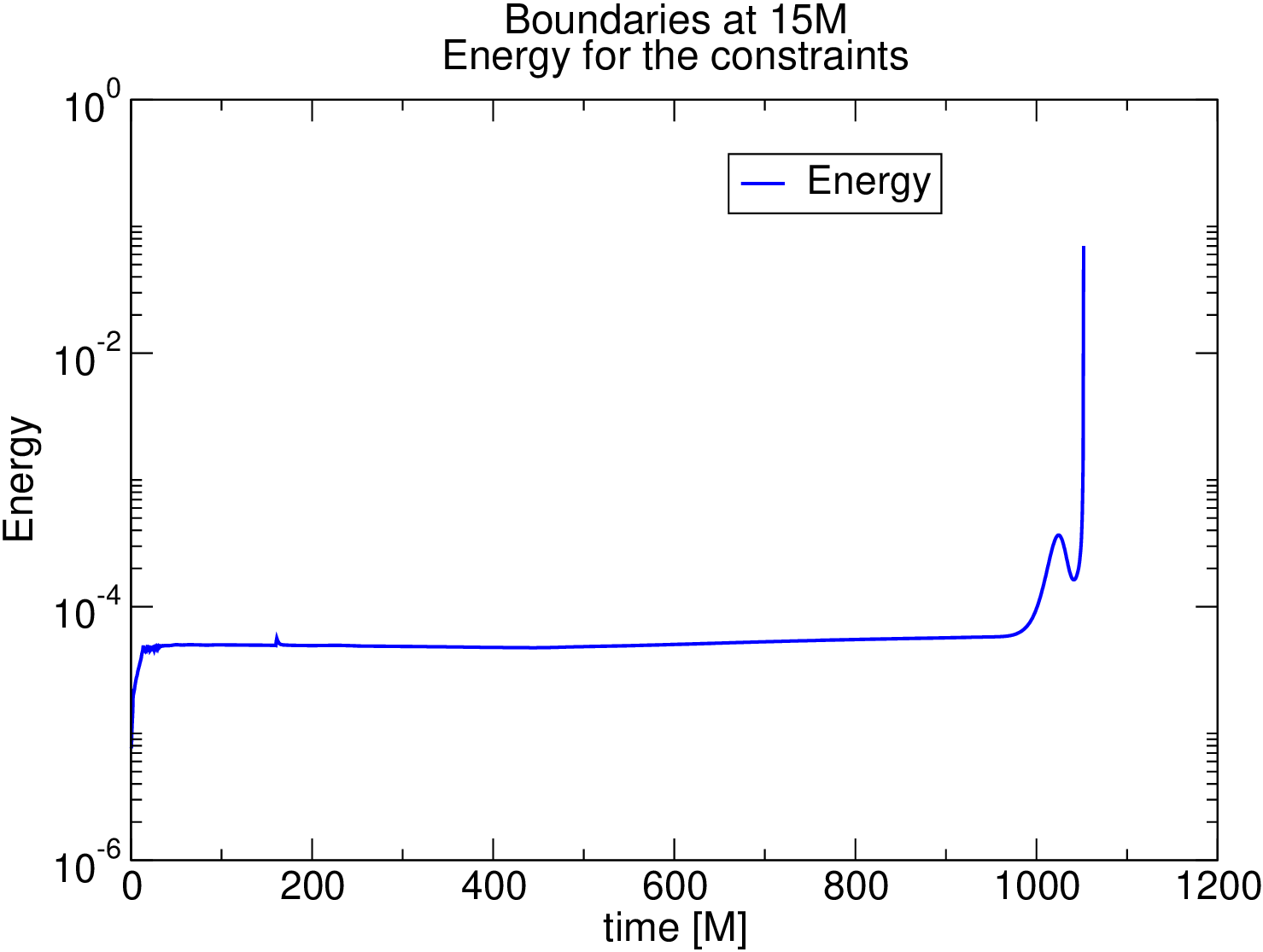}
\includegraphics*[height=6cm]{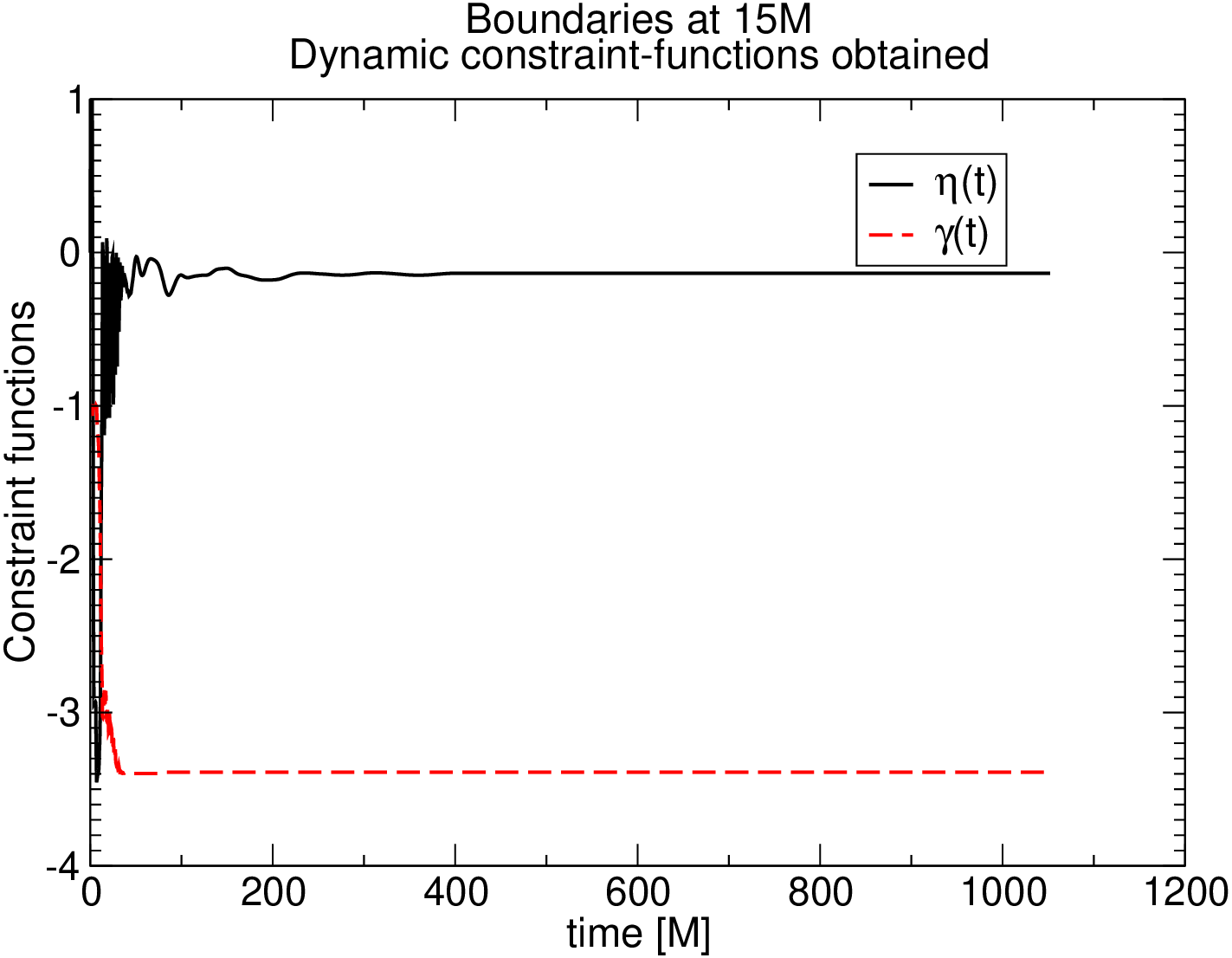}
\caption{Dynamic minimization done with boundaries at $15M$, $\triangle=M/5$,
$T=10^{-5}$, and $n_a=10^2$. The constraint-functions are constant for $t \ge 450M$, 
where they are $\eta = -0.135, \gamma =  -3.389$; thus, there is no response
of the constraint functions when the code is about to crash.}
\label{bi_dyn_bound15m} 
\end{center}
\end{figure}

Finally, we consider a configuration with boundaries at $15M$, though with less detail as before. 
Figure~\ref{bi_dyn_bound15m} shows results data equivalent to those
discussed for Figures~\ref{bi_dyn_bound5m} and~\ref{bi_dyn_bound10m}, 
except now that the boundaries are at $15M$. The initial, discretization 
value for the energy is $7.6459 \times 10^{-6}$, and $T=10^{-5}$, $n_a=100$ was used.
The minimization of the constraint-functions is stopped at $450M$, at which point
the constraint-functions are approximately constant, and equal to
\begin{equation}
\eta = -1.35\times 10^{-1} \;\;\; , \;\;\; \gamma =  -3.39. \label{pars15m}
\end{equation}

Figure~\ref{bi_dyn_bound15m} shows that the dependence of
the lifetime on the location of the outer boundaries is not monotonic, as
for this case the code runs for, roughly, $1000M$, while with boundaries
at $10M$ and $5M$ it ran for around $700M$, and $800M$, respectively. A
detailed analysis of such dependence would be computationally expensive
and beyond the scope of this work, and may even depend on the details
of the constraint minimization, such as the values for $T$ and $n_a$.

\section{Final comments} \label{final}

This paper presents a number of new techniques applied to the simulation of
Einstein equations, namely:  (1) a symmetric hyperbolic formulation
with live gauges; (2) a numerical discretization based on the energy
method with difference operators that satisfy summation by parts and
a projection method to apply boundary conditions; (3) a constraint
minimization method for dynamically choosing constraint-functions that multiply
the constraints in the evolution equations without requiring special
knowledge of the solution.\footnote{Some alternatives to control the constraints include   
explicitly solving them instead of some of the evolution equations ({\it constrained
evolution}, see for example \cite{constraintevolv} for recent work), 
projecting the solution onto the constraint surface ({\it constraint projection} 
\cite{constraintproject}),
enlarging the system in a way so as to force the solution to decay
towards the constraint surface ({\it lambda systems} \cite{constraintlambda}), or adopting 
appropriate discretizations and/or gauge choices so that the discrete constraints 
are exactly preserved during evolution \cite{constraintgauge}.}

We use a generalization of the Bona-Masso slicing conditions, and to
date, this is the only formulation of Einstein's equations with this
slicing conditions known to be symmetric hyperbolic (for symmetric hyperbolic 
formulations with other dynamical gauge conditions see \cite{ls}). 
There are {\em strongly hyperbolic} formulations with this 
gauge (see, for example,  \cite{strong} and references therein), 
though,  recall that contrary to
some common belief, strong hyperbolicity does not automatically
define a well-posed initial value problem (IVP).  A well-posed IVP
for strongly hyperbolic can be found by requiring the existence of a
smooth symmetrizer.  However, this smoothness is a non-trivial condition
and it is usually not studied in formulations of Einstein's equations.
Some algebraic conditions do imply the existence of a smooth-symmetrizer \cite{rendall},
but for the Bona-Masso slicings these conditions can only be a 
priori guaranteed for the time-harmonic subcase~\cite{st}. In the presence of  
boundaries the situation is even more complicated: there are examples in the context of 
Einstein's equations explicitly showing ill posedness of certain strongly hyperbolic equations 
which do have smooth symmetrizers, when   maximally dissipative boundary conditions 
are used (while for symmetric systems such a problem is 
known to be well posed) \cite{ill}. 
While we use time harmonic slicing in this paper, the freedom to use other
slicing conditions could prove useful in other scenarios~\cite{gauge}.

Finite difference discretizations based on the energy
method~\cite{gko,exc,sbp2} exploit results that rigorously
guarantee linear numerical stability of IVPs as well as IBVPs.
In particular,  stable simulations of the gauge wave with periodic
boundary conditions are obtained for large values of amplitude, $A$,
for at least a thousand crossing times.  These simulations show that
the constraints remain well behaved throughout the evolution, indicating
that constraint violations, if any, grow very slowly in time.  For the
black hole cases, where both inner and outer boundaries are present,
these numerical techniques allow for a clean handling of particularly
delicate issues. For instance, defining the difference and dissipative
operators at these boundaries, and how boundary conditions are imposed
(in particular, in non-smooth boundaries) are addressed in a completely
systematic way.

Numerical stability guarantees that errors do go away with resolution,
but at fixed resolution they can still grow quickly in time.  These fast
growing errors can be introduced by the continuum instabilities,
triggered by numerical errors, by the numerical scheme, or any combination
of the above.
The technique explored here~\cite{dyn}, automatically adjusts the
formulation of Einstein's equations in such a way that the discrete
constraint violations follow some prescribed behavior (for example,
their norm remains close to its initial, discretization value). There are
a number of lessons learned from the application of this dynamic
minimization procedure in this paper, which are worth highlighting:
\begin{enumerate} 
\item The semi-discrete picture describes what happens
in the fully discrete case remarkably well, even for cases that are highly
non-stationary. Here by semi-discrete one means a picture that assumes
time to be continuous, but space to be discretized with an arbitrary
(not necessarily small) grid spacing.  
\item The technique of~\cite{dyn}
can be used not only as a practical tool for extending the lifetime of
the simulations, but also for gaining conceptual insight in the problem
of constraints violations in free evolutions. 
Sometimes very large adjustments must occur on short time scales 
in order to control the constraints, which may become too  
fast or large for a fixed Courant factor. 
There are two issues related to this observation:
\begin{enumerate} 
\item Very likely this same feature is present in many
other cases, where different formulations of the equations and numerical
techniques are used. It clearly points out a limitation in adjusting
the equations so as to minimize the constraints growth, independently
of what the adjustment technique is.  
\item Nevertheless, the identification of  this
limitation points out a way to proceed with the technique of
\cite{dyn}.  Namely, to take advantage of many-constraint-function formulations by
redefining the equations in a way such that not only a given behavior for
the constraints is achieved, but also the adjustment varies as little
as possible between two time steps, as done in Section \ref{runs}.
\end{enumerate} 
The results of Section \ref{runs} confirms to a good
extent the validity of the previous discussion.  
\end{enumerate}

\begin{figure}[ht]
\begin{center}
\includegraphics*[height=6cm]{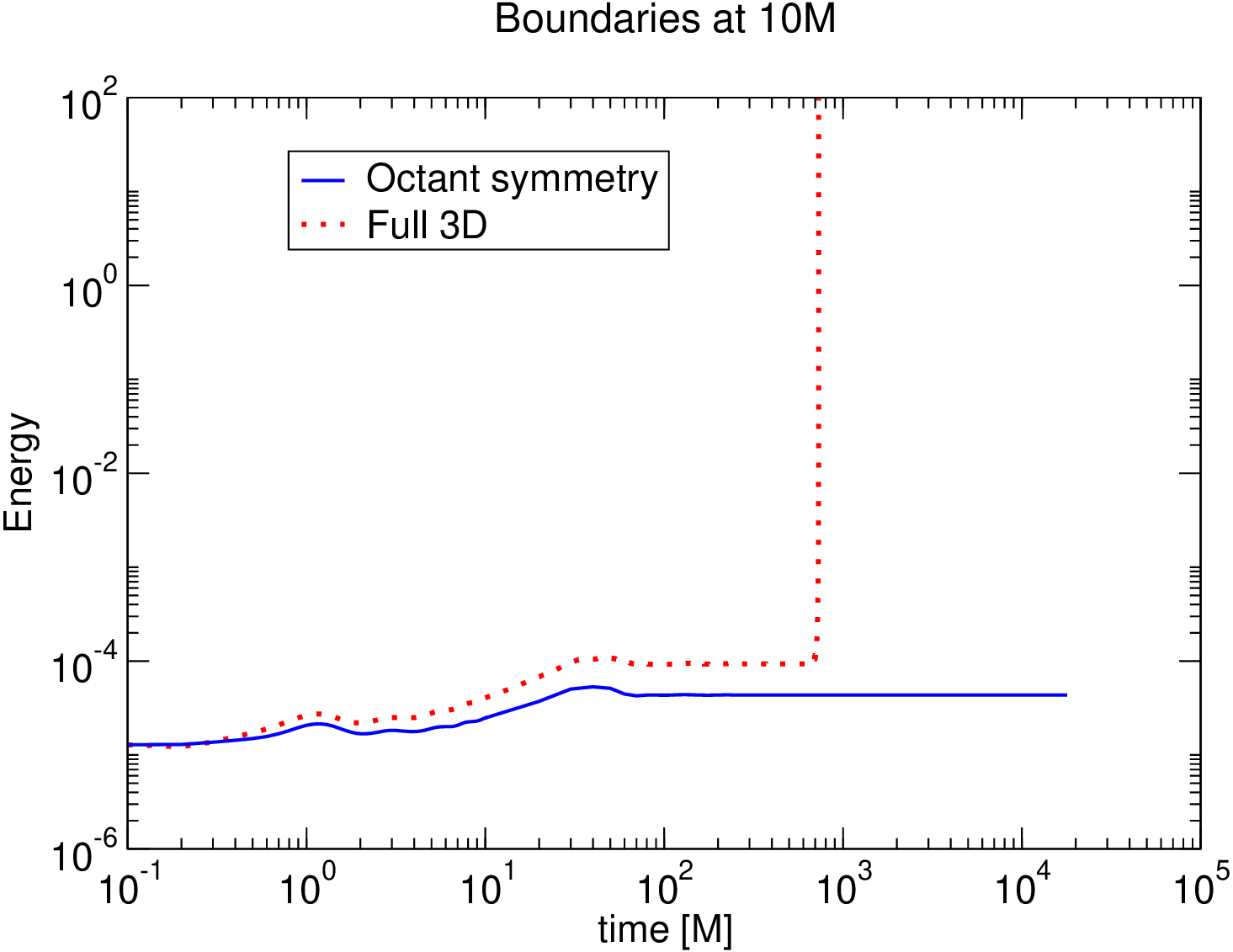}
\caption{This Figure shows a fully $3D$ run with fixed and fined-tuned parameters, given 
by Eq.(\ref{pars10m}), compared to the same run in octant symmetry, which was stopped at 
$18,000M$.}
\label{octant} 
\end{center}
\end{figure}

As a practical matter the lifetime of the full 3D black hole simulations
are extended from around $20M$  up to $700M-1000M$. This is
achieved without employing symmetry restrictions (like octant, bitant or
any  other), or previous knowledge of the expected solution. When employing symmetries 
the code actually runs much longer. For example, Figure \ref{octant} shows a fully $3D$ 
simulation with boundaries at $10M$, 
using the values around which the parameters settle down after a while, 
given by Eq.(\ref{pars10m}) (that is, one of the simulations of 
Fig.\ref{comparisonb10m}), compared to the same simulation in 
octant symmetry, which for convenience was stopped at $18,000M$.

At this point it is not clear why the code crashes after $700M-1000M$. One
possibility could be the presence of a numerical instability caused by
the inconsistency of having an inner boundary that is not purely outflow,
but is treated as if it was. Although we have done some convergence
tests, numerical instabilities are sometimes subtle, and very detailed
and careful convergence tests have to be done in order to detect them,
especially when the initial data has few and low frequencies (as is
here the case) \cite{stability}. However, an important feature of {\em
all} the simulations here presented,  {\em including those of Section
\ref{runs}}, is that large values of $n_a$ had to be used in order
to prevent large and quick variations in the constraint-functions. Therefore,
one {\em is not} completely controlling the constraints---for that a
value of $n_a$ of order one would have to be used---and they {\em do}
grow. Therefore, one possibility for achieving small values of $n_a$
without large and quick variations in the constraint-functions would be to introduce
more free constraint-functions and to make use of this extra freedom as in Section
\ref{runs}. The results of Section~\ref{runs} strongly suggest  that
this should extend the lifetime even more, but more work must 
be done in order to explicitly study this.
Finally, the constraint minimization method is designed to dynamically
control the constraints growth without any a priori knowledge of the
solution. Therefore, a natural next step also seems to be an application
to dynamical spacetimes.

\section{Acknowledgments}
This research was supported in part by the NSF under grant numbers
PHY0244335, PHY0326311, PHY0244699, PHY9800973 and
INT0204937 to Louisiana State University, PHY9907949
to the University of California at Santa Barbara and by the Horace Hearne
Jr. Laboratory for Theoretical Physics. The authors thank the KITP at the
University of California at Santa Barbara for hospitality. L.L. and
M.T. thank the Caltech Visitors Program in numerical relativity
for hospitality where the initial stages of this work began. 
L.L. has been partially supported by the Alfred P. Sloan Foundation. 
This work has been supported in part through LSU's Center for Applied
Information Technology and Learning, which is funded through Louisiana
legislative appropriation. 
Computations were done at LSU's Center for Applied Information Technology
and Learning, and parallelized with the Cactus Computational 
Toolkit~\cite{cactus}. The simulations presented also used the publicly available 
apparent horizon finder AHFinderDirect, from the CactusEinstein thorn. 

We thank Jorge Pullin and Olivier Sarbach for helpful discussions
throughout this project; Peter Diener for an interface with the
CactusEinstein thorn, and for help in running the apparent horizon
finder; Jonathan Thornburg for discussions and
help with AHFinderDirect; and Monika Lee, Joel Tohline and Ed Seidel 
for their support from LSU's Center for Computation and Technology (CCT). In addition, 
this paper made use of several analytical results, due to Olivier Sarbach,
regarding symmetrizability of the equations and characteristic variables
decomposition, for the formulation presented in Ref. \cite{st}, that had not been worked
out in that reference; we thank him for working out those results and
making them available to us for this work. Finally, we thank Ian Hawke, Carlos Palenzuela, Jorge Pullin, Oscar Reula, Olivier Sarbach, 
 and Erik Schnetter for comments on the manuscript.


\end{document}